# Laser engineering of biomimetic surfaces

E. Stratakis[1,2], J. Bonse[3], J. Heitz[4], J. Siegel[5], G. D. Tsibidis[1], E. Skoulas[1,6], A. Papadopoulos[1,6], A. Mimidis[1,6], A.-C. Joel[7], P. Comanns[7], J. Krüger[3], C. Florian[5], Y. Fuentes-Edfuf[5], J. Solis[5], W. Baumgartner[8]

[1] *Institute of Electronic Structure and Laser, Foundation for Research and Technology - Hellas, Heraklion, 71110 Crete, Greece*

[2] *Department of Physics, University of Crete, 70013 Heraklion, Greece*

[3] *Bundesanstalt für Materialforschung und −prüfung (BAM), 12205 Berlin, Germany*

[4] *Institute of Applied Physics, Johannes Kepler University Linz, 4040 Linz, Austria*

[5] *Laser Processing Group, Instituto de Optica (IO-CSIC), Consejo Superior de Investigaciones Científicas, CSIC, 28006 Madrid, Spain*

[6] *Department of Materials Science and Technology, University of Crete, 70013 Heraklion, Greece*

[7] *Institute for Zoology, RWTH Aachen University, 52074 Aachen, Germany*

[8] *Institute of Biomedical Mechatronics, Johannes Kepler University Linz, 4040 Linz, Austria*

## Abstract

The exciting properties of micro- and nano-patterned surfaces found in natural species hide a virtually endless potential of technological ideas, opening new opportunities for innovation and exploitation in materials science and engineering. Due to the diversity of biomimetic surface functionalities, inspirations from natural surfaces are interesting for a broad range of applications in engineering, including phenomena of adhesion, friction, wear, lubrication, wetting phenomena, self-cleaning, antifouling, antibacterial phenomena, thermoregulation and optics. Lasers are increasingly proving to be promising tools for the precise and controlled structuring of materials at micro- and nano-scales. When ultrashort-pulsed lasers are used, the optimal interplay between laser and material parameters enables structuring down to the nanometer scale. Besides this, a unique aspect of laser processing technology is the possibility for material modifications at multiple (hierarchical) length scales, leading to the complex biomimetic micro- and nano-scale patterns, while adding a new dimension to structure optimization. This article reviews the current state of the art of laser processing methodologies, which are being used for the fabrication of bioinspired artificial surfaces to realize extraordinary wetting, optical, mechanical, and biological-active properties for numerous applications. The innovative aspect of laser functionalized biomimetic surfaces for a wide variety of current and future applications is particularly demonstrated and discussed. The article concludes with illustrating the wealth of arising possibilities and the number of new laser micro/nano fabrication approaches for obtaining complex high-resolution features, which prescribe a future where control of structures and subsequent functionalities are beyond our current imagination.

# 1 Introduction

The study and replication of biological systems is popularly known as biomimetics - a combination of the Greek words ‘bios’, meaning life, and ‘mimesis’, meaning to imitate. Nature offers a diverse wealth of functional surfaces, which properties are unmatched in today’s artificial materials. Such solutions came as a direct consequence of evolutionary pressure, which forces natural species to become highly optimized and efficient. The adaptation of natural methods and systems into synthetic constructs is therefore desirable, and nature provides a unique source of working solutions, which can serve as models of inspiration for synthetic paradigms. In this context, a highly interdisciplinary field of research developed concerning the design, synthesis, and fabrication of biomimetic structures, based on the ideas, concepts, and underlying principles developed by nature. Biomimetic materials provide innovative solutions for the design of a new generation of functional materials and can lead to novel material design principles.

Currently, a large area of biomimetic research deals with water repellency, self-cleaning, drag and friction reduction in fluid flow, energy conversion and conservation, adhesion, aerodynamic lift, composite materials with high mechanical strength, antireflection, structural coloration, thermal insulation, antifouling, antibacterial and self-healing properties. All these exceptional functionalities are demonstrated by natural systems and are based on a variety of ingenious designs of biological surfaces, achieved through a sophisticated control of structural features at all length scales, starting from the macroscopic world down to the finest detail, right down to the level of atom. Therefore, natural surfaces are organized in a rather complex manner, exhibiting hierarchical structuring at all length scales.

In this context, several methodologies have been developed to facilitate the formation of bioinspired surfaces exhibiting hierarchical structuring at length scales ranging from hundreds of nanometers to several microns. Laser processing excels over mechanical, chemical, and electric discharge texturing as it allows local modifications with a large degree of control over the shape and size of the features, which are formed, and a broader range of sizes, which can be fabricated. Besides this, laser structuring techniques can be readily incorporated to computer aided design and manufacture systems for complex and customized surface texture designs and subsequently reproducible and cost-effective fabrication. This can give rise to a versatile class of laser-based rapid prototyping texturing systems that could potentially be commercialized for mass production and thus attract considerable attention in the following years.

This article reviews the current state of the art of laser processing methodologies used for the fabrication and engineering of biomimetic surfaces to realize extraordinary optical, mechanical, chemical, wetting, biological-active and tribological properties for numerous applications. In parallel, the biological principles behind the functionalities exhibited by the natural surface archetypes will be analyzed and discussed. Besides presenting the potential and significance of the laser based biomimetic surface structures, it will also delineate existing limitations and discuss emerging possibilities and future prospects.

# 2 Learning from nature: Design principles of natural surfaces

## 2.1 Optical properties

There are very few examples, where color plays no role in the life of an animal or plant. For example for troglobionts (i.e. cave dwellers) colors (as well as eyes) play no role as there is no or little light to reflect [1–3] .

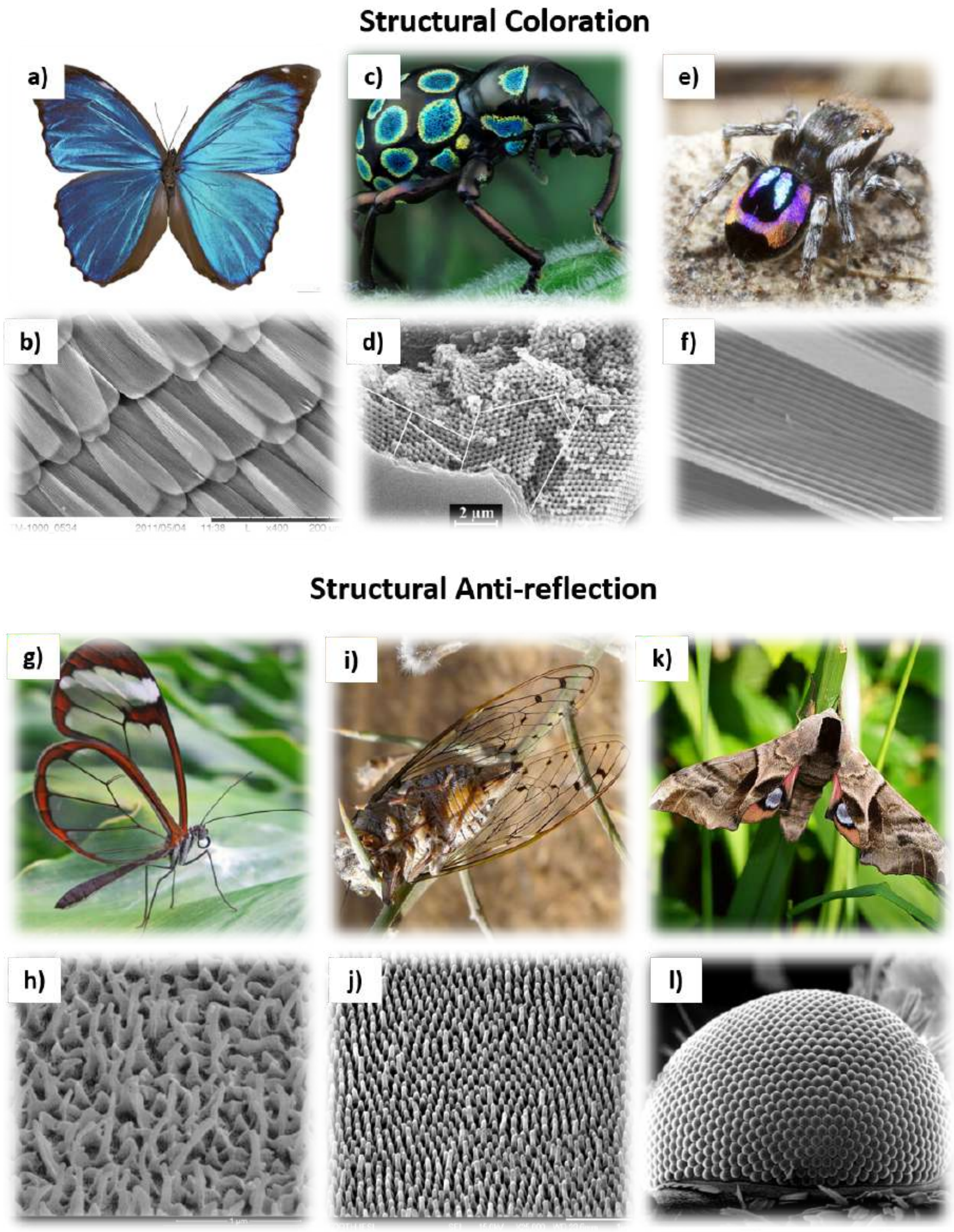

Fig. 1: Examples of biological systems with optical properties, photographs of the actual animal and the corresponding SEM image below. The *Morpho* butterfly with the scales of its wing (a-b), the snout weevil and the ultrastructure of the elytra (i.e. modified forewings) (c-d), as well as the peacock spider and its iridescent scales (e-f) show structural coloration. Structural antireflection was proven for the glasswing butterfly (g-h), a cicada with the nanopillars on its wing (i-j), and the moth's eyes (k-l). Sources: (a), licensed by Didier Descouens / CC-BY-SA-4.0 (Wikimedia Commons). (b), reprint permission from Rashmi Nanjundaswamy / Lawrence Hall of Science CC BY-NC-SA 3.0 US. (c), printed with permission from Javier Rupérez, Spanish photographer specializing in extreme macro photography (www.javier-ruperez.com), (d),reprinted from ref [4] with permission from American Physical Society. (e) and (f), reprinted with permission from [5] *Springer Nature* (Creative Commons

CC BY 4.0). (g), licensed by David Tiller CC BY-SA 3.0 (Wikimedia Commons). (h) reprint with permission from ref [6], Licensed by Radwanul H., CC BY-SA 3.0 (Wikimedia Commons). (i), licensed by Gail Hampshire CC BY-SA 2.0 (Wikimedia Commons). (j) reprint with permission from ref [7] by *John Wiley & Sons.* (k), licensed by Ben Sale CC BY-SA 2.0 (Wikimedia Commons). (l), Dartmouth college.

On the contrary, through natural selection and evolutionary pressure nature offers multifarious sub-micrometer surface morphologies producing colorful structures playing a crucial role on the species survival [8]. Numerous of biological systems have been studied from the ancient times until the present for their ability to utilize light for their own advantage. Coloration arising from living organisms was mentioned ~2000 years ago from Aristotle [9], intrigued Robert Hooke in 1665 to study insects, plants and cells in "Micrographia" [10], while the peacock's feather coloration is mentioned in 1704 Newton's "Opticks" for reflection, infections and colors of light [11] and finally the "Animal coloration" in 1892 from F.E. Beddard was the first dedicated study to deal with the ways that biological systems produce color [12]. The most prominent functionalities of biological light manipulation are signaling, e.g. to attract conspecifics [13] and pollinators or to scare off predators [14–18], drably colors are often used as disguise and camouflage [19,20], while depending of the ambient characteristics their "optical necessity" may vary and can be attributed as extreme light absorption [21], anti-reflection [22] or selective reflection (iridescence) [23,24].

The ability of living organisms to produce color can be attributed to light interference, scattering (structural coloration), selective wavelength absorption (pigmentation), or both. Structural coloration can be produced by the interaction between light and nanometer-scale variation in the integumentary tissues. The most prominent mechanisms responsible for producing structural color or anti-reflection have been summarized in several reviews [25–29]. Tyndall or Mie scattering, which favors the redirection of short-wavelength radiation, is an example of such filtering by extremely small (subwavelength) boundaries. Interference from variation of the optical path due to reflections on periodic or pseudo-periodic tissue formation is responsible for iridescence similar to grating effects. Furthermore, the gradient refractive index that can be produced collectively from subwavelength three-dimensional nanostructures in the main physical phenomena that nature utilizes for extremely low light reflection and high transmissivity [30]. Fig. 1 presents some of the most important biological paradigms for structural coloration and structural antireflective optical properties, which are discussed in the following sections.

The physical mechanisms, which describe most of the optical phenomena found in natural organisms, are summarized in Fig. 2. Most anti-reflective architectures in nature are not attributed to a layered thin film with a refractive index ($n$) on a substrate with ($n_s$), where $n_s > n$ (Fig. 2A-a). On the contrary, species utilize surface morphological features like barbule microstructures on the feather birds of paradise (Fig. 2A-b) to induce light trapping by multiple reflections, leading to a very black color [31,32]. Micro- and nanostructure arrays like the moth's eye or cicada's wing can gradually reduce the refractive index from the material (wing) ($n_2$) to the refractive index of air ($n_1$) resulting on extremely low reflectance values (Fig. 2A-c-d). J.C.M. Garnett [33] and D.A.G. Bruggeman [34] considered simple models like "effective medium theory" to analyze rough surfaces. A layer with microscopic surface roughness can be considered as multiple layers of the ''effective medium'' having refractive index in the limit of the substrate and ambient (air in most cases). The "effective refractive index" ($n$) of the ''effective medium'' can be approximated from the volume fractions ($f$) of the individual rough layers. Assuming the ''effective medium'' as a set of two layers, the Maxwell Garnett model, predicts that the effective ($n$) for a layer with $n_2$ surrounded by the other layer with $n_1$ will be given by:

$$\frac{(n^2 - n_1^2)}{(n^2 + 2n_1^2)} = (1 - f_1)\frac{(n_2^2 - n_1^2)}{(n_2^2 + 2n_1^2)} \qquad \text{Eq. 1}$$

where $n_1$ and $n_2$ are the refractive indexes of two constituent layers, $f_1$ and $f_2$ $(= 1 - f_1)$ are the corresponding volume fractions. While at the Bruggeman model, ''effective medium'' is assumed a homogeneous mixture of two constituent layers, where:

$$f_1\frac{(n_1^2 - n^2)}{(n_1^2 + 2n^2)} + f_2\frac{(n_2^2 - n^2)}{(n_2^2 + 2n^2)} = 0 \qquad \text{Eq. 2}$$

Other optical processes related to structural colors associate with single and multiple film interference. Let a plane wave of light be incident on a thin film with thickness $d$ and refractive index $n_b$ and the angles of incidence and refraction as $\theta_a$ and $\theta_b$ (Fig. 2B). Using the soap-bubble case, the condition for constructive interference becomes:

$$2n_b d cos\theta_b = (m - 1/2)\lambda \qquad \text{Eq. 3}$$

where $\lambda$ is the wavelength with the maximum reflectivity and $m$ is an integer. Moreover, multilayer thin film interference is a pair of thin layers stacked periodically. Assuming we have two layers, A and B with the corresponding thicknesses $d_A$ and $d_B$, and refractive indexes, $n_A > n_B$, respectively, as shown in Fig. 2C. If we consider a certain pair of AB layers, the phases of the reflected light both at the upper and lower B–A interfaces change by 180°. Thus, a relation similar to the anti-reflective coating of eq. (3) is applicable as:

$$2(n_A d_A \cos\theta_A + n_B d_B \cos\theta_B) = m\lambda \qquad \text{Eq. 4}$$

for constructive interference with the angles of refraction in the A and B layers as $\theta_A$ and $\theta_B$.

Fig. 2E presents schematic for incoherent and coherent scattering [35]. Incoherent scattering refers to individual scatterers, which are independent of the phases of the scattered waves. Coloration is affected from the size, refractive index, and shape of the scatterers, as well as the average refractive index of the medium. Rayleigh, Tyndall and Mie scattering mechanisms are prominent examples of incoherent scattering. On the other hand, colors can be produced by coherent scattering with optical phenomena such as constructive interference, reinforcement, and diffraction. The coloration is determined by the refractive index, spatial distribution, and size of light scatterers and are unlike the incoherent scattering it is directly dependent upon the phases of the scattered waves.

Natural photonic crystals are optical materials that manipulate the flow of electromagnetic waves by multiple Bragg scattered interferences defined by Bloch modes. In most species, like butterflies, they are typically composed of a matrix of chitin (high refractive index material) containing regularly arranged spherical air spaces. A number of artificial photonic crystals, including multi-dimensional architectures, have been designed and developed during the past more than two decades [36,37]. Examples that can be used to compare with the complex biological models mentioned in the section below are of one-, two-, and three-dimensional photonic crystal structures as shown in Fig. 2F.

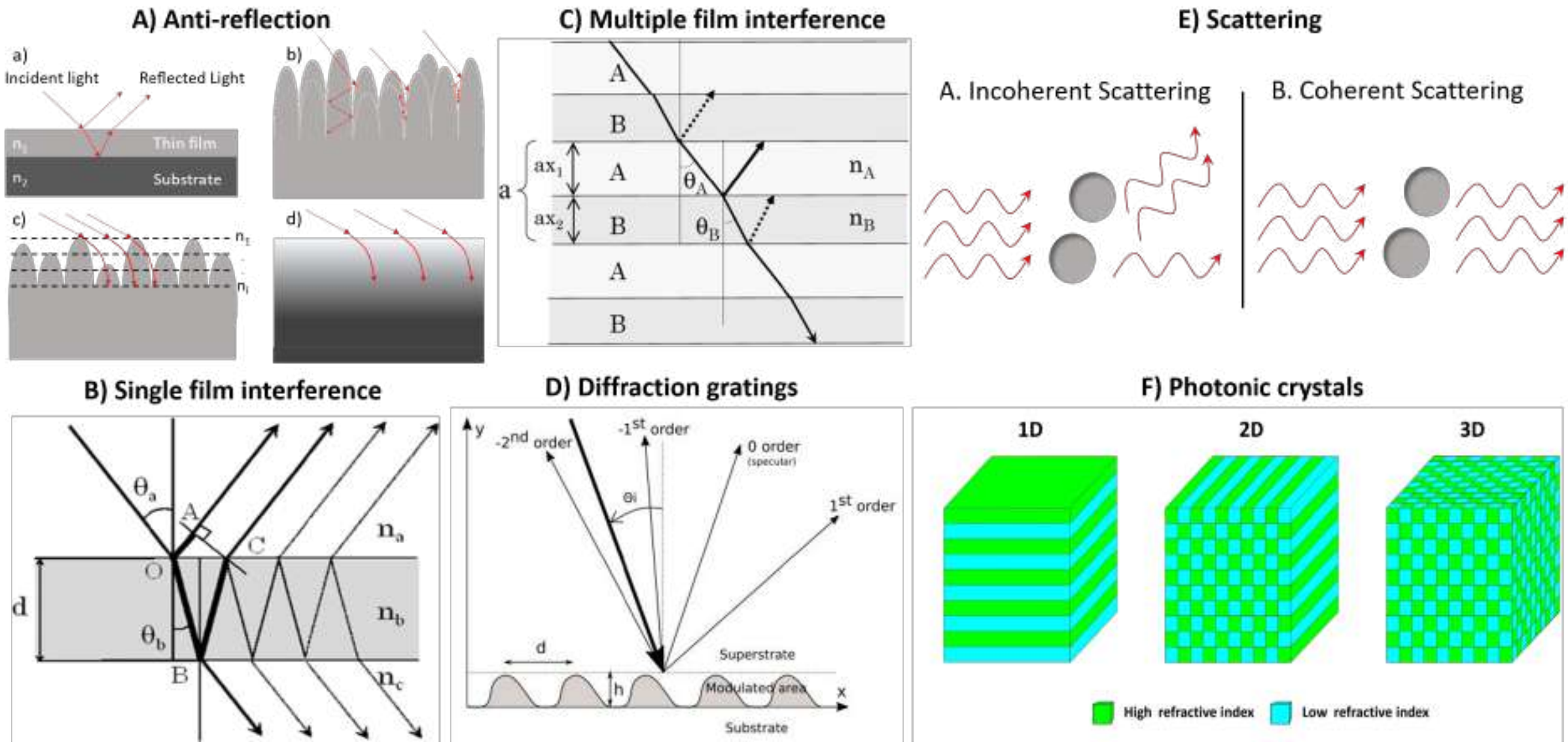

Fig. 2: Physical mechanisms responsible for anti-reflection and structural coloration. (A): Describing the light propagation through a single layer film on a substrate with ($n_s > n$), (a), multiple internal reflections of incident light in a microstructure array (b), the interaction of incident light with the subwavelength-size structures (c), gradient refractive index change (d). Image (A) taken from ref. [38]. (B) & (C): Single and multiple film interference. Images taken from [28] reprinted with permission from *IOP Publishing*. (D): Diffraction grating interaction with incident light. Image reproduced from ref. [39]. (E): Light coherent and incoherent scattering from ref. [35]. (F): Multi-dimensional photonic crystal structures image reprinted from ref [40] with permission from *Elsevier*.

### *2.1.1 Structural coloration*

The dragonfly *Orthetrum cledonicum* is a fine example of coloration via light scattering in the visible, as it contains sub-wavelength particles located in the cytoplasm on their epidermal cells resulting in a permanent blue coloration and red when viewed in transmission [27]. The exact shade of blue dependents only on the particle size. Another case can be observed when studying the *Morpho* butterfly, one of the best studied biological models for structural coloration produced by multi-layer reflections on the 3D structures of its scales [41]. However, sophisticated diffraction gratings are remarkably widespread throughout nature. In latest studies, peacock spiders such as *Maratus robinsoni* and *Maratus chrysomelas* have hairs with 2D nanogratings on microscale 3D convex surfaces, which can yield at least twice the resolving power of a conventional 2D diffraction grating with the same characteristics [5]. Remarkable tuning of structural colors are also produced due to a 3D photonic crystal network of chitin in air with a single diamond (Fd-3m) symmetry found on the elytra (i.e. modified forewings) of snout weevil, *Pachyrrhynchus congestus pavonius* [42].

Besides tunable structural coloring effects, non-iridescent colors in nature are produced by coherent scattering of light by quasi-ordered, amorphous photonic structures can be found on tarantulas *Poecilotheria metallica* and *Lampropelma violaceopes* [43]. Simple photonic structures attained by biological systems such as the sea mouse Aphroditidae (Polychaeta) reflect the complete visible spectrum over a range of small incident angles with a reflectivity of 100 % to the human eye [23]. Much more complex photonic architectures have been studied in butterflies, which have a three-dimensional periodicity, frequently enhanced by secondary structures which perform a cruder, more generalized, light-scattering or directing role and, thereby, modify the photonic crystal's effects [44]. The diversity and functionality of structures responsible for all these effects is a remarkable example of photonic engineering by living organisms.

### ***2.1.2*** *Structural anti-reflection*

Although most of the natural systems are associated with vivid color or broad angle reflectivity, nature has also provided nanostructures for extremely low reflectivity without transmission loses over broad angles or frequency ranges, or even both. The majority of biological systems gifted with these abilities are insects, which benefit from anti-reflective surfaces either on their eyes for night vision or on their wings to eliminate reflections for the purpose of camouflage and keep their body warm during the day.

Most prominent examples are the moth's and butterfly's eyes and the transparent wings of cicadas. In these cases, antireflection is achieved by using subwavelength, nanosized three-dimensional surface architectures. These structure architectures form a gradient refractive index and transmits light with extremely low loses. For instance, the surface of a moth's eye comprises conical nodules with rounded tips arranged in a hexagonal array with variable spatial characteristics depending on the species [16].

Cicada's wing structure is different to those of the moth's eyes, as they resemble nanopillars. The cuticles of such insects consist of self-assembled polysaccharides (i.e. chitin) and proteinaceous materials chitin and its derivatives. Their spatial characteristics range in height from approximately 100 to 340 nm, depending on the species and the location of the structure on the wing [20]. However, their anti-reflective properties are remarkable and for a wide range of angles of incidence [22]. Another profound example of nature anti-reflective structures in the glass wing butterfly *Greta oto* [45]. The random nanostructures on the transparent part of its wing average in distance between the pillars at 120 ± 20 nm and typical height ranging from 160 to 200 nm. Moreover the optical properties of the wing exhibit a stunning low haze and reflectance over the whole visible spectral range even for large view angles of 80° [6].

## 2.2 Wetting

Many animals or plants exhibit surfaces with specialized wetting properties. A combination of surface chemistry and structure (or roughness) causes a spreading or repelling of a liquid. Spreading occurs particularly in animals, which can use their body surface to collect water passively from various sources. This includes arthropods, amphibians, reptiles, birds and even mammals [46–53]. Some of the corresponding surface structures have been transferred to artificial materials using lasers (cf. Table 1).

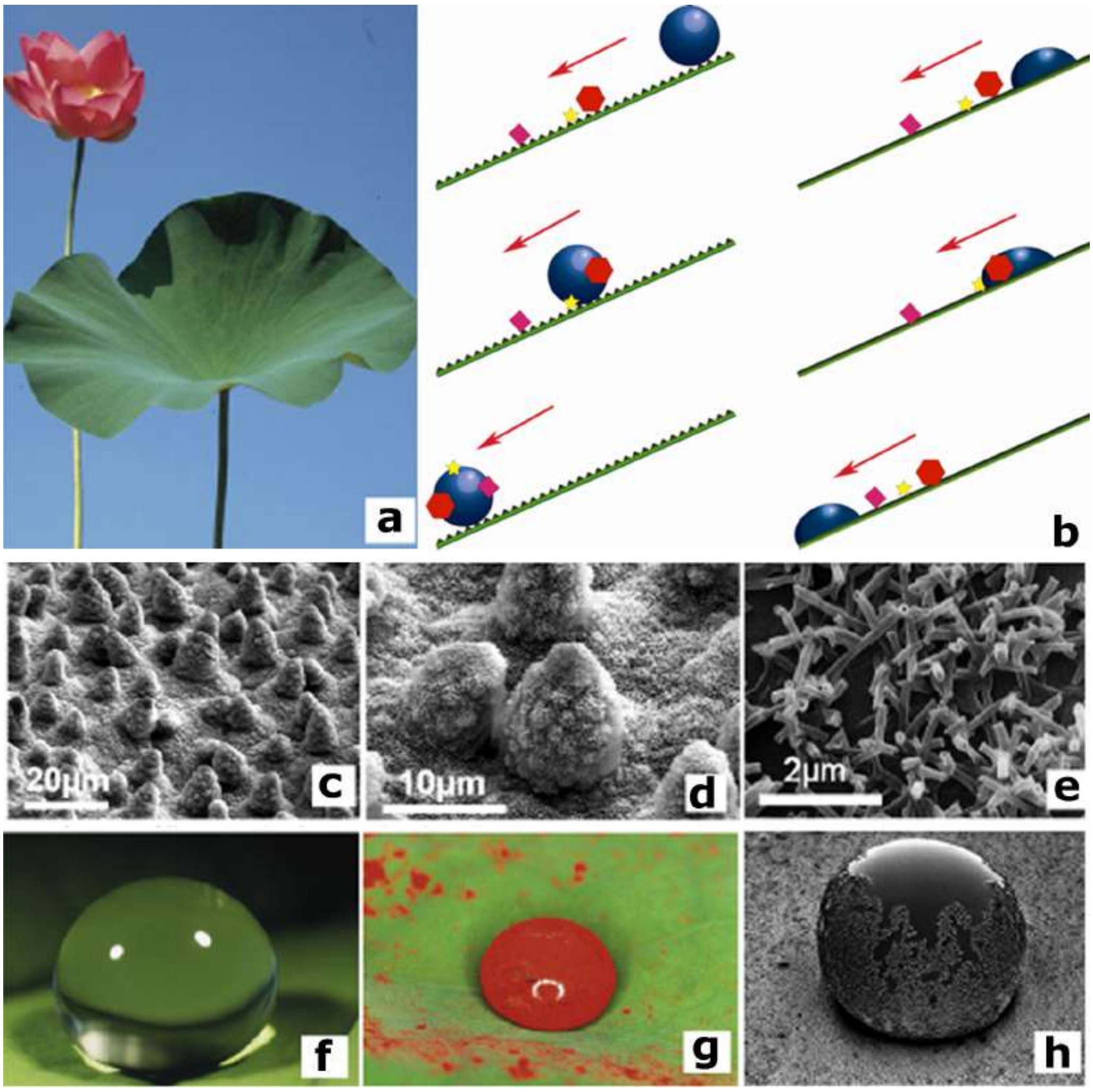

Fig. 3: Water repellence and self-cleaning property of the Lotus surface. A flowering plant of Lotus (*Nelumbo nucifera*) is shown in (a); (b) Schematic representation of the motion of a droplet on an inclined nanostructured superhydrophobic surface covered with contaminating particles (lotus effect. As the droplet rolls off the surface it picks up the particles and hence cleans it (left). On the contrary, in the case of a smooth surface the particles are only redistributed by the moving droplet (right). The SEM micrographs (c–e) show the Lotus leaf surface in different magnifications: (c) randomly distributed microsized cell papilla; (d) a detail of the cell papilla and (e) the epicuticular nanosized wax tubules on the cells. In (f), a spherical water droplet on a superhydrophobic leaf is shown. In (g) lipophilic particles (Sudan-red) adhere on the surface of a water droplet, rolling over the Lotus leaf. The SEM micrograph of a droplet illustrates the superhydrophobic property of the leaf surface (h) (reproduced *from* [54])

In the leaves of water-repellent plants, the cuticle is a composite material mainly built up by a hydrophobic polymeric matrix, called cutin and superimposed waxes. Water repellence has been qualitatively and sometimes quantitatively attributed to not only the chemical constituency of the cuticle covering their surface, but, even more importantly, to the specially textured topography of the surface. It is understood that the micro- and nano-structured rough surface enhances the effect of surface chemistry into super-hydrophobicity and water repellency. The super-hydrophobic property of such leaves is also related to reduced particle adhesion, namely the ability to remain clean after being immersed into dirty water, known as the self-cleaning property. This ability is excellently demonstrated by the leaves of the Lotus (*Nelumbo nucifera*) plant, which are untouched by the pollution or contaminants although it grows in muddy waters. Hence, in several oriental cultures the Lotus plant is considered as "sacred" and is a symbol of purity. Scanning electron microscopy (SEM) images of the Lotus leaf surface, shown in Figs. 3c-e, reveal a dual scale roughness created by papillose epidermal cells and an additional layer of epicuticular waxes. The roughness of the papillae leads to a reduced contact area between the surface and a liquid drop (or a particle), with droplets residing only on the tips of the epicuticular wax crystals on the top of the papillose epidermal cells [54]. Thus, droplets cannot penetrate into the structure grooves, and air pockets are formed between the water and the plant's surface. Contaminating particles can thus be picked up by the liquid and carried away as the droplet rolls off the leaf. This was coined the "Lotus-Effect". A schematic representation of this effect is shown in Fig. 3b, while images of water droplets with contaminants are presented in Figs 3f-h.

As far as biological implications of the Lotus effect, it is suggested that self-cleaning plays an important role in the defense against pathogens bounding to the leaf surface. Many fungal spores and bacteria require water for germination and can infect leaves in the presence of water. Therefore, water removal minimizes the chances of infection[54]. In addition, dust particle removal from leaf surfaces minimizes the changes of, for example, the plant overheating or salt injury. Although the Lotus leaf has been used as a model surface for water repellence and self-cleaning, many other biological surfaces are found to exhibit similar properties belonging both in flora and fauna families[54]. A common feature among those surfaces is that the special wetting characteristics come as a direct consequence of the synergy of micro- and nano-structured morphology and hydrophobic surface chemistry.

Flat bugs live on bark of trees and highly rely on their camouflage appearance. Some species, living in tropical South America, can cope with moisture induced color change of the bark. Unlike most other insects, they have a highly wettable body surface and change color when wetted. This unique mechanism of camouflage, i.e. a rapid and passive spreading of water over the body surface, results from almost superhydrophilic wetting properties. Here, hydrophilicity is enabled mainly by significant amounts of the chemical component erucamide in the surface wax layer and is enhanced by pillar-like surface microstructures [55–60].

Two lizard species of so-called moisture harvesting lizards have been found to transport water directionally over their skin surface to the mouth for drinking [56,57]. The skin of all lizards is covered by small scales, and moisture harvesting lizards exhibit capillary channels between these scales for water transport (Fig 4a-b). Here, directionality is enabled by an asymmetric geometry and specific interconnections that form a network of channels [56] (Fig 4c,d). Abstracted designs of such lizard skin structures have been transferred to metal surfaces using laser ablation [56,61] and polymer surfaces have been structured using a $CO_2$-laser [62]. Several studies have analyzed the capillary structure and resulting fluid transport [61–63].

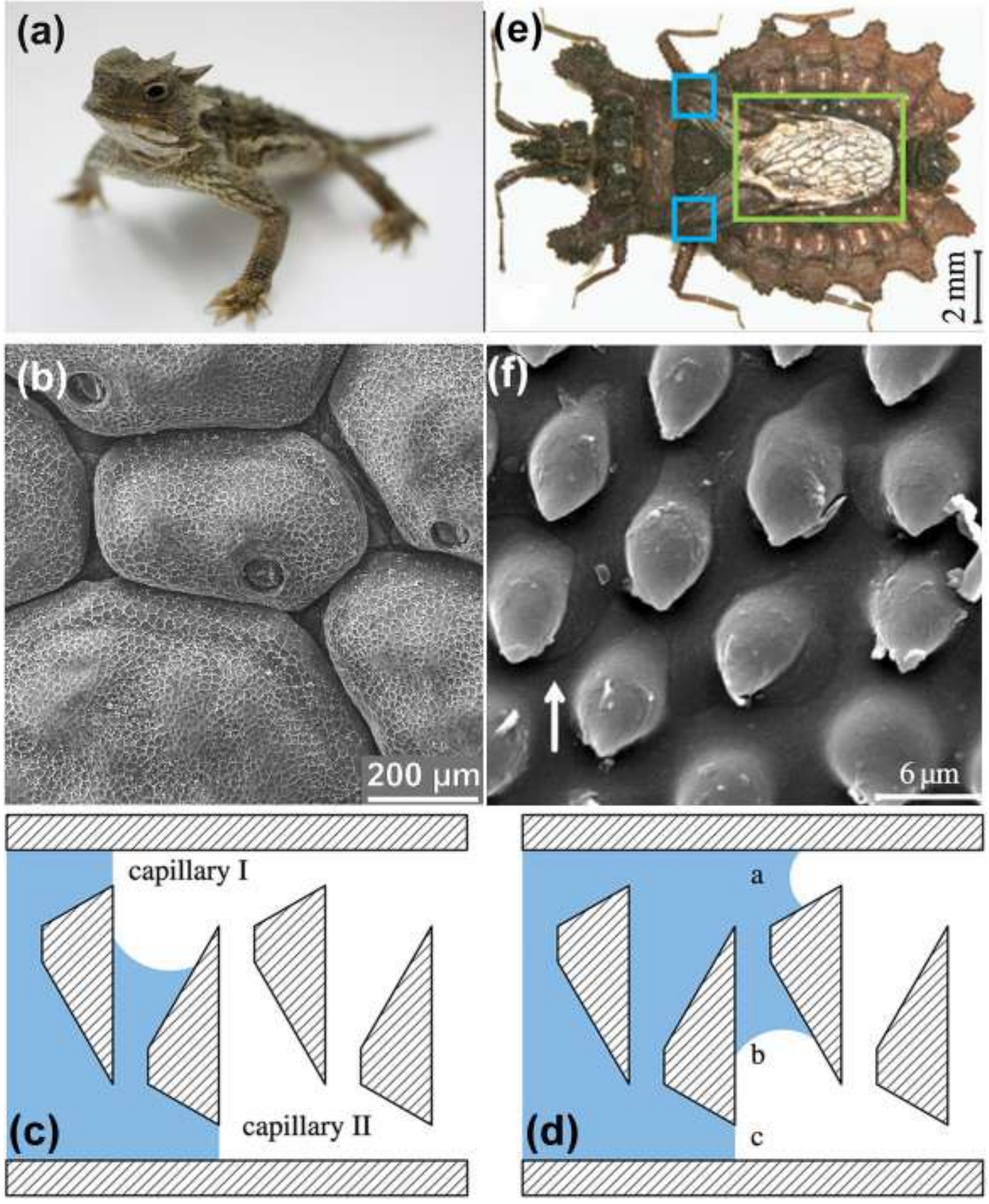

Fig. 4: (a) The moisture harvesting lizard *Phrynosoma platyrhinos*; (b) SEM image of dorsal scales from *P. platyrhinos*; (c,d) Principle of 'interconnection' for two saw-tooth-shaped capillary channels

(flow direction: from top to bottom). (c) An applied droplet is soaked into the structure by capillary forces. The liquid front stops at the sharp edges in capillary I, while it is transported farther into capillary II. As the liquid front in capillary II reaches the nearby interconnection, the liquid is transported into capillary I; (d) The liquid coming through the interconnection picks up the stopped liquid in capillary I and forms a new free liquid front. Thereafter, the liquid is transported through a second interconnection into capillary II, where the stopped liquid is picked up. (e) Optical image of the flat bug, *Dysodius lunatus*. The green rectangle highlights the wings and blue squares the region where an oily defensive liquid is thought to evaporate; (f) The micro-structured region under the wings, where drop-like microstructures can be observed (white arrow shows the preferential flow direction). Images (a), (b) are reprinted with permission from Ref. [64]. (c), (f) are licensed under CC-BY 4.0

Directional fluid transport also takes place in some bugs that use a defensive secretion to scare off their predators (Fig. 4e). The fluid is secreted in scent glands under their wings and then transported directionally in small channels to evaporation sides [58]. In the channels, directionality is facilitated by asymmetric protrusions with specific droplet shape and tips pointing in the direction of fluid transport (Fig. 4f). The distance between these protrusions also contributes to the directional fluid transport as a backflow is inhibited [58]. Recently, Plamadeala et al. fabricated such microstructures using two-photon polymerization [65].

Directionality can also be found in closed channels. For example, a backflow of fluid is omitted in the sperm storage chamber (spermatheca) of some female fleas. Again, asymmetry plays a central role. The shape of the spermatheca and the connection points to adjacent tubes facilitate the directional fluid transport [59]. A scaled-up transfer of abstracted spermatheca structures to polymer surfaces using laser technologies revealed that directionality is maintained also in open channels. Despite regular capillary transport, the transport velocity is relatively constant, based on parts that function as reservoirs [59,66].

### ***2.2.1*** *Surface tension*

The main driving force of all wetting and capillary transport phenomena is the surface energy and respective surface tension. Surface tension is a fundamental property of liquids and solids and described as tension acting perpendicularly to a line in the surface. Surface tension could be defined as the energy required to increase the surface area by one unit. However, the surface tension is defined as force per unit length of a line in the surface is a tensor, whereas specific surface free energy is a scalar thermodynamic property of an area of the surface [67]. For liquids in equilibrium, surface tension is numerically equal and physically equivalent to the specific surface free energy [67].

It is important noticing that the resulting force acting on a molecule in the bulk and at the interface equals zero as both are in equilibrium [68]. However, an increase in the liquid/gas surface causes an increase of the number of interface-molecules leading to a growth of the surface energy. Consequently, in order to diminish the number of interface molecules, i.e. to minimize the energy in the system, liquids tend to minimize the free surface.

Let the potential describing the pair intermolecular interaction in the liquid be *U(r)*. The surface tension, $\gamma$, could be estimated assuming the local radius of curvature of the liquid surface to be much larger than the molecules diameter $d_m$ to be

$$\gamma \approx \frac{N}{2}\frac{|U(d_m)|}{d_m^2} \qquad \text{Eq. 1}$$

where *N* is the number of directly neighboring molecules (the factor ½ is due to the lack of neighbors outside the liquid, i.e. in the gas phase) and $d_m$ is the diameter of a liquid molecule. Obviously, the intermolecular potential *U(r)*, especially $U(d_m)$, is a crucial parameter. There are three main kinds of intermolecular interactions influencing this potential: The Keesom-interaction, describing the interaction of permanent dipoles, the Debye-attraction between dipolar molecules and induced dipoles, and London-dispersion interaction. The London dispersion interaction results when the electrons in two adjacent atoms occupy positions by quantum mechanical means that make the atoms form temporary dipoles. These three interactions (Keesom, Debye and London) are collectively termed van der Waals interactions [69].

In solid state the surface tension is not necessarily equal to the surface free energy as the surface stretching tension (or surface stress) is defined as the external force per unit length that must be applied to retain the atoms or molecules in their initial equilibrium positions (equivalent to the work spent in stretching the solid surface in a two-dimensional plane), whereas a specific surface free energy is the work spent in forming a unit area of a solid surface. The relation between surface free energy and stretching tension is shown for example in [69].

### 2.2.2 *Surface Tension*

Wettability is the ability of a liquid to maintain contact with a solid surface, resulting from intermolecular interactions when the two are brought together. Let us assume $\gamma_{SA}$ , $\gamma_{SL}$ , $\gamma = \gamma_{LG}$ are the surface tensions at the solid/gas, solid/liquid and liquid/gas interfaces respectively. When the droplet on a flat surface forms a cap, the line at which solid, liquid and gaseous phases meet is called the triple or (three phase) line.

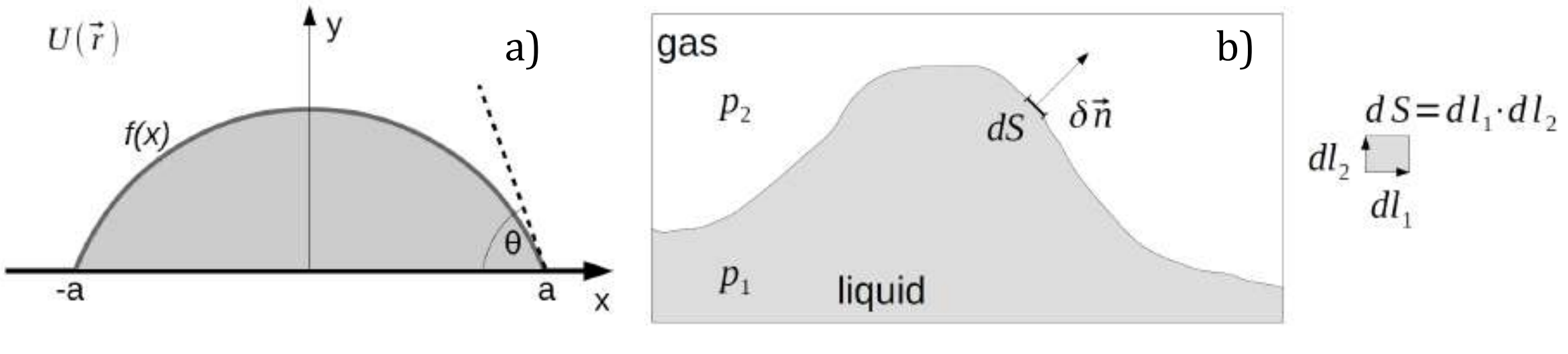

Fig. 5: (a) Cross-section of a spherical droplet deposited on an ideal smooth solid surface. An external field $U(\overrightarrow{r})$ might act on the liquid. The contact angle $\theta$ is measured between the solid surface and the tangent to the liquid, where a liquid–gas interface meets a solid surface. $f$ expresses mathematically the boundary position; (b) Schematic of the interface between a liquid (medium 1) and a gas (medium 2). An infinitesimal surface element $dS$ is observed which undergoes an infinitesimal (virtual) displacement $\delta n$ normal to the surface element $dS$.

For the sake of simplicity, a liquid droplet is considered on an ideal, i.e. atomic flat, chemically homogeneous, isotropic, insoluble, nonreactive and non-deformed solid surface (Fig. 5). It is also assumed that the volume of the droplet remains constant and evaporation does not occur. Then, the following Young equation correlates the above surface tensions with $\theta$ that denotes the contact angle in equilibrium (Young-contact angle) [70]

$$\cos\theta = \frac{\gamma_{SA} - \gamma_{SL}}{\gamma} \qquad \text{Eq. 5}$$

It is noted that this expression is valid in specific conditions. Appropriate modifications in the expression are required if surface inhomogeneity or surface roughness are considered [69]. The effect of the macroscopic roughness on the wettability of surfaces has been theoretically approached by two different models: In the Wenzel model [71], the liquid is assumed to wet the entire rough surface, without leaving any air pockets underneath it. The apparent contact angle, $\theta_w$, is given by the following equation:

$$cos\theta_w = r\cos\theta_o \qquad \text{Eq. 6}$$

where $r$ is the ratio of the unfolded surface to the apparent area of contact under the droplet, and $\theta o$ is the contact angle on a flat surface of the same nature as the rough surface. Since $r > 1$, this model predicts that the contact angle will decrease/increase with surface roughness for an initially hydrophilic ($\theta_o < 90^o$) respectively hydrophobic ($\theta_o > 90^o$) surface. In contrast, Cassie and Baxter (CB) [72], assumed that the liquid does not completely wet the rough surface and air is trapped underneath the liquid. As result, a droplet will form a composite solid–liquid/air–liquid interface with the sample in contact (Fig. 3f,g,h) In order to calculate the contact angle for this heterogeneous interface, Wenzel's equation can be modified by combining the contribution of the fractional area of wet surface and the fractional area with air pockets ($\theta = 180^o$). In this case, the apparent contact angle, $\theta_{CB}$, is an average of the flat surface, $\theta o$, and the value for perfect hydrophobicity (that is, $180^o$) and is given by the equation:

$$\cos\theta_{CB} = r\cos\theta_o - f_{la}\,(1 + r\cos\theta_o) \qquad \text{Eq.}$$

7

where $f_{la}$ is the fractional flat geometric area of the liquid–air interfaces under the droplet. As $f_{la}$ is always lower than unity ($f_{la}+f_{ls}= 1$), this model always predicts enhancement of the hydrophobicity, independently of the value of the initial contact angle $\theta_o$. Thus, even for a hydrophilic surface, the contact angle increases with an increase of $f_{la}$.

The contact angle hysteresis is another important characteristic of a solid-liquid interface that determines the self-cleaning properties. When a droplet sits on a tilted surface (Fig. 3b) the contact angles at the front and back of the droplet correspond to the advancing, $\theta_{adv}$, and receding, $\theta_{rec}$, contact angle, respectively. The advancing angle is greater than the receding angle, which results in contact angle hysteresis occurring due to surface roughness and heterogeneity. Contact angle hysteresis is a measure of energy dissipation during the flow of a droplet along a solid surface. Surfaces with low contact angle hysteresis have a very low water roll-off angle, which is the angle to which a surface must be tilted for a droplet to roll off it. A relationship for contact angle hysteresis as a function of roughness has been derived, given as[73]:

$$\theta_{adv} - \theta_{rec} = \sqrt{(f_{la} - 1)}\, r \frac{\cos\theta_{adv,o} - \cos\theta_{rec,o}}{\sqrt{2r\cos\vartheta_o + 1}} \qquad \text{Eq. 8}$$

For a homogeneous interface $f_{la} = 0$, whereas for a composite interface $f_{la}$ is a non-zero number. It is observed from eq. 7 that, for a homogeneous interface, increasing roughness (high $r$) leads to increasing contact angle hysteresis (high values of $\theta_{adv}$- $\theta_{rec}$), while for a composite interface, an approach of $f_{la}$ to unity provides both a high contact angle and a low contact angle hysteresis. Therefore, a heterogeneous interface is desirable for superhydrophobicity and self-cleaning as it dramatically reduces the area of solid-liquid contact and, therefore, reduces adhesion of a liquid droplet to the solid surface and contact angle hysteresis.

Formation of a composite interface is a multiscale phenomenon that depends on the relative sizes of the liquid droplet and roughness details. Such interface is metastable and can be irreversibly transformed into a homogeneous one, thus damaging superhydrophobicity. Even though it may be geometrically possible for the system to become heterogeneous, it may be energetically profitable for the liquid to penetrate into the valleys between asperities and form a homogeneous interface. Destabilizing factors, such as capillary waves, nanodroplet condensation, surface inhomogeneities and liquid pressure can be responsible for this transition. It has been demonstrated that the mechanisms involved into the superhydrophobicity are scale-dependent with effects at various scale ranges acting simultaneously. Thus, a multiscale, hierarchical, roughness can help to resist the destabilization. High $r$ can be achieved by both micro- and nano-patterns. For high $f_{la}$, nano-patterns are desirable because whether the liquid–air interface is generated depends on the ratio of the distance between two adjacent asperities and droplet radius. Furthermore, nanoscale asperities can pin the liquid-air interface and thus prevent liquid from filling the valleys between the micro-asperities even in the case of a hydrophilic material. Despite numerous experimental and theoretical studies, the effect of the hierarchical roughness on wettability remains a non-clarified issue and a subject of intense scientific discussions.

### 2.2.3 Laplace pressure

One interesting feature that characterizes droplets or bubbles is the pressure difference between their inner and outer regions [69,74,75]. This gives rise to several phenomena including capillary transport of liquids. In Fig. 5b, the mechanism is illustrated in which the media (1 and 2), in this case, a liquid and a gas are separated by a curved interface.

According to Fig.5b, an infinitesimal surface element *ds* is (virtually) displaced by an amount equal to $\delta n$ perpendicularly to the unit vector along the surface. Then, the work which is necessary to perform this displacement is equal to

$$\delta W = \delta W_v + \delta W_s = (p_2 - p_1)\delta n \cdot dS + \gamma \delta S \qquad \text{Eq. 9}$$

where $p_1$ and $p_2$ are the pressure of material in liquid and gas phases, respectively. The work for volume change $\delta W_v = p \cdot dV = (p_2 - p_1)\delta n \cdot dS$ and the work for the surface change are related through the following equation

$$\delta W_s = \gamma \delta S = \gamma \cdot \delta n dS \left(\frac{1}{r_1} + \frac{1}{r_2}\right) \qquad \text{Eq. 10}$$

with $r_1$ and $r_2$ being the main radii of curvature of the surface. In thermal equilibrium the virtual work $\delta W$ must be equal zero, i.e.,

$$\delta W = \delta n \cdot \left[(p_2 - p_1) + \gamma \cdot \left(\frac{1}{r_1} + \frac{1}{r_2}\right)\right] dS = 0 \qquad \text{Eq. 11}$$

that leads to the famous Laplace-formula

$$p_2 - p_1 = p_L = \gamma \cdot \left(\frac{1}{r_1} + \frac{1}{r_2}\right) \qquad \text{Eq. 12}$$

It is noted that in the above approach, the displacement vectors are positive when pointing from medium 1 into medium 2 and the radii are positive when oriented towards the first medium. This means that if a liquid surface is curved in a convex way, a negative pressure (pointing from medium 2 to medium 1) is obtained while for a concave curvature, the positive pressure forces the liquid (medium 1) towards the gas (medium 2). Thus, the Laplace pressure $p_L$ is generally working in a way to flatten the surface if there are no other restrictions.

#### *2.2.4 Capillarity*

Capillarity (capillary motion, capillary effect, or wicking) [69] is the ability of a liquid to flow in narrow spaces without the assistance of, or even in opposition to, external forces like gravity. This can be observed when the combination of surface tension (caused by cohesion within the liquid, leading to a certain Laplace pressure) and adhesive forces between the liquid and container wall (leading to a certain contact angle) act in concert to propel the liquid.

### 2.3 Mechanical

The diverse and often harsh natural habitat populated by many natural organisms has sparked the need for exceptional mechanical properties required to support their body functions. Constituent synthetic materials usually have inferior mechanical response, but the clever combination of materials into composites [76] and the complexity of engineered body patterns [77] vastly improve its properties of interest which helps species survive and thrive in their environment.

#### *2.3.1 Wet and dry adhesion*

Adhesion is a vital property that a plethora of species require to sustain their body weight during anchoring and locomotion on a variety of surfaces [78]. Dry adhesion, in nature, usually involves hierarchical micro- to nanoscale filamentous structures, which decorate many insects and some reptiles enabling them to rapidly climb surfaces. For example, the skin of the *Tokay gecko* (Fig. 6 (1a)) is comprised of a complex hierarchical structure of millimeter long ridges, the lamellae, which are located on its toes and allow compression against rough surfaces [79]. Extending from the lamellae are micrometer sized, densely packed (~14.000/mm$^2$), curved hairs called setae [80] (Fig. 6 (1b) and (1c)). Each seta is further decorated with 100 to 1000 nanometer sized spatulae that emanate from its tip (Fig. 6 (1d)). The attachment pads on the feet of the *Tokay gecko* possess a combined area of approximately 220 mm$^2$ where ~$3\times10^6$ branch out from [81,82]. Eventually, these spatulae reach and adhere via van der Waals bonds [83] even to rough surfaces since they are able to explore their geometry. Owing to these intricate multiscale formations and their large numbers, a clinging ability of approximately 20 N can be reached from the gecko's pads [78]. The interpretation of the strong dry adhesion of the geckos is based on the contacts mechanics Johnson-Kendall-Roberts (JKR) model introduced by [84]. According to this model, the splitting of a single contact into multiple smaller ones will always result in a stronger overall adhesion. More specifically, if a seta is considered having a hemispherical tip with radius $R$, and the adhesion work per unit area is $W_a$, then the predicted adhesion force will be:

$$F_a = \frac{3}{2}\pi W_a R \quad \text{Eq. 53}$$

According to eq. 13, the adhesion force per contact area (i.e. seta) is proportional its radius. If a seta's contact area is divided into a number of smaller and equally sized spatulae $n$, the radius of each of the spatulae $R_n$ will be called as $R_n = \frac{R}{\sqrt{n}}$ (self-similar scaling). Therefore, the total adhesion force in this case will transform into $F_{an} = \sqrt{n}F_a$ [85]. However, this model considers that the adhesion occurs onto flat surfaces, which is not the case in natural surfaces. On natural rough surfaces the compliance and adaptability of setae are the primary sources of high adhesion enabling them to conform to the rough surface's contours and increase contact [86]. There are also other models developed over the years trying to introduce a saturation in adhesion force, which is not evident in this model, since the force increases indefinitely for very large values of $n$ [80].

Tree frogs and some insects on the other hand, make use of wet adhesion where adherence occurs via a thin film of liquid between contact areas [87]. On insects, wet adhesion can be attained on both smooth (ants, bees, cockroaches, and grasshoppers) and hairy (beetles, and flies) pads. Smooth toe pads consist of a dense fibrous material that is soft in compression while strong in tension. Furthermore, it possesses functional material properties including adaptability, viscoelasticity, and pressure sensitivity [88,89]. Hairy pads, on the other hand, are composed of a diverse density of setae which range in length from few micrometers to several millimeters [89–91]. It has also been reported that the density of the hairs increases with increasing body weight, thus increasing the number of single contact points and produce stronger adhesion [85,90,92]. By measuring the single-pad frictional and adhesional forces in a sample of hairy *Gastrophysa viridula* and a smooth *Carausius morosus* pad, it has been reported that the force per unit pad area was similar between the two configurations and that both types adhered via a thin liquid film [93]. To sustain this liquid film the hairy pads of reduviid bugs, flies, coccinellid beetles and chrysomelid beetles secrete fluids directed in the contact area to enhance the adhesion [94–98]. Moreover, it has been shown that the adhesion force depends on the fluid wetting the contact areas and a reduction of the volume of the fluid hinders adhesion [88].

Tree frogs depicted in Fig. 6 (2a) take advantage of wet adhesion to cling to leaves and other smooth surfaces [99]. Remarkably, they are able to attach to almost any surface [100,101]. Their attachment pads possess a complex hierarchical geometry that facilitates their adhesive strength. Blunt surfaced, soft epidermal cells in hexagonal formations (Fig. 6 (2b) and (2c)), separated by 1 μm channels used mucus glands ornament their toe pads [102]. On the surface of each cell thousands of nanocolumns branch out in a similar manner as the foot of the gecko, forming a hierarchical geometry [99] illustrated in SEM micrographs Fig. 6 (2d) and in higher magnification (2e). Strong attachment is then formed when there is a thin fluid film between contact areas [99,103]. Wet adhesion is believed to emerge from the interplay between physical contact between the two mating surfaces and the capillary forces resulting from the mucus which fills the remaining cavities [90,104]. The contribution of capillarity, however, is more pronounced in the case of hairy insects, where air-water interface (meniscus) is assumed to surround each small area of contact [90,97,101]. In the case of the blunt shaped epithelial cells of the tree frogs, direct contact with the mating surface is achieved by squeezing out the fluid film that keeps the two surfaces from interacting, thus achieving a form of "frictional adhesion" [93,99,101,102].

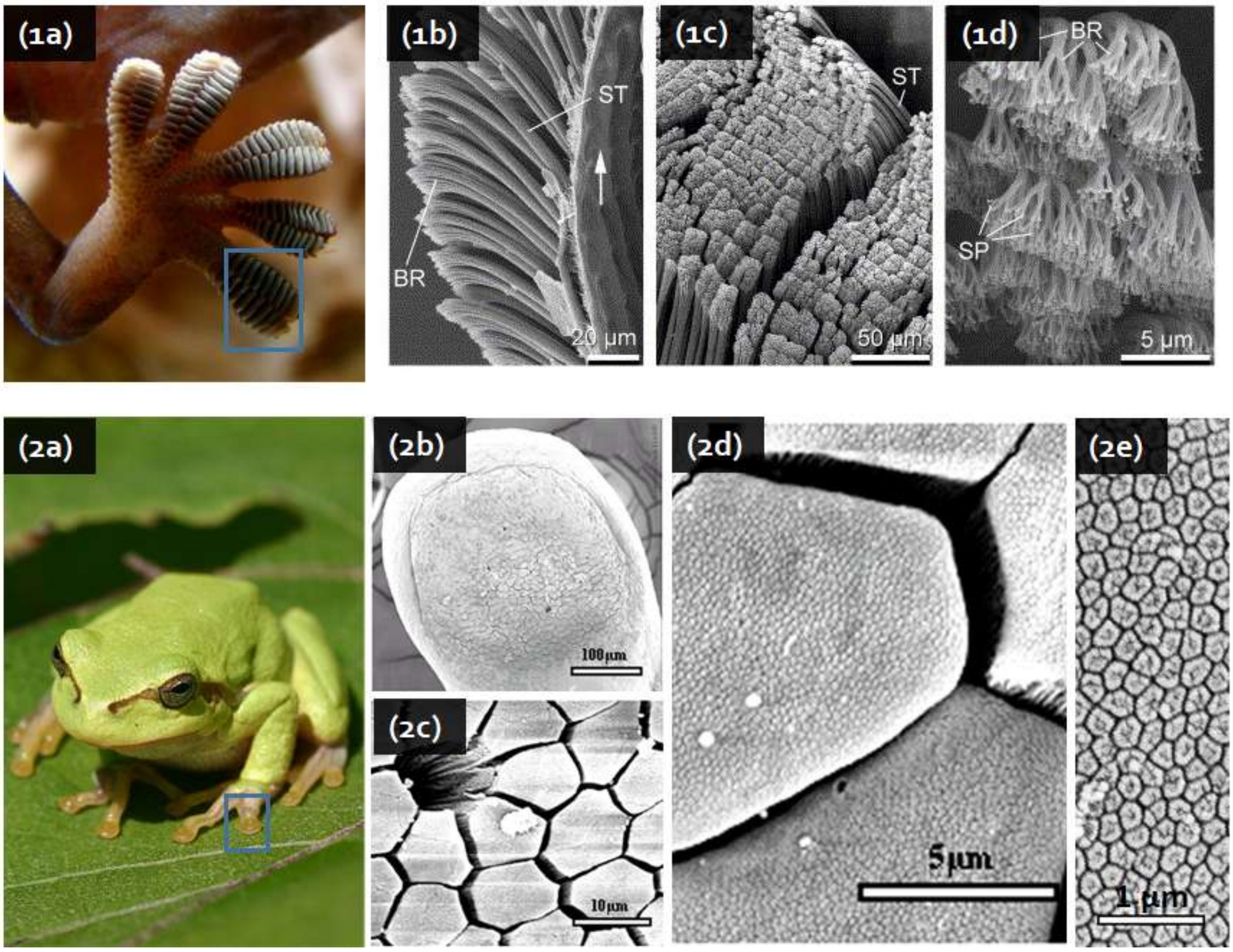

Fig. 6: (1a): Close-up photograph of the underside of a gecko's (Gecko gecko) foot as it walks on vertical glass. (1b)-(1d) SEM micrographs of the hierarchical structures on the foot of the gecko that provide it's supreme adhesion. (1b) and (1c) are different magnifications of rows of setae, (1d) higher magnification SEM image where spatulae can be seen branching from the tip of each seta. ST: seta; SP: spatula; BR: branch. (Image (1a) was adopted from Bjørn Christian Tørrissen / CC BY-SA 3.0. Images (1b), (1c) and (1d) were reproduced with permission from H. Gao et., al [80] *Copyright © 2004 Elsevier Ltd. All rights reserved*). (2a): Photograph of European tree frog (*Hyla arborea*). (2b)-(2e) present SEM images of a frog toe pad (2b), with hexagonally aligned epithelial cells (2c) and high magnification of a single hexagonal cell decorated with nanocolumns in (2d). Higher magnification of the nanoculmns is presented in (2e). (Image (2a) provided by Christoph Leeb / CC BY-SA 3.0. Images (2b)-(2e) were taken from [101]. Reprint permission granted from *Experimental Biology.*)

### *2.3.2 Friction reduction*

Friction reduction is of paramount importance in nature, as a means to conserve energy or protect against abrasion for both land and marine species. Especially for legless reptiles (snakes and even a few lizard species), but also for other reptiles living in hostile and challenging tribological environments close to the ground, functional adaptation has manifested in optimized surface designs and locomotion that is distinguished by economy of effort [105]. Few examples, like the skin of snakes or the desert lizards *Laudakin stoliczkana* and *Scincus scinus*, show a passive abrasion resistance, which is due to clever material combination (soft layer beneath a rather hard, but flexible surface) and, in case of the sandfish, also due to a highly specialized surface chemistry [106–108]. The frictional performance especially of snakes has been thoroughly investigated and attributed to its ventral skin [105,108–110] and specific skin formations (scales and structures onto the scales) to friction reduction [106,111]. Fig. 7 (A) depicts a photograph of a California King Snake (*Lampropeltis getula californiae*), in (B) an SEM image of its intricate skin pattern is illustrated. A complex interplay between multiple functional skin layers and epidermal microtexture allows their optimized tribological response [111–114]. In particular, the nanoridge microfibrillar geometry on its skin provides ideal conditions for sliding in a forward direction with minimum energy consumption. Furthermore, the highly asymmetric profile of the microfibrillar ending with a radius of curvature of 20 to 40 nm may induce friction anisotropy along the longitudinal body axis and functions as a kind of stopper for backward motion, while simultaneously providing low friction for forward motion [106]. Additionally, the system of micropores penetrating the snakeskin may serve as a delivery system for a lubrication or anti adhesive polar macromolecular lipid mixture, which provides boundary lubrication of the skin. This mixture may also aid locomotion by removing debris that contaminate the skin during contact [105]. More recently the effect of mechanical properties on friction and wear behavior on different counter surfaces have been investigated, revealing a link between material properties, abrasion and the counter surface which in nature is mitigated by specific locomotion pattern and adaptability of different species to their natural habitat [108,115]. Recent works have modelled the frictional forces of snake skin by applying a modification of Tomlinson-Prandtl (TP) model which physically interprets the interaction of nanostructure arrays of the ventral surface of the snake skin with variously sized asperities, confirming the frictional anisotropy between forward and backwards motion [116].

Marine species such as fast-swimming sharks like the Galapagos shark in Fig. 7 (C) possess skin textures that are known to reduce skin friction drag in the turbulent-flow regime while also protecting against biofouling [117]. The skin of the shark is covered with tiny scales shapes in small riblets which are aligned along the body of the shark in the direction of fluid flow [118]. Riblet formations on the skin of various shark species can be seen in Fig. 7 (D). Anti-biofouling is facilitated by the spacing between neighboring riblets, which is such that is able to hinder the adhesion of aquatic organisms. Drag reduction works by impeding the cross-stream translation of the streamwise vortices that occur in the viscous sublayer [117,119]. This reduces the occurrence of vortex ejection into the outer layer and the momentum transfer that this effect is correlated with. More specifically, in turbulent flow regime, vortices form above the riblet surface and remain confined there [120]. Therefore, they interact only with the tips and rarely cause any high-velocity flow located in the valleys of the riblets. This minimization of the surface area interacting with the vortices, also minimizes the high-shear stresses only to a small fraction of the hole surface area, (i.e.., the tips) since the low-velocity fluid flow into the valleys of the riblets produces very low shear stresses on the majority of the surface [121,122]. The exact mechanism explaining the impediment of vortex translation is not yet fully elaborated [117].

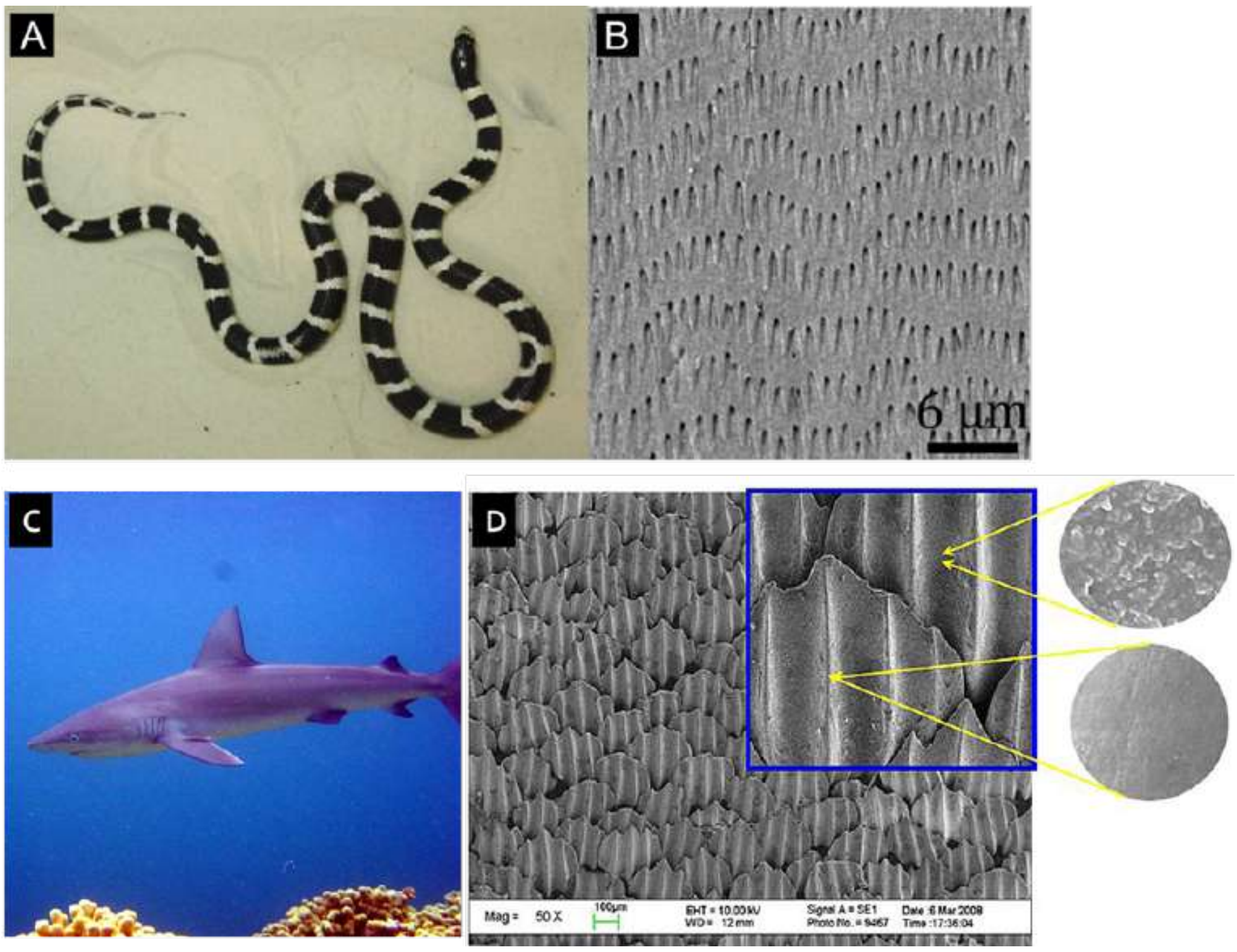

Fig. 7: (A) Picture of California King Snake (*Lampropelt getula californiae*, A) and SEM-image of the ventral scales of its skin (B). Picture of Galapagos shark (*Carcharhinus galapagensis*) (C). (D) Scale patterns on various fast-swimming sharks (the scale bar is 100 µm). (Images (A) and (B) were taken from [111] licensed under Creative Commons 2.0 from *Beilstein Journal of Nanotechnology*. Image (D) was adopted from "X. Pu, G. Li, and H. Huang, "Preparation, anti-biofouling and drag-reduction properties of a biomimetic shark skin surface," Biol. Open, vol. 5, no. 4, pp. 389–396, Apr. 2016." Under CC BY 3.0 https://creativecommons.org/licenses/by/3.0/).

#### *2.3.3 Strength, stiffness and stretching*

Steel is a strong and flaw tolerant material, ideal for structural applications. On the contrary, ceramics and polymers and less favorable as structural materials. Although ceramics are also strong, they are also very prone to crack formation and propagation. Polymers on the other hand are crack tolerant but deform easily under applied stresses. In nature, though clever design has resolved the issue of material selection in designing materials with high tensile strength and hardness. For example, the nacreous layer of mollusk shells, like the one found on *Cypraea chinensis* (Fig. 8 (a)), shows remarkable crack-resistance and stiffness, although it is made up from relatively weak components [123–125]. In fact, 95 % of its volume is brittle aragonite platelets and 5 % is biological macromolecules [126]. In spite of this, the mechanical design found in nacres, combines platelet-like building blocks with polymeric matrices between forming a complex material. Ultimately this intricate brick-and-mortar nanostructure is the reason of its unusual mechanical response [127–129]. A schematic of the "brick-and-mortar" architecture is illustrated in Fig. 8 (b), demonstrating the alternating layers between hard material layers glued together by layers of soft material. An SEM image of this structure is demonstrated in Fig. 8 (c) with the visible consecutive layers. The interpretation of the mechanical properties of composites made up from a matrix and a filler material is usually treated in two limiting configurations expressed by the Voigt or the Reuss models. In the former, all constituents are treated as being parallel, thus they experience the same strain. In this case the Young's modulus is calculates via the expression

$$E_c = V_f E_f + (1 - V_f) E_m \quad \text{Eq. 14}$$

In the latter the constituents are placed in layers and consequently undergo the same stress, in this case the composite Young's modulus is expressed by a reciprocal rule of mixtures

$$\frac{1}{E_c} = \frac{V_f}{E_f} + \frac{(1-V_f)}{E_m} \quad \text{Eq. 15}$$

*V* stands for volume fraction, *E* is the Young's modulus and the underscores *c*, *m* and *f* denote composite, matrix and filler respectively. Several other models have also been introduced with mixed success in predicting the Young's modulus of a natural nacre mainly due to the contribution of the stresses along the length of the filler (i.e., organic) material relative to its dimensions and the interaction between consecutive layers [123].

Another example of smart composite natural material can be found looking at mussels. Mussels are aquatic sessile organisms, whose survival in exposed habitats depends in large part on an adaptive holdfast that secures the animal to a solid surface (byssus) and dissipates the shock of wave impact through repeated large strains [130]. A typical byssus consists of hundreds of complex threads with a collagenous interior and a thin (2 to 4 μm) protein cuticle with highly repetitive sequences. Intertidal mussels, like *Mytilus galloprovincialis*, are located predominantly on turbulent wave-swept seashores thus require greater capacity for energy dissipation by its byssal threads. Their cuticle consists of distinct biphasic granules in a homogeneous matrix, which are not evident on mussels living in stiller waters, like *Perna canaliculus*. In fact, it has been shown that only this structural feature distinguishes these species. The granules are typically 0.8 µm in diameter and comprise about 50 % of the cuticle volume. Additionally, they show an intriguing phase-separated morphology with a domain size of 20 to 40 nm. Strain tests reveal a striking difference between the cuticles of *M. galloprovincialis* and *P. canaliculus*. When the threads of *P. canaliculus* are stretched 30 % beyond its length, the cuticle began cracking. The cracks propagated following their formation through the entire thickness of the cuticle until they were deflected from the interface with the collagenous core. This exposes the thread core making it prone to microbial attack and abrasion compromising its performance. On the contrary, stretching the threads of *M. galloprovincialis* reveals that cracks first formed within the matrix of the composite cuticle but did not propagate through the entire cuticle. Cuticle rupture only occurred when strain levels reached 70 % [130]. This striking difference is due to the existence of the granules, which seemed to arrest impinging cracks. The eventual cracking manifested when a large number of microcracks coalesced.

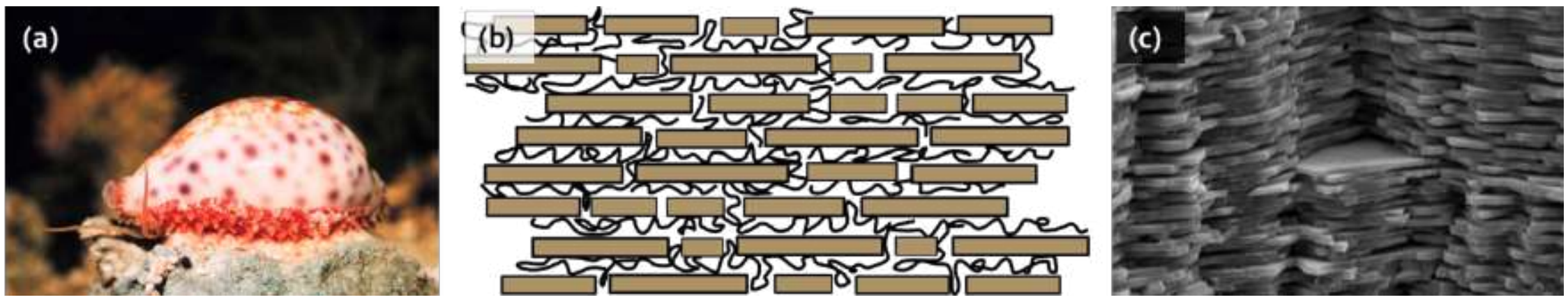

Fig. 8: (a) Cowrie (*Cypraea chinensis*), a member of small to large marine gastropod mollusks in the family Cypraeidae. (b) Schematic of the microscopic structure of nacre layers with alternating layers of hard aragonite platelets and biological macromolecules. (c) Electron microscopy image of a fractured surface of nacre. (Image (b) was taken from © User:Kebes / Wikimedia Commons / CC-BY-SA-3.0).

## 2.4 Biological

For numerous applications, surface properties must be sustained to maintain their function. In most cases that means to avoid or reduce biological fouling. The Lotus leaf surface is a prominent example [60]. Here, a double-hierarchical microstructure (contains micrometer and nanometer range) enhances the hydrophobic surface properties to superhydrophobic. Superhydrophobicity requires surface structures or specific roughness additionally to a hydrophobic surface chemistry. In case of the Lotus plant the combination of surface chemistry and structure leads to a contact area of 0.7 % for water droplets – perfect conditions for self-cleaning properties [66]. A transfer to artificial surfaces has been made using various techniques, including laser technologies [131]. All such techniques contribute to the functional principle of hierarchical surface structures that enable the Lotus inspired surface properties.

### *2.4.1 Anti-fouling*

Turning to the aquatic environment, surfaces in contact with water often undergo fouling contamination with particles and living organisms [132]. Unwanted fouling contamination may affect the performance of a surface or other component of a machine. The most prominent examples are the water vessels where the contamination of their surface reduce their hydrodynamic performance and cause malfunction to the propulsion system [133,134]. Remarkably, various biological species have solved this problem through sophisticated surface texture as presented in Fig. 9. The Taro leaf with hierarchical micro-bumps and nanostructures on its surface is forcing particles and microorganisms to adhere only in the areas between the bumps [135]. The anti-fouling effect is more pronounced in the areas with nanostructures while the wetting properties of the surface does not influence the outcome. In the other hand, the water repellent surface of Lotus and rice leaves were also found to be resistant to contamination in order to avoid surface overheating and the demotion of photosynthesis [60,136,137]. The mechanism where surface structures promote the self-cleaning effect was recently examined [138]. Furthermore, sea animals have introduced similar anti-fouling strategies by employing surface texture, in an effort to preserve their fitness. Mytilid shells [139,140], echinoderm skin [141] and shark skin [142–144] are some of the examples with resistance to contamination. Surface structures like ripples for the mytilid shell and dermal denticles covered with riblets for the sharkskin, are considered to prevent biofouling among others. In fact, for the shark skin it is believed that the combination of surface topography and a mucous coating is responsible for the anti-fouling properties of sharks and other fishes [136]. Altogether, nature's concepts could inspire and impact future fabrication technologies for various applications.

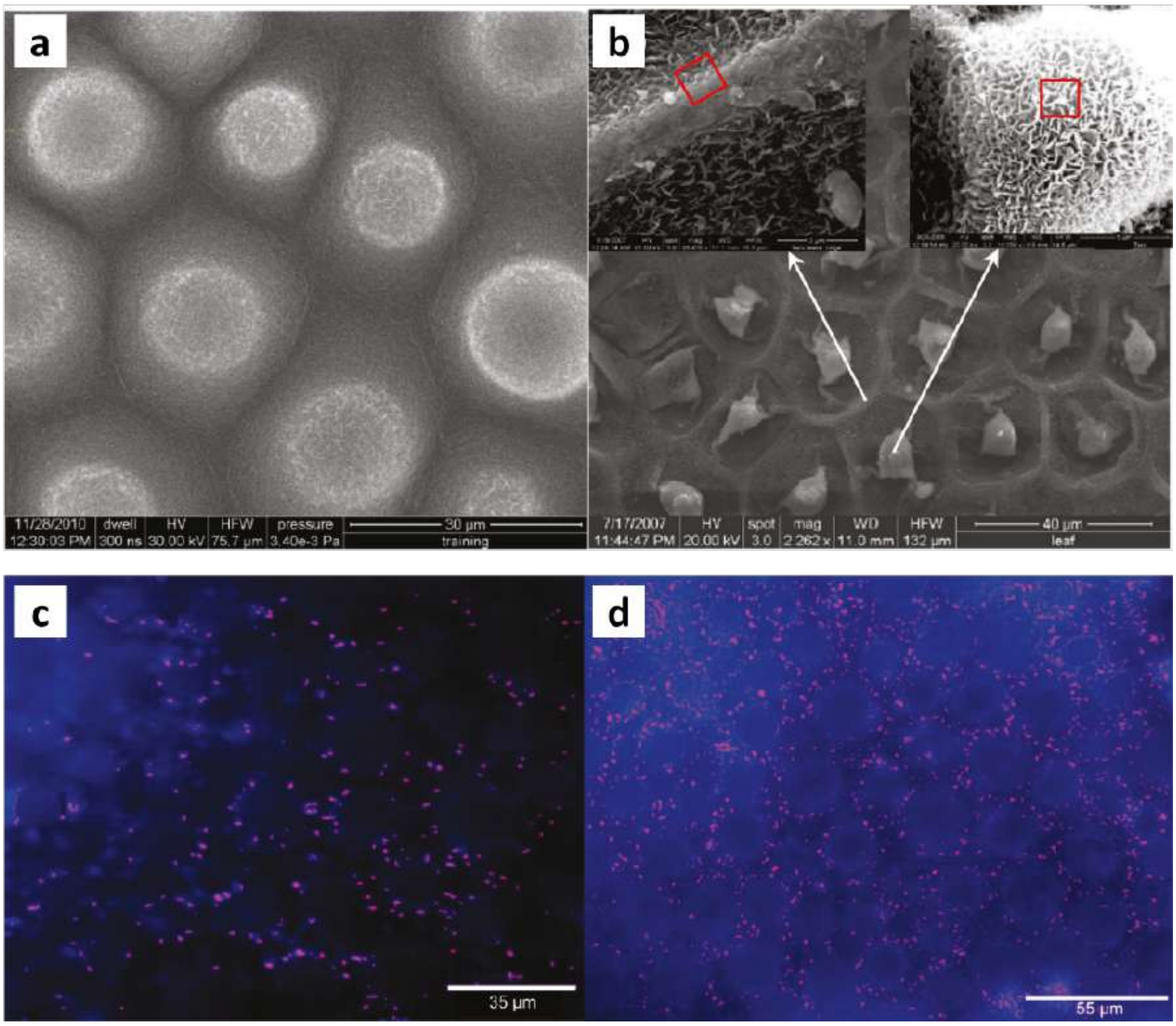

Fig. 9: SEM images of the Taro leaf surface after a) liquid substitution and b) air-dry. Fluorescence microscope images of stained pseudomonas aeruginosa bacterium on taro leaf at c) dry and d) wet conditions. Reprinted with permission from [135]. Copyright (2011) American Chemical Society

### *2.4.2 Anti-bacterial*

A well-known biomimetic example is the Lotus leaf where the structures on the leaf surface lend the remarkable properties of self-cleaning and contamination resistant surfaces. Taking a step forward, some organisms have evolved and presents anti-bacterial properties. When the wings of dragonfly and Cicada are contaminated with bacteria, their aerodynamic performance will be affected which is a critical element for their survival [145,146]. Also, the contamination of a reptile's skin can cause infection and diseases [147]. In order to protect their functionalities, dragonflies [145,146,148], cicadas [148–151], planthoppers [152] and geckos [153,154] have developed a nano-pillar like texture on the surface of interest. It seems like the nano-pillars are able to penetrate the membrane and hence the bacteria are rupture. The process of bacterial rupture is under investigation in an effort to reveal the detailed mechanism. A combination of adhesion and shear force, the important role of nanostructure geometry and generally the physio-mechanical nature of the process rather than chemical are some of the important findings of the anti-bacterial investigation of dragonfly's and cicada's nanostructure wings (Fig. 10). Notably, the resistance of bacteria to traditional chemical antibiotic [155] could be overcome by the physio-mechanical nature of the nanostructure bactericidal properties [156].

Following the great potential of nanotextured surfaces with bactericidal properties, theoretical calculations have been conducted for an in-depth analysis in an effort to better understand the mechanism and apply a biomimetic approach to artificial surfaces [157–160]. The main parameters that affect the bactericidal property of a nanostructure surface have been investigated [157,159,160]. To begin with, the adhesion force is translated to bending and stretching forces on the membrane of the bacterial. The biochemical properties of the surface can tune the amount of adhesion force [161]. In the case of a nanostructure surface, the adhesion force leads to higher stretching and bending values due to the increase of the contact area compared to flat surfaces. Interestingly, Pogodin et al. suggested that the membrane is more likely to rupture at the areas between the pillars where the stretching forces are higher [158]. The bactericidal property of a nanostructure surface strongly depends on the surface roughness where for instance higher surface roughness produces higher bending and stretching forces. Besides, the anti-bacterial surfaces of the dragonfly and the cicada are tailored with structures at least 10 times larger than the bacterial membrane. The membranes of different bacteria have different stretching modulus and thus can handle the respective amount of stretching and bending forces. Consequently, membranes with lower stretching modulus are more vulnerable to rupture on a nanostructure surface [158]. Additionally, the mechanical properties of bacterial membrane change depending on the environmental conditions [162].

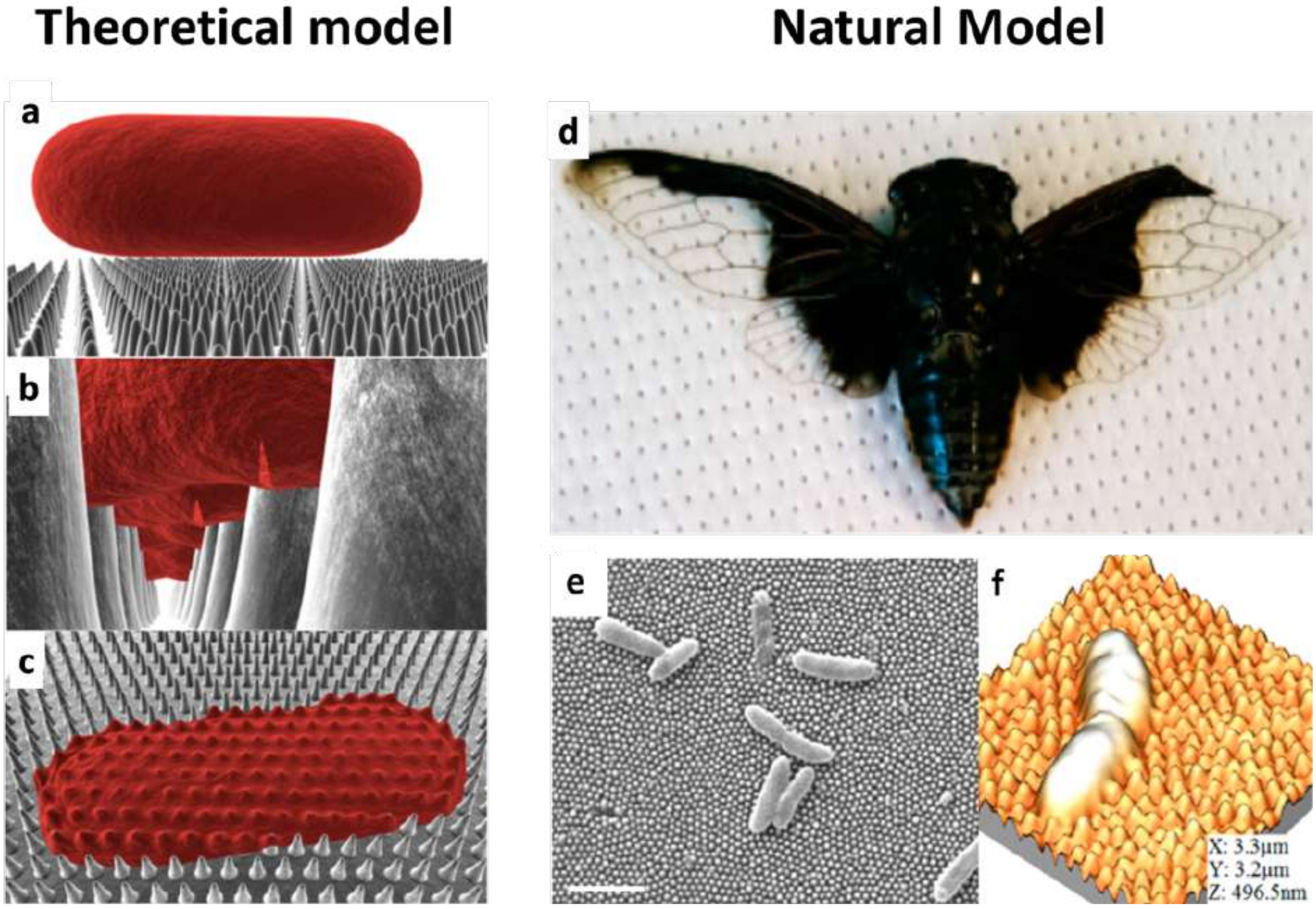

Fig. 10: a-c) Schematic representation of the proposed mechanism which a bacteria rupture on a nanostructure surface. Reprinted from [158], Copyright (2013), with permission from Elsevier. d) An image of a cicada (C. aguila). e) SEM images of the cicada's (C. aguila) wing with bacterial cells. f) AFM image of cicada's (C. aguila) wing with rupture bacterial cells. d-f) Reprinted with permission from [151]. Copyright (2016) American Chemical Society.

# 3 Laser fabrication of biomimetic surfaces on hard materials and related applications

Table 1*: Biomimetic examples showing the capabilities for laser manufacturing of hard material surfaces.*

| Material | Natural archetype and functionality (-ies) | Laser parameters (Wavelength - Pulse duration, Repetition rate) | Fabrication Parameters (Polarization – Effective number of pulses – Fluence) | Structural features (type, periodicity, density, geometric characteristics) | Functionality (-ies) of processed material | Ref |
|---|---|---|---|---|---|---|
| $SiO_2$ - Sapphire | *Greta oto, Cicada cretensis* | 1026 nm, 170 fs, 1 kHz | Circular, 6–50 pulses, fluence 5.7–8.3 J/cm$^2$ | Pillar-like nanostructures, 70 – 100 nm radius, 224 ± 41 nm height, 200–400 nm periodicity | Anti-reflection | [7] |
| Silicon | *Nelumbo nucifera* (the sacred Lotus) | 1064 nm, 1 ns, 100 kHz | Unreported polarization 4.2 W, 100 mm/s | Spikes, Silicon nanowires, heights: 1 mm to 30 mm | Anti-reflection | [163] |
| Silicon | *Nelumbo nucifera* (the sacred Lotus) | 800 nm, 180 fs, 1 kHz | Linearly polarized, 300 to 3000 pulses, fluence = 0.37 to 2.47 J/cm$^2$, | Spikes 10 nm size and aspect ratio of ~ 4 | Water repellency | [164,165] |
| Steel alloy | *Dysodius lunatus,* Texas horned lizard, *Phyton regius* snake, Western diamondback snake | 1030 nm, 340 fs, 1 kHz – 2 MHz | Linear polarization, 0.12 J/cm$^2$, $N_{eff}$: 150000 | ripples, grooves, spikes | Water repellency | [166] |
| Nickel | Shark skin | 1026 nm, 170 fs, 1 kHz, | Linear, radial and azimuthal polarization, Pulse energy: 2.4 mJ, $N_{eff}$=36 - 207 | Complex ripples, Period ~1μm, Dual scale hierarchical structures, Ripples 615 nm period | Water repellency, iridescence | [167] |
| Aluminum | *Nelumbo nucifera* (the sacred Lotus) | 355 nm, several ns, 100 kHz | 2.8 J/cm$^2$, Unknown polarization, Unknown effective No. of pulses | Line and grid texturing with 10μm, 15 μm, 20 μm and 25 μm spacing, Ripples, Grooves, Spikes | Water repellency | [168] |
| Magnesium alloy | Petal of red rose | Ct-200ii engraving machine | Unknown parameters | Nanoscale flakes, flower like microstructures 100 nm – 1 μm. | Water repellency, corrosion resistance | [169] |
| Copper | *Morpho* Butterfly, *Trogonoptera brookiana* | 800 nm, 100 fs, 1 kHz, 1064 nm, 1 μs, 30 kHz | Linear polarization, Pulse energy: 2.4mJ, Unknown No. Of pulses | Ripples 615 nm period | Hydrophobicity, structural coloring | [170] |
| Steel alloy | *Dysodius magnus,* Texas horned lizard | 1) 790 nm, 0.03 ps, 1 kHz<br>2) 532 nm, 9 ps, 1000 kHz<br>3) 1026 nm,0.17 ps, 1 kHz<br>4) 1030 nm, 0.5 ps, 10-1000 kHz | Linear polarization, Fluence: 0.15 – 2.5 J/cm$^2$, $N_{eff}$: 5 - 1200 | Ripples, rooves, spikes, line scanning with spacing 20 – 100 μm | Corrosion resistance, hydrophobicity, unidirectional fluid transport | [64,171] |
| 100Cr6 steel, $Al_2O_3$ | Snakes and Sand fish lizards | 1064 nm, 2 μs, 41 kHz, | 11 W, unknown polarization and number of pulses | Scale-like structures, 13 and 150 μm scale diameter, 6 ± 1 μm height | Friction reduction | [172] |
| 100Cr6 steel | Snake *Phyton regius* | 1064 nm, 1μs, 30 kHz | Unknown parameters | Scale-like structures, lateral size 50 μm, 5 μm height | Friction reduction | [173] |
| Stainless steel | *Nelumbo nucifera* (the sacred Lotus) | 355 nm, 25ns, 6.8 Hz | 6.7 & 18 mJ/mm$^2$, 146 J/cm$^2$, unknown polarization, pulse overlapping from 20 – 80% | Line scanning with spacing 20 – 100μm | Corrosion resistance, hydrophobicity | [174] |
| Silicon | *Nelumbo nucifera* (the sacred Lotus) | 1026 nm, 170 fs, 1 kHz | Linearly polarized, 300 to 3000 pulses, fluence = 0.73 J/cm$^2$, 50 W, 500 mm/s scanning speed. | Spikes, 6.5 – 14.2 μm height, 2.1 – 2.3 roughness ratios, round humps *r* = 50 μm, square protuberance *L* =50 μm, mountain range-like structures *L* =50 μm; covered by nano-scale mastoid | Corrosion resistance, hydrophilicity | [175] |
| | | | | structure | | |
| Graphite cast iron | Dung beetles | 1064 nm, 5 ms, 14 Hz | Fluence 25, 75, 125 and 175 J/cm$^2$, | Microstructures, depth 0.141 - 0.389 mm, width 0.568 - 1.023 mm, hierarchical structures. Spikes of 10-20 μm size and ~200 nm undulation onto spikes | Wear resistance | [176] |
| H13 steel | Dung beetle, tree leaf | 1060 nm, 6 ms, 4 Hz | 146 J/cm$^2$, unknown polarization, unknown number of pulses | Spikes with 20 - 40 μm period, ripples with 0.5 - 0.9 μm period (linear). Nano-pillars with 0.8 – 1.3 μm period (azimuthal)random micro structuring | Mechanical/ mechanical/ fatigue resistance | [177,178] |
| Aluminum alloy | *Nelumbo nucifera* (the sacred Lotus), rose petals, water striders, butterfly wings | 520 nm, 380 fs, 200 ns, 20 kHz | 50 W, 500 mm/s scanning speed. | Round humps *r*=50 μm, square protuberance *L*=50 μm, mountain range-like structures *L*=50 μm; covered by nano-scale mastoid structure | Anti-icing | [179] |
| Titanium | *Nelumbo nucifera* (the sacred Lotus) | 800 nm, 50 fs, 1 kHz | Circular polarization, 20-100 J/cm$^2$ | Hierarchical structures. Spikes of 10-20 μm size and ~200 nm undulation onto spikes | Bactericidal | [180] |
| 316 L stainless steel | *Nelumbo nucifera* (the sacred Lotus) | 1030 nm, 350 fs, 1 MHz, 250 kHz | Linear (20.2 – 1910 J/cm$^2$), Azimuthal (12.2 J/cm$^2$) | Spikes with 20 - 40 μm period, ripples with 0.5 - 0.9 μm period (Linear). Nano-pillars with 0.8 – 1.3 μm period (Azimuthal) Ripples, microgratings multi-scale hierarchical structures | Bactericidal | [181] |
| Pyrex | Namib desert beetle | 520 nm, 380 fs, 200 kHz, 1032 nm, 310 fs, 1 MHz | Linear, 2.1 - 6.2 J/cm$^2$, unknown number of pulses | Double-hierarchical surface structures with 16 μm hatching distance | Fog collection | [182] |
| Titanium | Cactus, Namib desert beetles | 1030 nm, 800 fs, 400 kHz | Unknown parameters | Hierarchical micro/nanostructures | Water-collection | [183] |
| Zns | Structural color of butterflies | 800 nm, 100 fs, 1–1000 Hz kHz | Linearly polarized, 0.16 – 1 mJ/cm$^2$ | Ripples, microgratings multi-scale hierarchical structure, Low, medium, high aspect ratio Spikes, Density: ~2.5-9.5x10$^6$/cm$^2$, Height: 1.2-8.6 μm, Period: 2.3-6.5μm | Iridescence, water repellency | [184] |
| Titanium alloy (ti-6al-4v), stainless steel (x6cr17) | Collembola | 1032 nm, 310 fs, 1 MHz | Circular polarization, 54-100 mJ/cm$^2$ 118 – 147 number of pulses | Triangular LIPSS | Uniform iridescence | [185] |
| Aluminum | *Argyroneta aquatic* | 800 nm, 65 fs 1 kHz | Linear polarization, 0.8 mJ, 1 mm/s, | Line scans, 100 μm spacing | Water repellency | [186] |
| Silicon316 L stainless steel | *Nelumbo nucifera* (the sacred Lotus) Pangolin | 800 nm, 150 fs, 1 kHz1, 070 nm, 500 μs, 1 kHz | 0.68 -1.50Linear polarization, Fluence 2000 to 20000 J/cm$^2$, Linearly polarized | Low, medium, high aspect ratio Spikes, Density: ~2.5-9.5x10$^6$/cm$^2$, Height: 1.2-8.6 μm, Period: 2.3-6.5μm, Overlapping microdots with micro-riblets | Friction reduction, anti-adhesion, Neuronal outgrowth | [187,188] |
| Aluminum alloy, Silicon | Structural complexity of tissues (extracellular matrix), Leaf with venation network | 1080 nm, 800 nm, 150 fs, 1 kHz, and 12 ns, 532 nm, | Linear polarization, Fluence 0.20 to 0.97 J/cm$^2$, Polarization | Ripples 146 nm periodicity, grooves 152 – 146 nm periodicity and 2.4 – 11 μm height, fs-laser irradiation, followed by hydrophobic chemical surface modification followed by ns-laser processing of venation network | Tissue engineering, Water transport | [189,190] |
| Silicon | *Nelumbo nucifera* (the sacred lotus), *Cicada orni*, Rhinotermitidae | 800 nm, 150 fs, 1 kHz | Linear Polarization, 500 pulses, Fluence 0.34 to 0.69 J/cm$^2$ | Spikes, aspect ratio 2.6 – 6.9 | Tissue Engineering | [191] |

## 3.1 Self-organized surface structures

Laser surface patterning can be classified in approaches based on *self-organized* laser-irradiated structures and *direct laser-inscribed* structures (see Fig. 11) [192]. Self-organized means here that although the surface is irradiated using a homogeneous spatial beam profile in a spot or scanning geometry, the resulting surface topography features characteristic (quasi-)periodic surface morphologies. The self-organized surface structures may consist of microstructures, nanostructures, or hybrid variants.

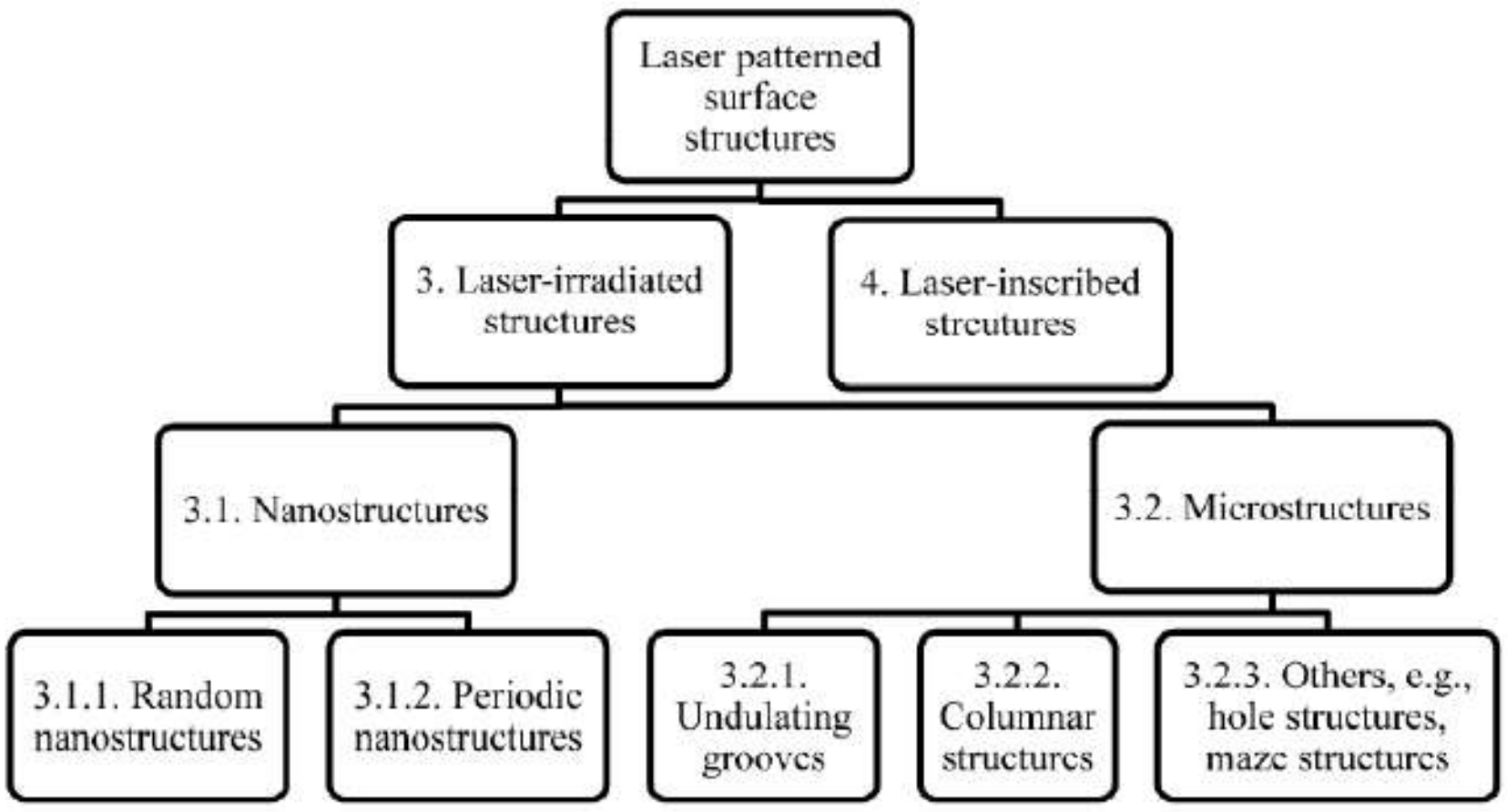

Fig. 11: Classification of laser-patterned surface structures. Reprinted from [192].

The periodic self-organized surface structures are usually classified as nanometric *laser-induced periodic surface structures* (LIPSS, ripples), and micrometric *grooves* and *spikes* (Fig. 12). Ripples are observed as *high spatial frequency LIPSS* (HSFL, Fig. 12(a)) featuring periods significantly smaller than the irradiation wavelength ($\Lambda < \lambda/2$) or as *low spatial frequency LIPSS* (LSFL, Fig. 12(b)) showing periods of the order of the laser wavelength [193–195]. Additionally, grooves (Fig. 12(c)) as transitory morphology between LIPSS and micrometric spikes (Fig. 12(d)) are seen. LIPSS and grooves structures were not only observed in conductive and semiconductor materials, but also in dielectrics [196,197]. LIPSS and grooves show a well-defined orientation with respect to a linear polarization state of the incident laser light and are distorted or even absent for other polarization states. All these different structures resemble surface morphologies found in nature and, therefore, can be considered as "biomimetic" [198–202].

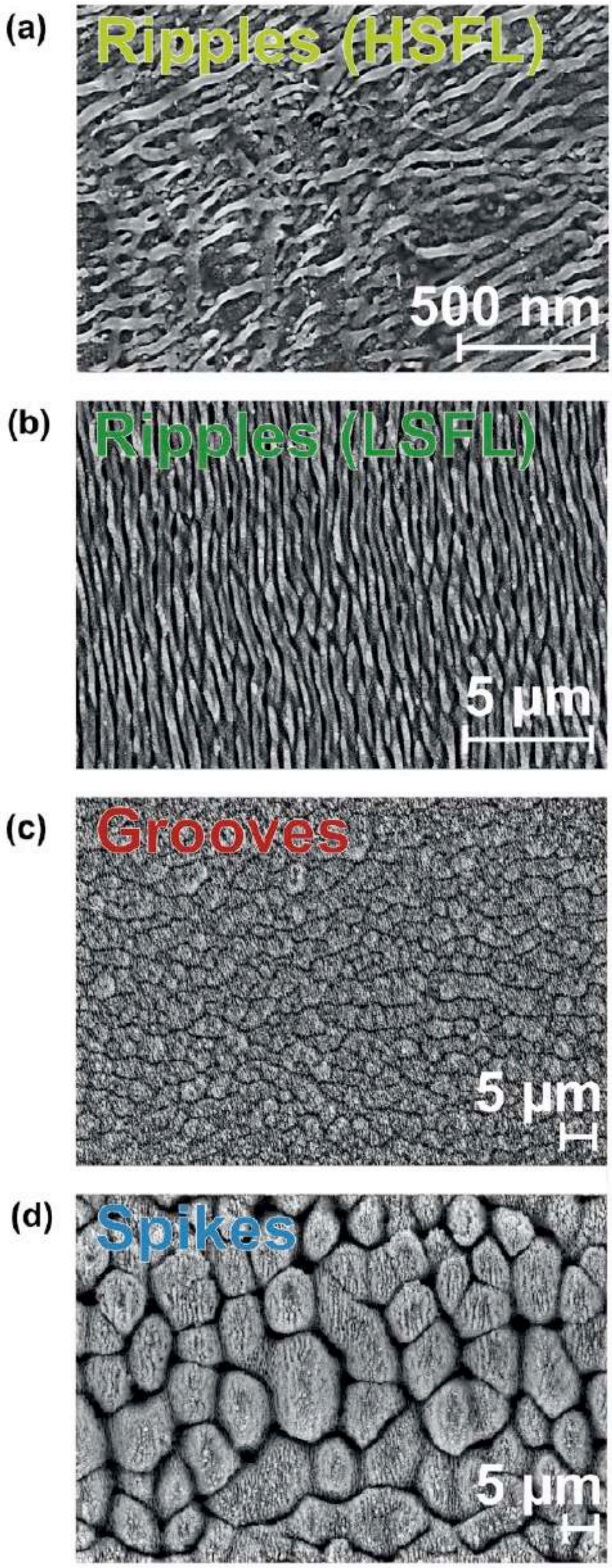

Fig. 12: Top-view scanning electron microscopy images of four characteristic surface morphologies observed upon femtosecond laser scan processing of a steel surface [790 nm, 30 fs, 1 kHz]. (a) High spatial frequency LIPSS (HSFL), (b) Low spatial frequency LIPSS (LSFL), (c) Grooves, (d) Spikes. In all cases the linear polarization is horizontal. Note the different magnifications. (Reproduced with permission from Hermens et al [64]. Copyright (2017) reprinted with permission from Elsevier)

The proper selection of laser processing parameters (peak fluence $\phi_0$, effective number of laser pulses $N_{eff}$) allows to control the surface morphology as demonstrated for steel in Fig. 13.

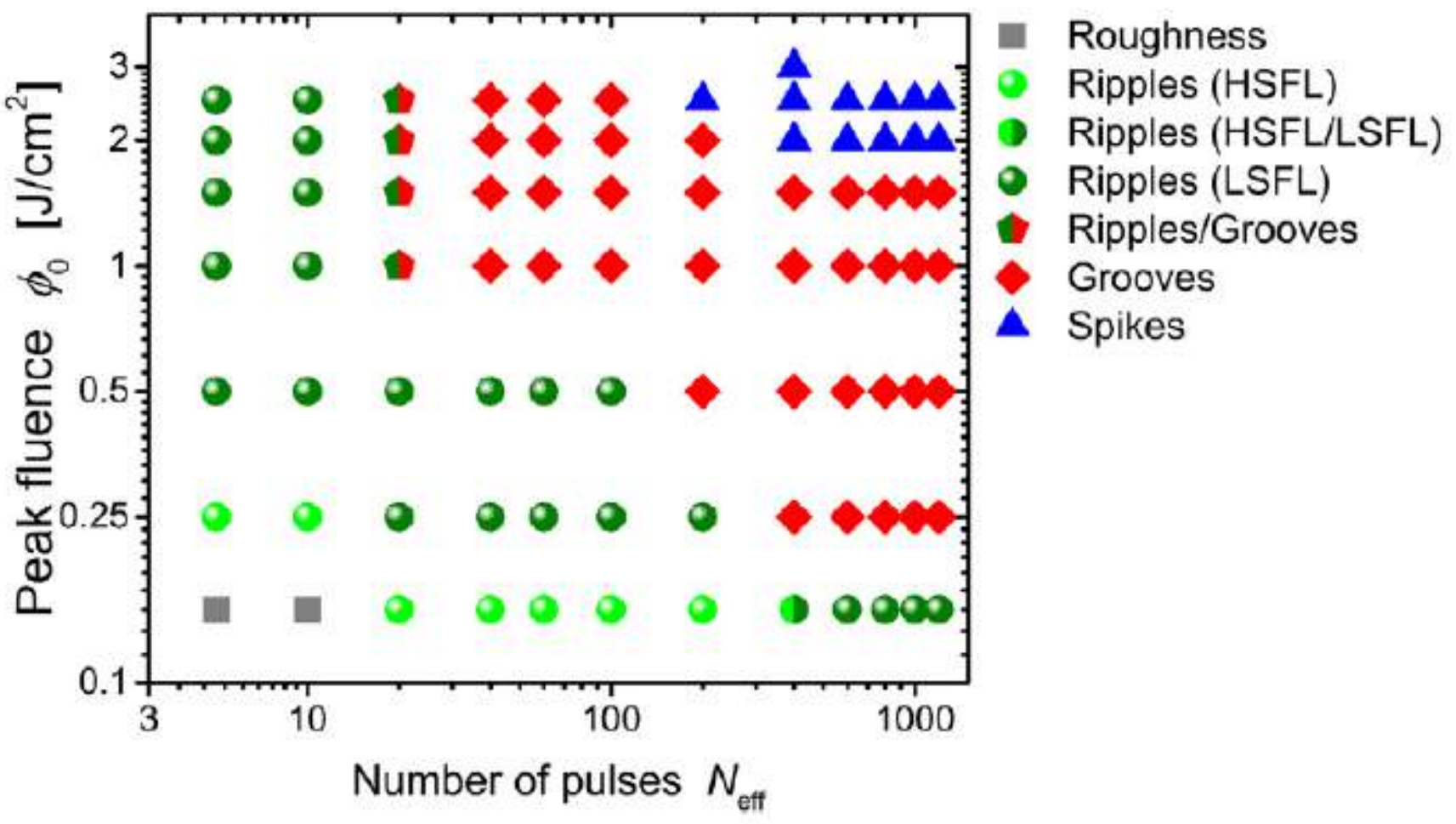

Fig. 13: Morphological map of femtosecond line processing of steel ordering the surface characteristics according to the laser peak fluence $\phi_0$ and the number of laser pulses per beam spot diameter $N_{eff}$. [790 nm, 30 fs, 1 kHz] (Reproduced with permission from Hermens et al [64]. Copyright (2017) reprinted with permission from Elsevier)

In the following sections, different types of laser-induced periodic surface structures are discussed in more detail.

### 3.1.1 *Ripples*

The characteristics of laser-induced periodic surface structures (LIPSS, ripples) strongly depends on the irradiated material [194,203]. In addition to the above given classification in HSFL and LSFL, a further distinction can be made (Fig. 14). For strongly absorbing materials (metals), near-wavelength sized LSFL (type LSFL-I) are usually generated with an orientation perpendicular to the laser beam polarization. The periods are close to the laser wavelength and their specific value has been found to depend on the dielectric function of the material [203–206], the pulse number [207], and the surface roughness [205,208–210]. These structures are observed in the ablative regime for fluences up to several times the ablation threshold (Fig. 13). For fluences very close to the ablation threshold, HSFL (type HSFL-II, Fig. 14) can be formed [194]. Note that semiconductors often behave like metals as they can be transiently made metallic upon high-intensity ultrashort laser pulse irradiation [204,211]. Semiconductors in addition may show the peculiarity of enabling the formation of amorphous-crystalline LSFL-LIPSS in a narrow fluence interval below the ablation threshold [212,213].

For weakly absorbing materials (dielectrics), sub-wavelength-sized LSFL (type LSFL-II) are found mostly with an orientation parallel to the laser polarization. The periods are close to $\lambda/n$, with $n$ being the refractive index of the material [214,215]. The corresponding HSFL (type HSFL-I) on dielectrics normally are oriented perpendicular to the polarization [196,216] of the laser beam but sometimes also parallel [217].

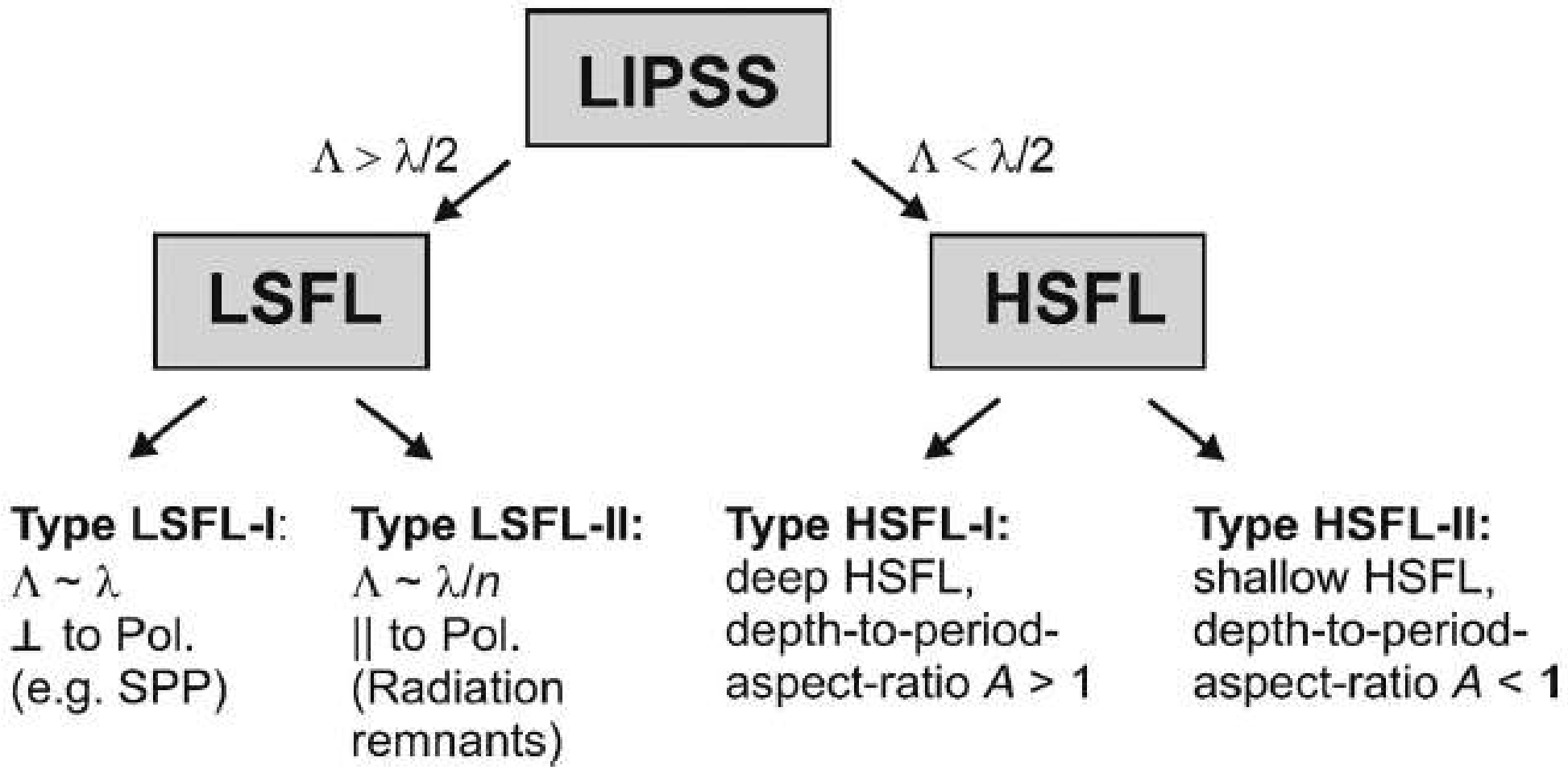

Fig. 14: Classification scheme of fs-laser-induced periodic surface structures. © 2017 IEEE. Reprinted, with permission, from [194]

Since both the LSFL and HSFL orientation always depend on the polarization direction it can be inferred that they originate from an electromagnetic mechanism leading to a spatially modulated deposition (absorption) of the laser pulse energy [194,203,216]. The absorption channels, however, can be different among the material classes. While for metals and semiconductors usually the excitation and interference of *Surface Plasmon Polaritons* (SPP) with the incident laser beam are involved leading to the formation of LSFL-I [203,204,211,218,219], in dielectrics another absorption channel is triggered, via so-called *Radiation Remnants* [196,203,220]. Recent approaches attribute the formation of HSFL to the interference between the incident beam and the radiation scattered at the rough surface while the proposed mechanism for LSFL-II formation is associated to the coupling of the incident light with the far-field scattering of the rough dielectric surface [216]. In most models, excitation of modes and generation of conditions that lead to LIPSS formation is closely related to the carrier density levels upon irradiation with high intensity fields [196,197,204,216,218–223].

To provide a detailed description of the physical origin of LIPSS formation as well as the quantitative features of the induced self-assembled structures on the surface of the irradiated material upon excitation with ultrashort pulsed lasers, a thorough investigation of the underlying multiscale phenomena that take place is required. While the precise physical mechanism for the origin of LIPSS is still debatable (but as mentioned above appears to be of electromagnetic origin), one process that undoubtedly occurs is a transient phase transition, i.e., melting, that eventually leads to a surface modification. Other processes might involve ablation (i.e., mass removal), spallation or even thermomechanical effects. In principle, the laser beam parameters (wavelength, pulse duration, fluence, number of pulses, angle of incidence, and beam polarization state) determine the onset of the surface modification as energy absorption, electrodynamical effects and relaxation processes are critical to the material heating. Femtosecond laser interaction also involves several complex phenomena, including energy absorption, photo-ionization processes, electron excitation and electron-relaxation processes.

Various theoretical approaches or experimental observations were performed in a variety of conditions to describe the evolution of the surface morphology and LIPSS formation. One approach was based on Kuramoto-Sivashinsky that assumed self-organization processes [224], however, the underlying theoretical approach was not sufficient to correlate the laser conditions with the induced surface morphological features.

A systematic study to couple lattice thermal response with electrodynamical effects to describe the formation of LSFL-I assumed firstly the interference of the SPP with the incident beam [203,204,211,218,225,226] to determine the orientation of the periodic structures. A unified model [219] comprises: (i) an electromagnetics component that describes SPP excitation and interference with the incident beam that leads to a periodic modulation of the laser field energy density, (ii) a heat transfer component that accounts for carrier-lattice thermalization through particle dynamics and heat conduction and carrier-phonon coupling, and (iii) a hydrodynamics component that describes the dynamics of the molten material and the subsequent resolidification process assuming an incompressible Newtonian fluid (melt) flow (Fig. 15a) that includes recoil pressure and surface tension contributions as well as Marangoni effects and hydrothermal convection. Due to the need for the consideration of a phase transition for the description of an induced morphological change, fluid dynamics is introduced. The material that undergoes melting is assumed to be an incompressible Newtonian fluid and its dynamics is described by the following equations:

(i). for the mass conservation (incompressible fluid):

$$\vec{\nabla} \cdot \vec{u} = 0 \qquad \text{Eq. 6}$$

(ii). for the energy conservation:

$$C_L^{(m)} \left( \frac{\partial T_L^{(m)}}{\partial t} + \vec{\nabla} \cdot \left( \vec{u} T_L^{(m)} \right) \right) = \vec{\nabla} \cdot \left( K_L^{(m)} \vec{\nabla} T_L^{(m)} \right) + \frac{C_c}{\tau_e} (T_c - T_L) \qquad \text{Eq. 7}$$

The following Navier-Stokes equation is used to describe the movement of hydrothermal waves

$$\rho_L^{(m)} \left( \frac{\partial \vec{u}}{\partial t} + \vec{u} \cdot \vec{\nabla} \vec{u} \right) = \vec{\nabla} \cdot \left( -P + \mu \left( \vec{\nabla} \vec{u} \right) + \mu \left( \vec{\nabla} \vec{u} \right)^T \right) \qquad \text{Eq. 8}$$

where $\vec{u}$ is the velocity of the fluid, $\mu$ is the liquid viscosity, $P$ is the total pressure (hydrodynamical and recoil) and $C_L^{(m)}$ stands for the heat capacity of the liquid phase. $K^{(m)}{}_L$ stands for the heat conductivity of the molten material while $C_c$ is the heat capacity of carrier subsystem. Finally, $T_C$, $T_L$, correspond to the temperatures of the carrier and lattice system, respectively. On the other hand, $t_e$ is the carrier-phonon energy relaxation time (~0.5ps for Silicon).

The model predictions presented remarkable similarity of the simulation results with experimental data (Fig. 15c). Although the model presented here was applied for sub-ablation (or with little ablation) conditions in order to ensure the presence of surface roughness that is a significant ingredient for LIPSS generation, other scenario were also investigated with good agreement with experimental observations in metals and semiconductors (plastic effects) [227,228]. The unified model also provided satisfactory results for different irradiation conditions (i.e., radially polarized beam [228–230], irradiation at a non-zero incident angle [231]) or assuming different surface orientation of the irradiated materials [232]. The evolution of the ripple periodicity as a function of the energy dose (number of pulses) is derived by computing the effect of the corrugation features (i.e. shape and height) on the excited SPP when successive pulses irradiate the material. Studies show that there is a blue shift of the plasmon-grating resonant frequency to smaller SPP wavelengths upon increasing number of pulses as the surface grating profile becomes deeper [211,228,233,234] (Fig. 15d).

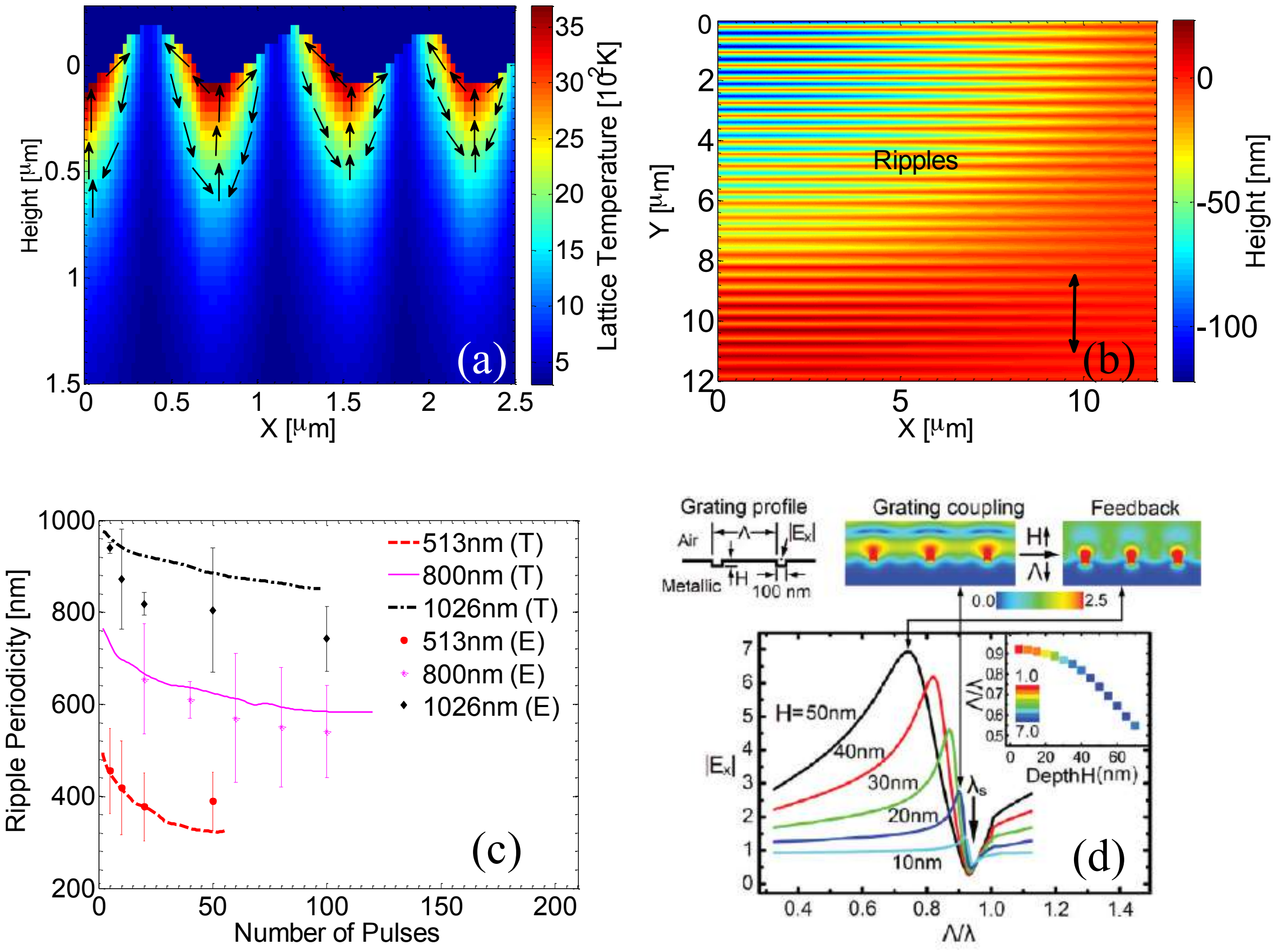

Fig. 15: (a) Simulated melt displacement (Si) (Reprinted with permission from Tsibidis et al [223]. Copyright (2015) by the American Physical Society), (b) Ripple profile (100Cr6), (c) Ripple periodicity vs. Number of pulses (100Cr6),: lines: simulations, data points: experiments (b,c) were reprinted from [222] and with permission from Springer Science and Business Media, (d) Finite Difference Time Domain simulation results for the picture of grating-assisted SPP-laser coupling

(ZnO). (Reprinted with permission from Huang et al [211]. Copyright (2009) American Chemical Society).

Recent approaches also attempted to unveil the role of local electromagnetic fields in the formation of HSFL and LSFL surface structures on dielectrics as well as the transition from one structure to another (Fig. 16a). *Finite Difference Time Domain* (FDTD) numerical schemes were used to solve Maxwell's Equations [216,235,236] to correlate laser-induced electron plasma structures originating from randomly distributed surface inhomogeneities. These inhomogeneities were found to interact strongly and to organize in regularly spaced sub-surface planes oriented perpendicularly to the laser polarization account for HSFL formation on dielectrics (Fig. 16a) [222]. The difference of the methodology compared to previous approaches is related to the derivation of a precise spatiotemporal distribution of the energy deposition which is very important at large number of pulses in which both the corrugation of the surface as well as the profile morphology is expected to influence significantly the amount of absorbed energy. In a recent report a model is also presented for metals that correlates morphological changes and LSFL-I formation according to laser parameters (e.g. the fluence) and relevant physical processes such as ablation, cavity-induced field-enhancement and spallation-related changes [236], see Fig. 16b.

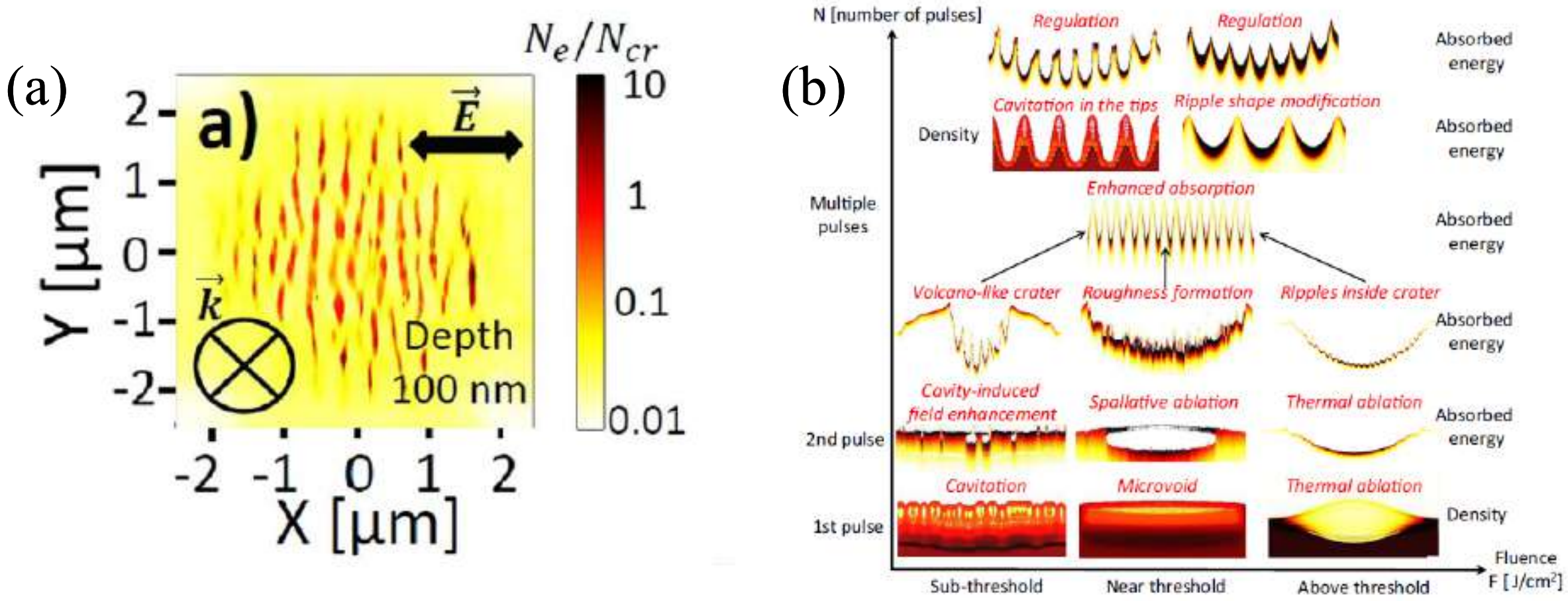

Fig. 16: (a) Top-view of electron density distribution calculated by 3D-FDTD calculations coupled with electron density equation (for $SiO_2$) (Reproduced with permission from Rudenko *et al* [216], and with permission from Springer Science and Business Media), (b) Schematics of coupled electromagnetic and hydrodynamic processes upon multipulse- femtosecond laser irradiation of metals as a function of laser fluence and number of applied pulses (Reprinted with permission from Rudenko *et al* [236]. Copyright (2019) by the American Physical Society).

### *3.1.2 Microgrooves*

Grooves manifest as quasi-periodic self-organized ablation lines with supra-wavelength periods. On metals and semiconductors they usually are aligned parallel to the linear laser beam polarization and feature periods of a few micrometers [206]. The origin of the grooves is not fully understood yet. Their formation is consistent with electromagnetic scattering/absorption [237], particularly if an additional reflection of the incident laser radiation at the wall of the emerging surface craters is present [206]. On the other hand, recent simulations point toward additional impact of hydrodynamic effects [222,223]. Although Marangoni convection and fluid transport allows a detailed description of LSFL structures, specific values of hydrodynamics-related parameters such as Prandtl and Marangoni numbers are considered to explain the creation of larger structures as well as their orientation [222,223]. A key feature for groove formation in metals and semiconductors is an originally surface profile covered with sub-wavelength ripples. The decreasing periodicity that occurs for LSFL ripples with increasing energy dose is not capable to explain the formation of grooves. While at a higher number of pulses, the ripple periodicity reaches a plateau, further irradiation leads to larger surface modulation depths (a result, predominantly, of ablation and mass displacement) that fails to yield a sufficient condition for SPP excitation and interference of the incident beam with the induced surface waves. In that case, enhanced energy accumulation due to a larger curvature and transport of the induced fluid volume along the walls of the ripples yields formation of counter-rotating convection rolls moving on a curved region (Fig. 17a) [197,222,223]. These are physical modes that are solutions of the Navier-Stokes equation, however, propagation of hydrothermal waves that lead to stable structures upon solidification are predicted for particular values of the frequency of the produced waves that may explain the supra-wavelength character of grooves (Fig. 17b). The proposed physical mechanism explains satisfactorily the groove formation for both metals and semiconductors, namely, the transition from ripples to grooves, the orientation of grooves and the periodicity (Fig. 17c,d) [197,222,223,238]. By contrast, simulations indicate that an initially formed crater with a large depth can rapidly lead to the excitation of sufficient hydrodynamical conditions that can induce groove formation (Fig. 17e,f). The requirement of an initial formation of LSFL structures for metals or semiconductors is also reflected on the presence of 'pseudoripples' on top of grooves as the amplitude of the hydrothermal waves during the dynamics of the instabilities is not sufficient to destroy the ripples (Fig. 17c,e). In contrast, in dielectrics, the fact that a rippled region is not required to excite convection rolls that leads to grooves, pseudoripples are not produced [197,239].

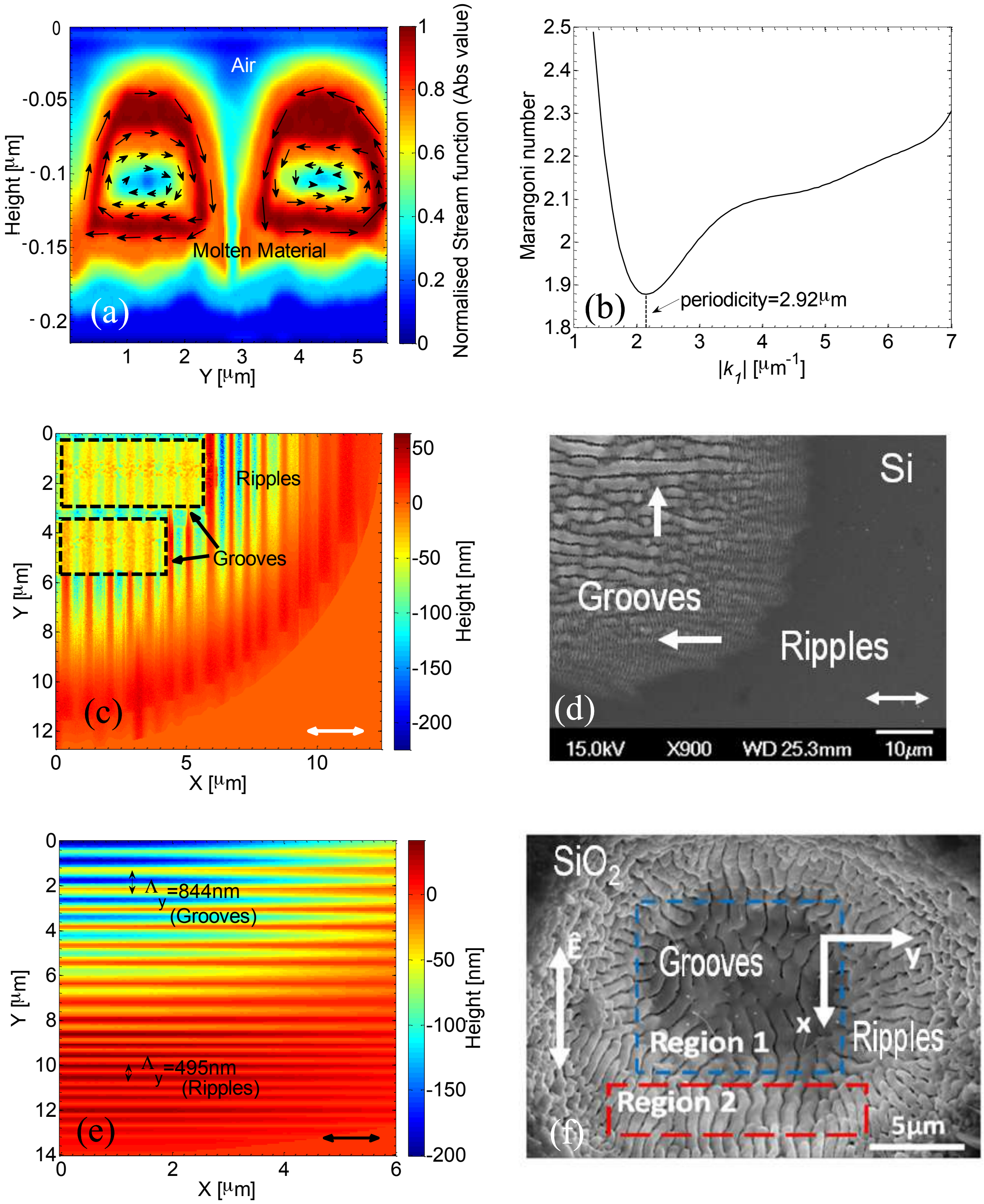

Fig. 17*:* (a) Convection roll and melt displacement, (b) Marangoni number and condition for stable solutions, (c,e) simulation results for grooves and ripples, (d,f) experimental results for Si and $SiO_2$. (Reprinted with permission from Tsibidis *et al* [197]. Copyright (2016) by the American Physical Society and Reprinted with permission from Tsibidis *et al* [223]. Copyright (2015) by the American Physical Society)

Furthermore, the enhanced role of the hydrodynamical factor as the curvature of the roughness of the irradiated material increases (i.e., at larger number of pulses) is also reflected on the increase of the groove periodicity at higher energy dose (Fig. 18a) [222]; this behavior is attributed to the dynamics of the molten material as a result of induced larger temperature gradients [197,222,223]. Similarly, groove periodicity variation with fluence increases as the energy deposition becomes larger (Fig. 18) [222].

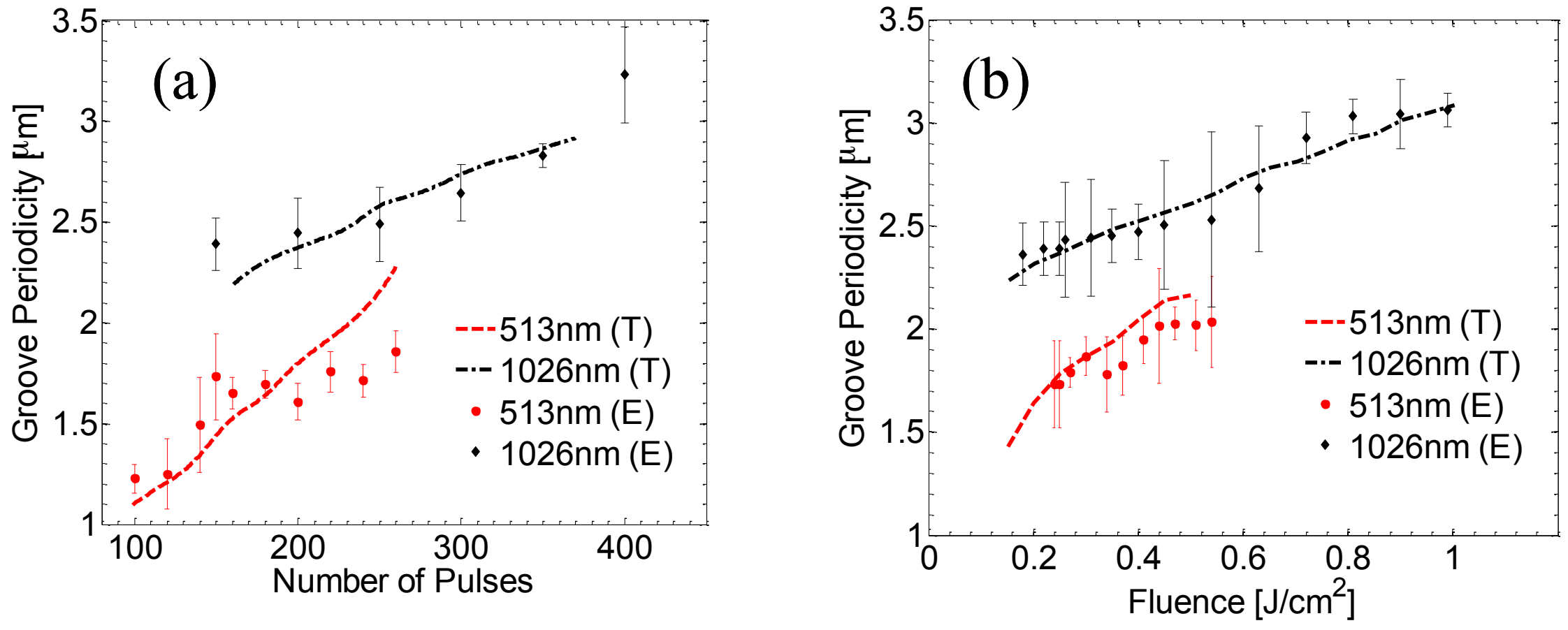

Fig. 18*:* Groove periodicity for 100Cr6 steel as a function of the number of pulses (a) and fluence (b) at laser wavelength $\lambda_L = 513$ nm and 1026 nm. ('*E*' and '*T*' stand for experimental and theoretical results). Figures were reprinted from [222] and with permission from Springer Science and Business Media

### *3.1.3 Microspikes*

For the laser processing of biomimetic spikes, two different processing strategies were suggested. The most straightforward approach (Strategy I) relies on a meandering line-wise movement of a focused laser beam across the surface at a certain (constant) velocity within each line, defining the effective number of pulses per beam spot diameter [64]. The chosen process parameters can be adjusted with the morphological map (Fig. 13) to result in the desired spikes morphology – typically for large $\phi_0$ and $N_{eff}$. In an extended approach (Strategy II) often used for high repetition rate lasers, this overscan process is repeated several times ($N_S > 1$) under identical conditions [166,240].

Figure 19 demonstrates the result of the two mentioned processing strategies (I & II) for fs-laser irradiation of stainless steel [240], i.e., the application of a single area scan at three different accumulated fluences ($F_{\Sigma line,max}$) after a single scan (top row, Strategy I) and for up to 24 more overscans (lower rows, all Strategy II). Note that both, $\phi_0$ and the number of laser pulses per beam spot area (*PPS*) were varied among the different columns in Fig. 19.

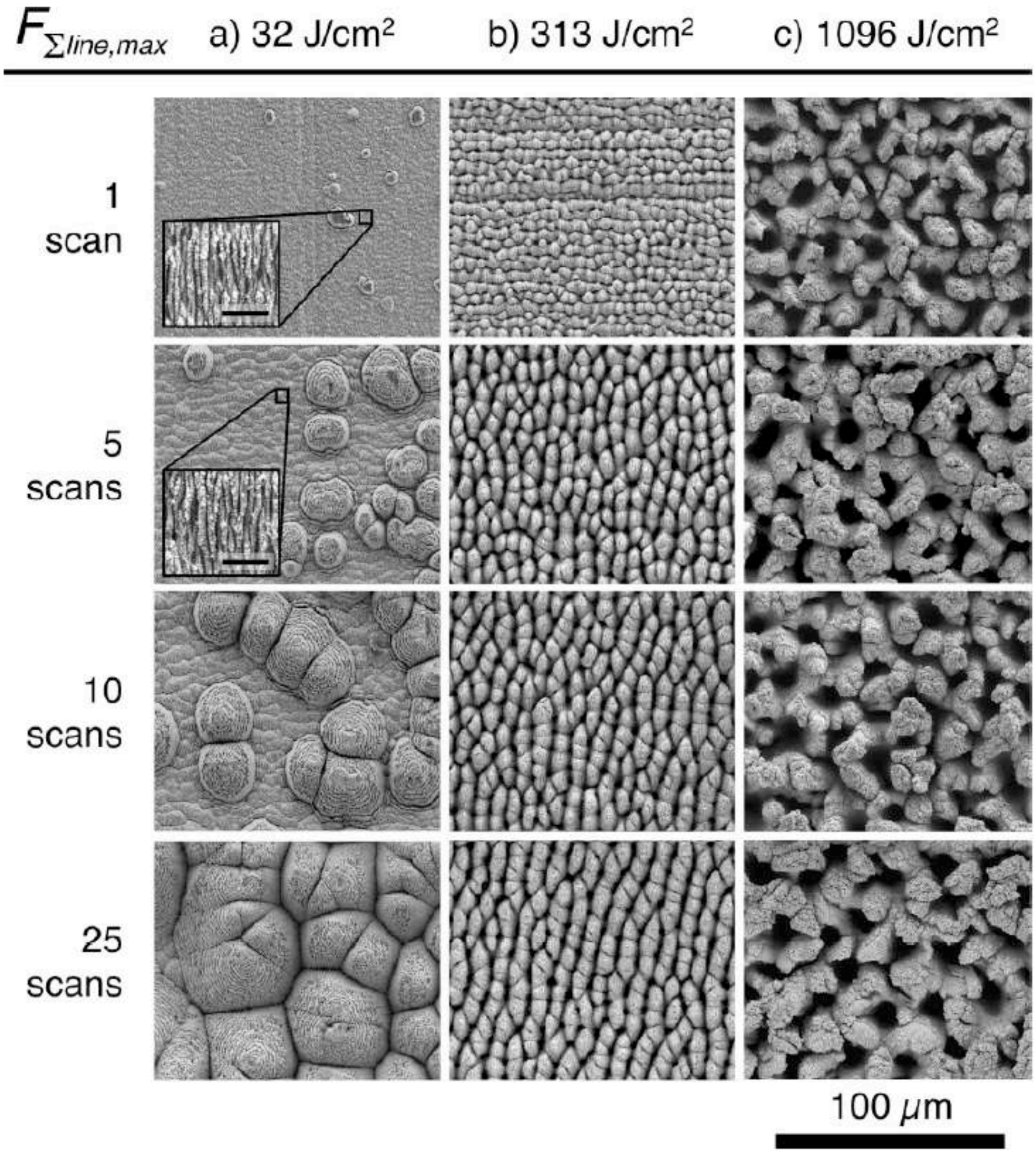

Fig. 19: Evolution of fs-laser processed surface morphologies on steel with an increasing number of overscans [800 nm, < 100 fs, 10 kHz]. (Reproduced with permission from Ling *et al* [240] . Copyright (2015) reprinted with permission from Elsevier)

At low accumulated fluences (Fig. 19, left column) processed at a peak fluence ($\phi_0$ = 0.11 J/cm$^2$) close to the ablation threshold and with a larger number of pulses (*PPS* = 1826), it becomes clear that the spikes "nucleate" here at specific surface sites. Their areal density and size increase with the number of overscans $N_S$, finally leading to their coalescence and a homogeneous coverage of the surface by coarse spikes. At medium accumulated fluences (Fig. 19, middle column), the surface is rather uniformly covered by finer microspikes ($\phi_0$ = 1.1 J/cm$^2$, *PPS* = 757), while at large accumulated fluences (Fig. 19, right column) the surface exhibits a rather chaotic morphological pattern ($\phi_0$ = 3.85 J/cm$^2$, *PPS* = 661) [122].

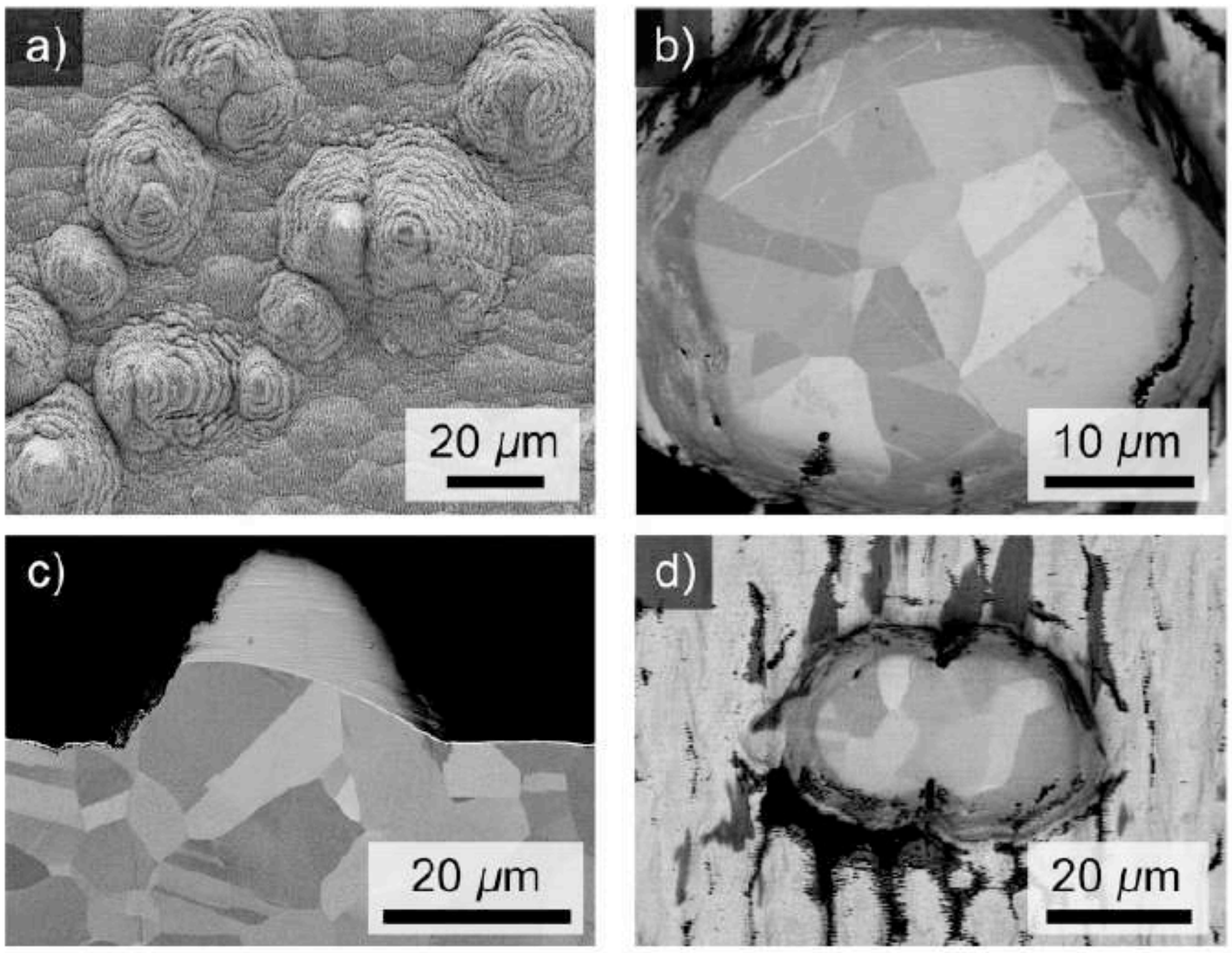

Fig. 20: Crystallographic analysis of spike structures. (a) Top-view SEM, (b)-(d) Electron channeling contrast imaging of cross sections. (Reproduced with permission from Ling et al [240]. Copyright (2015) reprinted with permission from Elsevier)

A crystallographic analysis performed at these spike structures ($N_S$ = 10, $\phi_0$ = 0.17 J/cm$^2$, *PPS* = 1400) is presented in Fig. 20 [240]. Electron channeling contrast imaging of cross sections through different spike structures revealed that the materials poly-crystalline structure remained unaltered, confirming that the spikes are not formed through structural or compositional alterations of the steel. The authors concluded that the spikes are composed of non-ablated steel covered by a layer of redeposited nanoparticles (debris) [240]. They summarized the laser processing according to the scheme reprinted in Fig. 21.

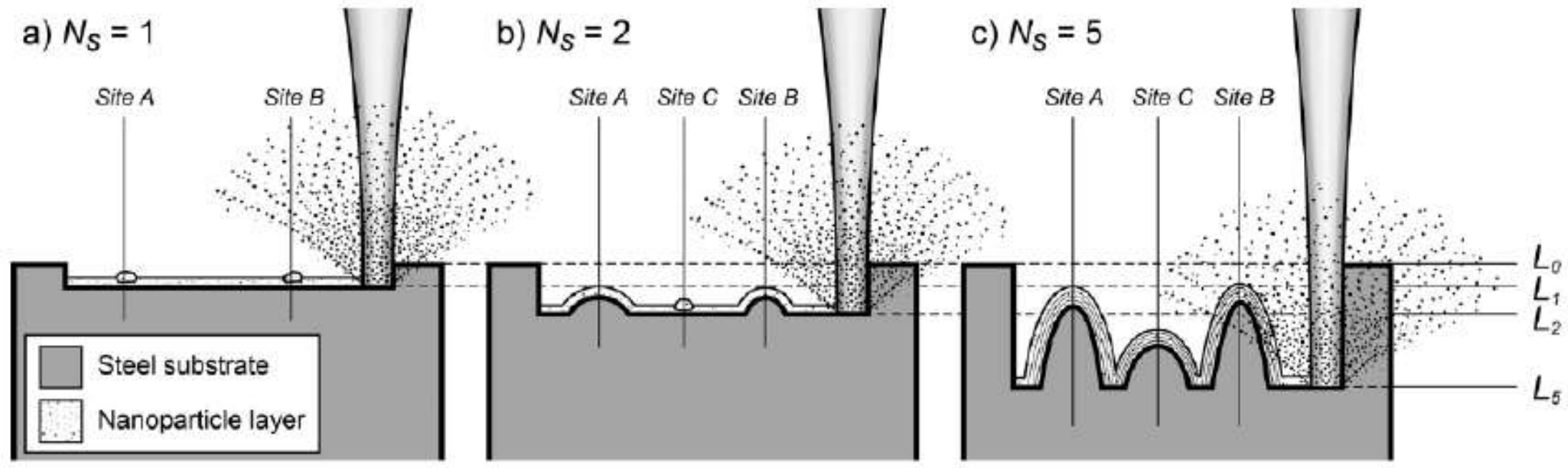

Fig. 21: Scheme of the formation of spikes structures upon multiple overscan fs-laser processing. (Reproduced with permission from Ling et al [240]. Copyright (2015) reprinted with permission from Elsevier)

The formation of spikes occurs here in the ablative regime, removing with every overscan a layer material from the surface. While most of the nano-scale debris are re-distributed homogeneously across the scanned area, some (nano-)particles agglomerate forming larger clusters. During the following overscan these clusters may gradually shield the underlying material from the incoming laser radiation, resulting in local protrusions at specific sites A, B, C, etc. The spike formation is further reinforced for increasing $N_S$ since at the edges of the protrusion the local angle of incidence of the radiation is increased, resulting in a reduced absorption and therefore reduced ablation. Moreover, the variations of local reflectivity/absorption caused by the polarization dependence for s- or p-polarized radiation (Fresnel) can lead to an elliptic shape of the bases of the spikes. Early variants of such particle initiated defect shielding mechanisms were proposed already for the ns-laser ablation of polymeric materials [241,242].

More general, Zuhlke et al. [243] summarized two distinct spikes formation scenarios through the following scheme (Fig. 22).

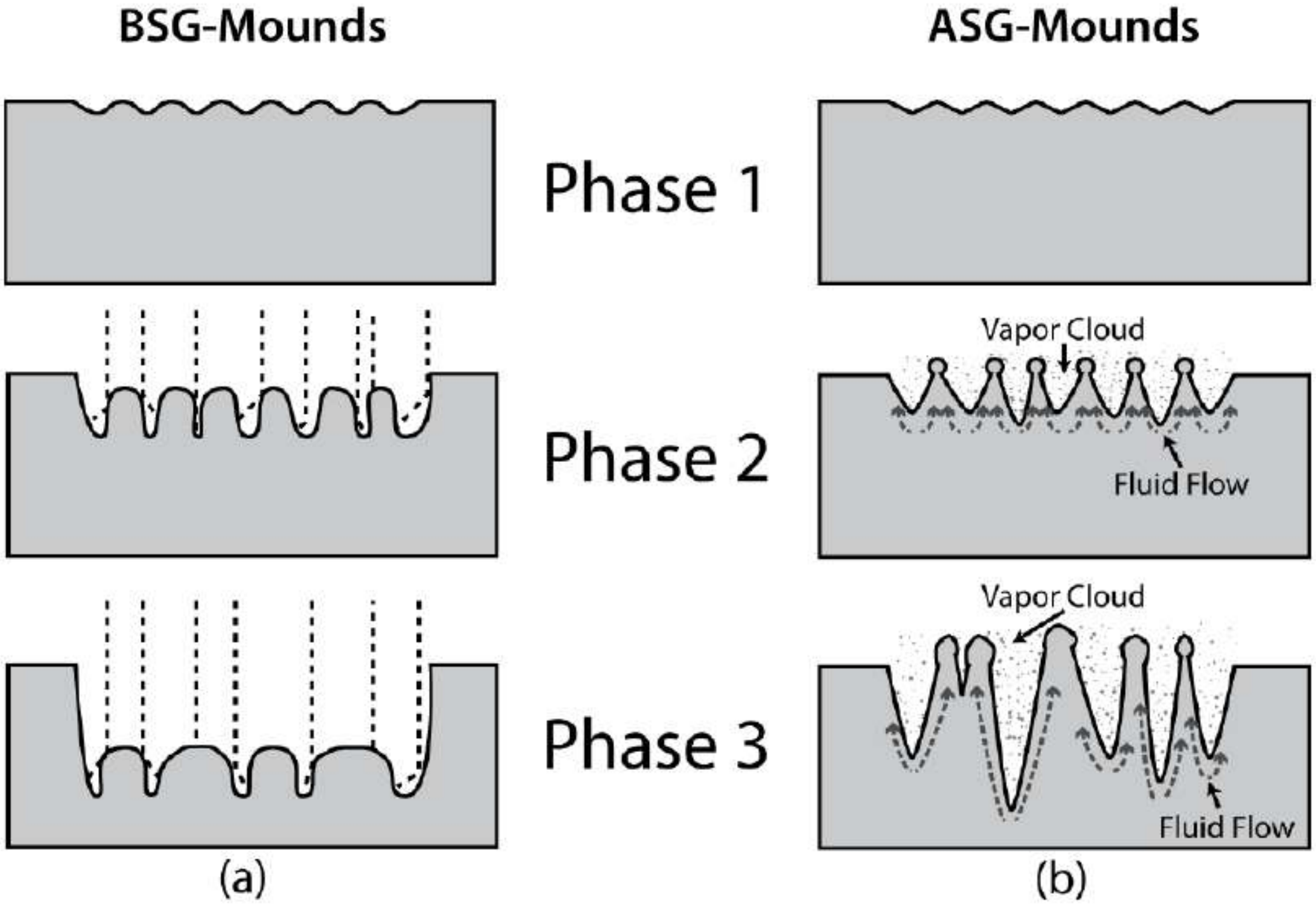

Fig. 22: Scheme of the formation of spikes through below surface growth (BSG-) mounds [(a), left column] or through above surface growth (ASG-) mounds [(b), right column]. Reprint permission text …[243]

In both scenarios (BSG & ASG) three different phases can be identified, i.e., the formation of precursor sites (Phase 1), followed by the development of multi-scale structures (Phase 2), before forming the final surface morphology (Phase 3) [243]. While the BSG scenario is mainly caused by electromagnetic scattering/absorption effects already discussed above, in the ASG scenario hydrodynamic melt-flows caused by thermocapillary [244] of chemocapillary effects and the recondensation of constituents of the ablation vapor cloud at protrusions of the surface may be involved [245]. Recent simulations come in agreement with the ASG scenario and attribute the microspikes formation to convection-roll driven hydrodynamic phenomena [223]. More specifically, simulations show that spike-related protrusions require an originally built groove-covered profile. For metals and semiconductors in which grooves are decorated with pseudoripples, further irradiation and increase of energy dose leads to gradual diminishing of the pseudoripples and increase of the height of grooves. This eventually leads to disappearance of pseudoripples that combined with the enhanced temperature gradients generated from the grooves' grating in this case drives predominantly the fluid flow in a direction perpendicularly to laser beam polarization vector. The enhanced surface tension gradients and temperature differences will produce an overall preferred Marangoni convection perpendicularly to the grooves and thus hydrothermal waves will be induced, following a similar process to the one described previously for grooves formation. The roll patterns that will be developed due to the convective instability evolve into supra-wavelength protruding structures upon melt solidification [223]. The height of the protruding structures progressively increases with increasing the number of pulses giving rise to spike-like assemblies upon further irradiation (Fig. 23). Although further theoretical investigation is required to establish a more enlightening picture of the process of spikes formation, comparison with experimental results that the proposed model underlines the predominant role of hydrodynamical processes (Fig. 23).

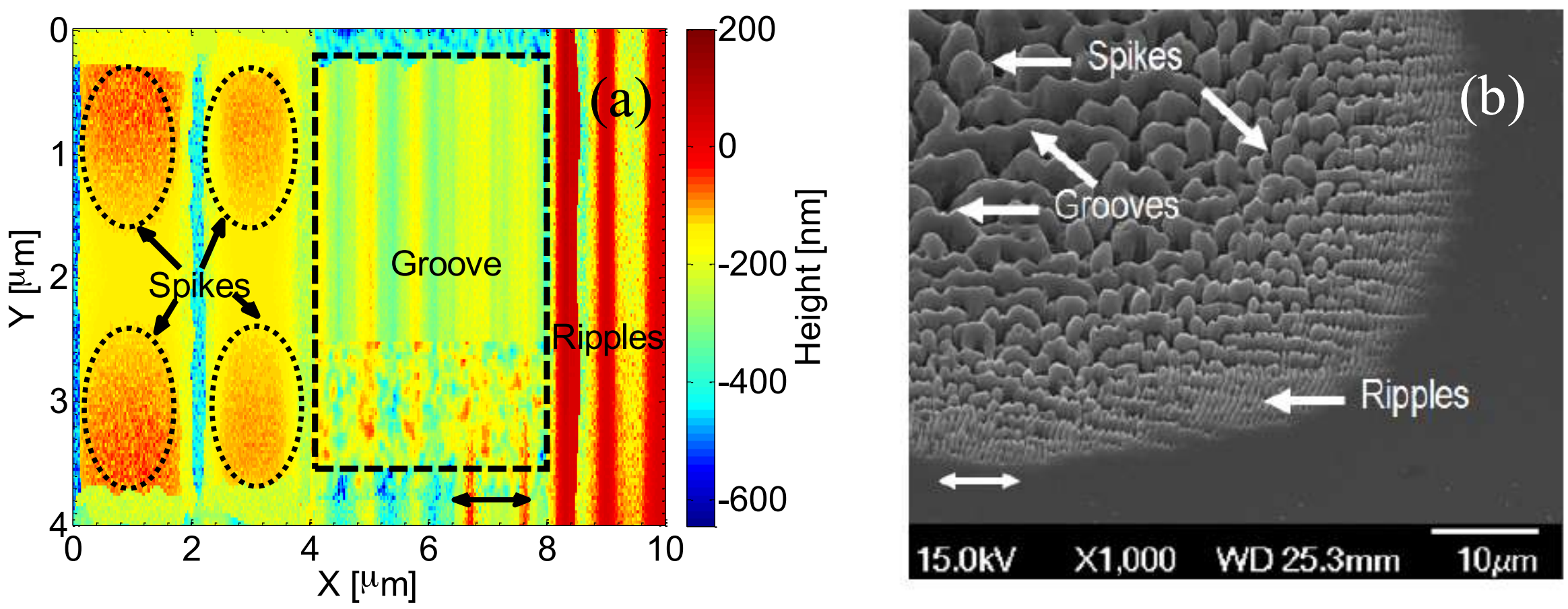

Fig. 23*:* (a) Surface pattern on Silicon following irradiation with 120 laser pulses, (b) Experimental results for 200 laser pulses (Double-ended arrow indicates the laser beam polarization). (Reprinted with permission from Tsibidis *et al* [223]. Copyright (2015) by the American Physical Society)

In Table 2, the main theoretical approaches towards modelling LIPSS formation are summarized:

Table 2: Literature survey of models proposed to explain the formation of LIPSS (LSFL, HSFL), Grooves, and Spikes

| **Work by** | **Type of structures** | **References** |
|---|---|---|
| Sipe et al. | LSFL (electrodynamics) | [203] |
| Huang et al. | LSFL (surface plasmon waves) | [211] |
| Bonse et al. | LSFL (electrodynamical models) | [204,206] |
| Reif et al. | LSFL (self-organization) | [224] |
| Rudenko et al. | LSFL, HSFL (electrodynamical + hydrodynamical models) | [216,236] |
| Amoruso et al. | LSFL, Grooves | [238] |
| Skolski et al. | LSFL, HSFL (electrodynamical models) | [221] |
| Tsibidis et al. | LSFL, Grooves, Spikes (transition between structures-hydrodynamical + plastic deformation) | [219]<br>[223]<br>[229]<br>[197,228] |
| Derrien et al. | LSFL (electrodynamical models) | [225,226] |
| Dolgaev et al. | Spikes (instabilities) | [245] |
| Zuhlke et al. | LSFL, Spikes | [243] |
| Fuentes-Edfuf et al. | LSFL (surface plasmon + roughness) | [208] |

### *3.1.4 Hierarchical structures and complex morphologies*

Despite the increasing scientific interest, regarding laser induced self-assembled structures, the majority of natural surfaces has yet been proven to be extremely difficult to mimic. This is a result of the pronounced morphological complexity that most natural surfaces exhibit. This structural diversity can be addressed regarding their spatial features, as a form of hierarchy and structural orientation which can be extremely complex. Due to this structural variety, over the years the scientific community has made some significant efforts following substantially different approaches for mimicking such hierarchical and complex surface structures.

So far, there have been three dominant approaches to mimic efficiently the natural structural hierarchies. Among the most popular techniques is the ultrafast laser direct writing method, which involves the use of reactive etching gas atmosphere to induce both micro and nano hierarchical structures (Fig. 24). This technique has been conducted successfully on silicon surfaces (black silicon) [164,246]. This method can be rapid; however, it is characterized by limitations to the size of the produced structures because the sample has to be exposed to a reactive gas which is in most of the cases $SF_6$ or in liquid environment [247,248]. The second approach is based on a two-step surface treatment [249,250] in which the laser beam is used to form an ablation pattern and in a second step the same beam with different parameters can decorate the unaffected regions with LIPSS. Finally, the last approach involves the use of a spot-by-spot irradiation approach for multi-scale structuring by exploiting the spatial distribution of cylindrical vector (CV) beams for the fabrication of precisely controlled surfaces consisting of structures with more than dual scale spatial frequencies [167]. This approach is characterized by an enhanced precision as it is possible to control the primary ablated micro-morphology with micrometer accuracy depending on the focusing conditions and also define the LIPSS spatial resolution as a secondary submicron morphology with the laser wavelength. However, this is a more time-consuming methodology.

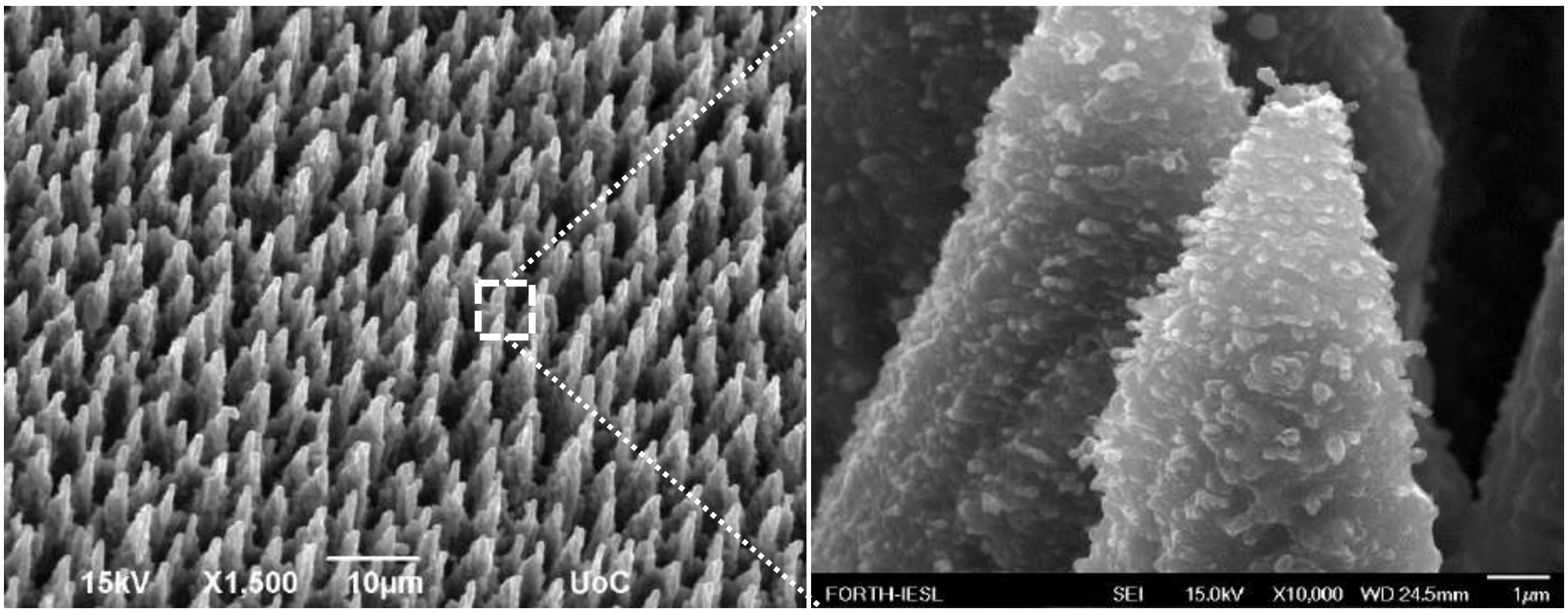

Fig. 24: SEM images of spike surface generated by fs-laser processing in $SF_6$ atmosphere to mimic the Lotus leaf on low magnification (left) and high magnification (right) (Reproduced with permission from Zorba *et al* [164], and with permission from Wiley)

### *3.1.5 Complex Polarization states*

Ultrafast laser processing has been proven a valuable method for the realization of biomimetic surfaces that can mimic both the surface morphological and functional features [199]. However, to date, laser fabrication of biomimetic structures has been mainly demonstrated mainly using laser beams with a Gaussian intensity spatial profile and spatially homogeneous linear polarization [194]. In this context, and based on the sensitivity of laser induced structures on laser polarization, it is possible to further advance the complexity of the fabricated structures via utilizing laser beams with a spatially inhomogeneous state of polarization. CV beams, exhibiting tangential polarization states, are some prominent examples. This prominent approach relies on the fundamental aspect of the fs laser material interactions that the spatial features of the surface structures orientation attained are strongly correlated with the laser beam polarization state. For example, laser-induced periodic surface structures (LIPSS) and quasi-periodic microgrooves, which are preferentially oriented perpendicular and parallel to the laser beam polarization respectively have been successfully produced by means of CV beams. In 2011, Hnatovsky et al. used complex polarization states including the presence of the longitudinal component of the electric field for sub-wavelength resolution diagnostic tool [251], which was lately used as a simple method to characterize intense Laguerre-Gauss vector vortex beams [252]. Due to the ability of the CV beams to produce complex morphologies the use of complex polarization states and cylindrically polarized laser pulses have widely been used to increase the directional complexity of LIPSS for various materials including metals [167,229,253,254] semiconductors [238,255–257] and dielectric surfaces [230]. Fig. 25 below illustrates laser induced morphological profiles for silicon, after static irradiations with cylindrically polarized optical vortexes generated via a q-plate.

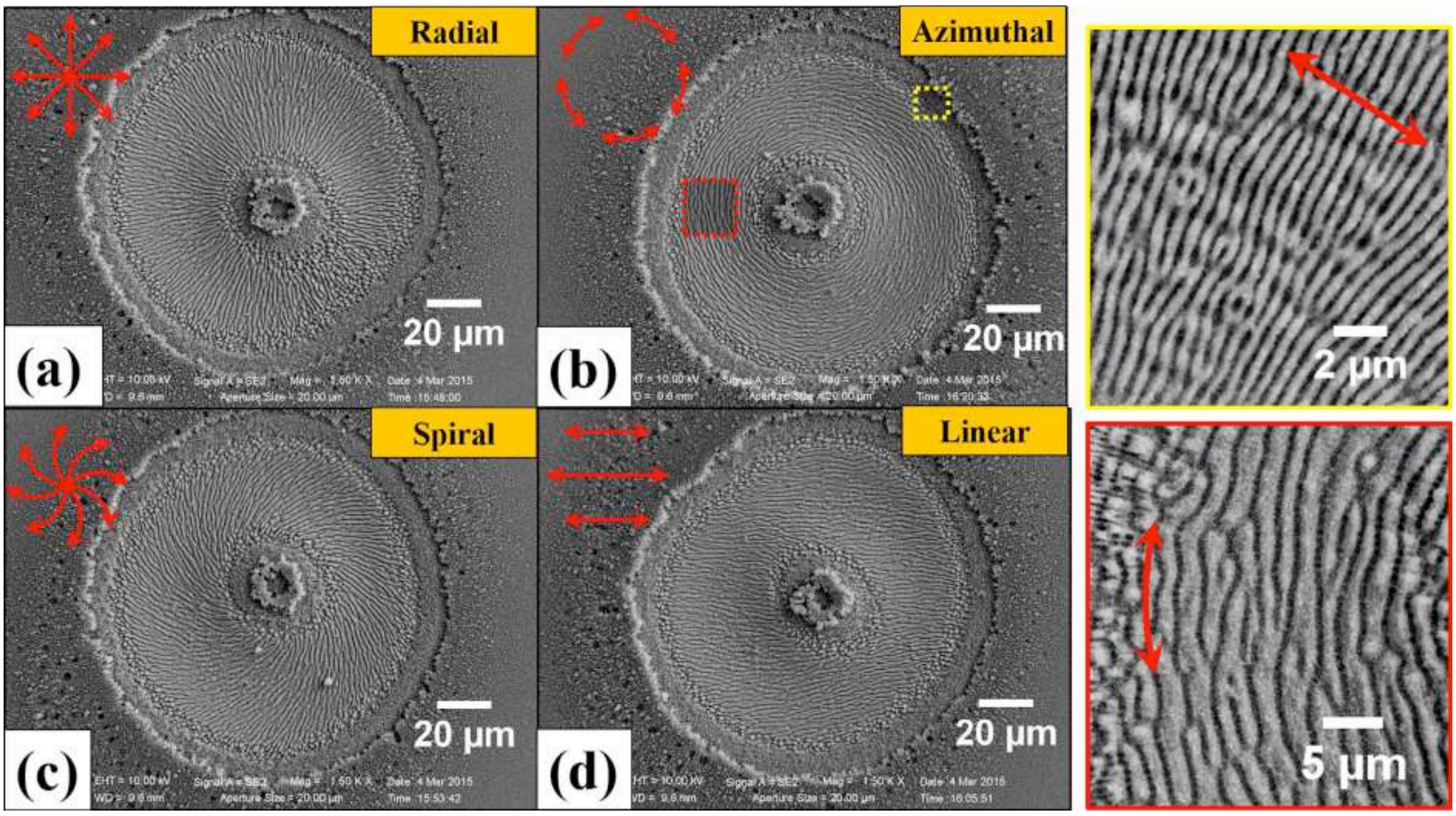

Fig. 25: Examples of the surface structures (LIPSS and grooves) developed on a silicon target after an irradiation sequence of N = 100 pulses at an energy $E_0$ = 48 μJ with: (a) radial, (b) azimuthal, (c) spiral and (d) linear polarization. (Reproduced with permission from Nivas *et al* [238], and with permission from Springer Science and Business Media)

Remarkably, the utilization of these donut-shaped beams with radial or azimuthal polarization states on dynamic surface structuring such as laser scanning has showed that the LIPSS morphological complexity and directionality can be further increased, resulting in rhombic, multi-ordered structures that could mimic the shark skin structural features [167]. This locally variant polarization can define dynamically, via raster scanning both shape and orientation of the self-assembled structures. The following Fig. 26 illustrates SEM images, depicting line scans produced by linearly (a,b), radially (c,d), and azimuthally polarized (e,f) beams, respectively. The precise control of the local electric field $\vec{E}$ changes within a laser pulse or within laser pulse cycles and in combination with line scanning could potentially produce even more complex and well-defined LIPSS.

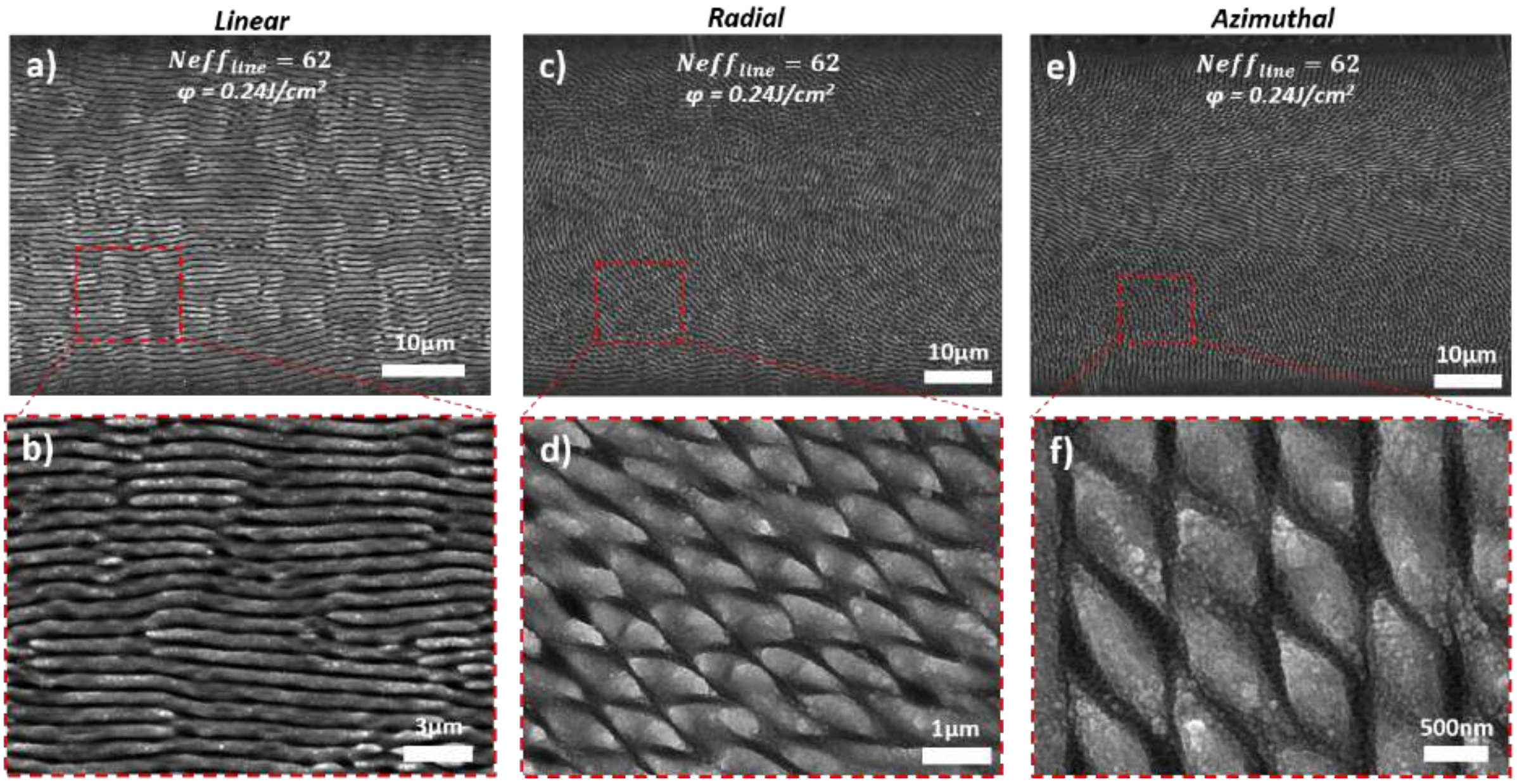

Fig. 26: Top-view SEM images depicting line scans produced by linearly (a,b), radially (c,d), and azimuthally polarized (e,f) beams, respectively, at v = 0.5 mm/s ($N_{effline} = 62$), and $\phi = 0.24\ J/cm^2$. The images (b,d,f) are higher magnifications of an area inside the red-dashed-squares and reveal the biomimetic shark skin-like morphology of the processed areas.) (Reproduced with permission from Skoulas *et al*, and with permission from Springer Science and Business Media),

Furthermore, circularly polarized ultrafast laser pulses are recently utilized to fabricate triangular periodic surface structures on metal surfaces [185,258] which can be used for inducing uniform iridescence of laser structured metallic surfaces.

### *3.1.6 Double-Pulse irradiations*

Another promising approach is the interplay between pulse delays in double-pulse irradiation experiments. Recently, many works, have used the combination of different polarization states [259–264] or different wavelengths [265–267] of two consecutive time delayed ultrashort laser pulses to fabricate 2D-LIPSS structures. The double-pulse interplays in combination with crossed polarization or counter rotating circularly polarized laser pulses can produce ordered 2-dimentional surface structures for specific pulse to pulse delays. The double-pulse approach is more versatile, since varying the delay and the properties of two consecutive laser pulses can potentially produce even more complex LIPSS and allows further control, although it is relatively difficult to execute and not so robust for industrial transfer. Fig. 27 from Fraggelakis et al. [265] shows the variation of surface morphologies of stainless steel after double-pulse irradiation with an inter-pulse delay ranging from 0 to 10 ps, for both polarization configurations: double cross-polarized (XP) and double circular counter rotating polarizations (CP). Finally, the combination of cylindrically polarized laser pulses with Gaussian ones on a pulse-to-pulse interplay is yet to be investigated as it can potentially lead to promising possibilities of complex bio-inspired surface structures.

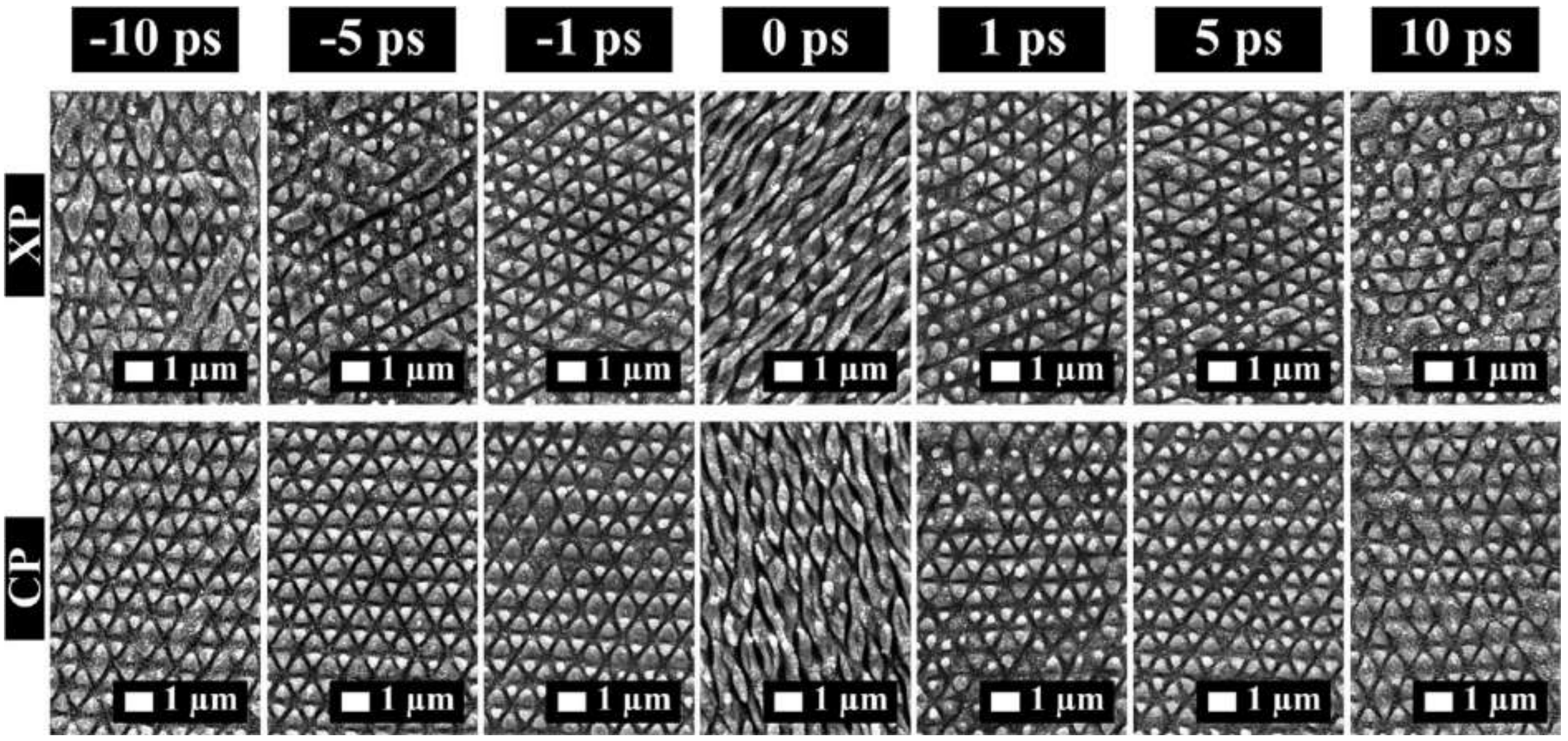

Fig. 27: SEM images of stainless-steel surfacea irradiated with two different polarization configurations (XP and CP) for differet double-pulses delays as indicated at the top. The rest of process parameters were fixed; pulses per laser spot (pps) = 10 and the distance between two scanning lines (H) = 1 while the fluence was $\Phi = 0.1$ J/cm$^2$. (Reproduced with permission from Fraggelakis *et al* [259]. Copyright (2019) reprinted with permission from Elsevier)

## 3.2 Directly-written surface structures

The use of focused laser beams to directly write complex structures on material surfaces is a powerful processing technique which allows high-precision, contact-less surface patterning over large areas, and which has already found strong uptake by industry for numerous applications. One of the key advantages is the inherent flexibility to change the pattern to be written via digital process design, without the need to fabricate physical masks or master samples. Using ultrashort laser pulses for direct writing adds further advantages to this technique: The high peak intensities of such pulses enables processing of transparent materials through strong field ionization, not only at the surface but also inside the material [268,269]. Moreover, their ultrashort pulse duration effectively reduces unwanted thermal effects that often cause collateral damage, thus greatly enhancing the precision and spatial resolution of the fabricated structures [270].

The main drawback of direct writing is the inherent incompatibility between the fabrication of very small features with high processing speed, since the writing of small structures requires tight laser focusing (small spot size) and sequential writing. An attractive alternative is Laser Interference

Lithography (LIL), sometimes also called Direct Laser Interference Patterning (DLIP), explained in more detail in one of the next sections, which allows parallel processing to produce patterns with very small features, but which requires a complex layout and lacks flexibility. Although the use of liquid-crystal based spatial light modulators for parallel processing of complex structures by beam multiplexing [271] provides an alternative in some cases, pattern refreshing rates and SLM´s damage thresholds at high average powers are still an important issue. These drawbacks are partly overcome by self-organization based processing, as discussed in the previous section, which allows the fabrication of nanoscale features despite using relatively large laser spot sizes. It should be emphasized here that in many cases hybrid processing strategies can be used, combining fast direct writing of low-resolution patterns with high-resolution nanostructures, formed by self-organization processes as substructures of the coarse pattern.

In the following sections, we briefly review different types of biomimetic structures that can be fabricated with direct writing approaches.

### *3.2.1 Simple structures*

The simplest 2D structures that can be produced consist in parallel or orthogonal lines, forming respectively a grating or grid structure [250]. The width of each line is approximately equal to the laser spot size used (typically from 5 - 50 μm) and the separation is chosen depending on the application. Even such simple structures show promise in several biomimetic applications, as recently reported by Florian et al. [272], who demonstrated high control over the contact angle (CA) of water on the structure in steel. By changing the grid spacing, the CA could be varied continuously from 100º up to 150º, and the use of a grating (parallel lines) instead of a grid (crossed lines) structure allowed achieving anisotropy in the wetting behavior (see Fig. 28). This study also demonstrated the importance of the nanoscale sub-structures formed by self-organization, which were shown to influence the CA.

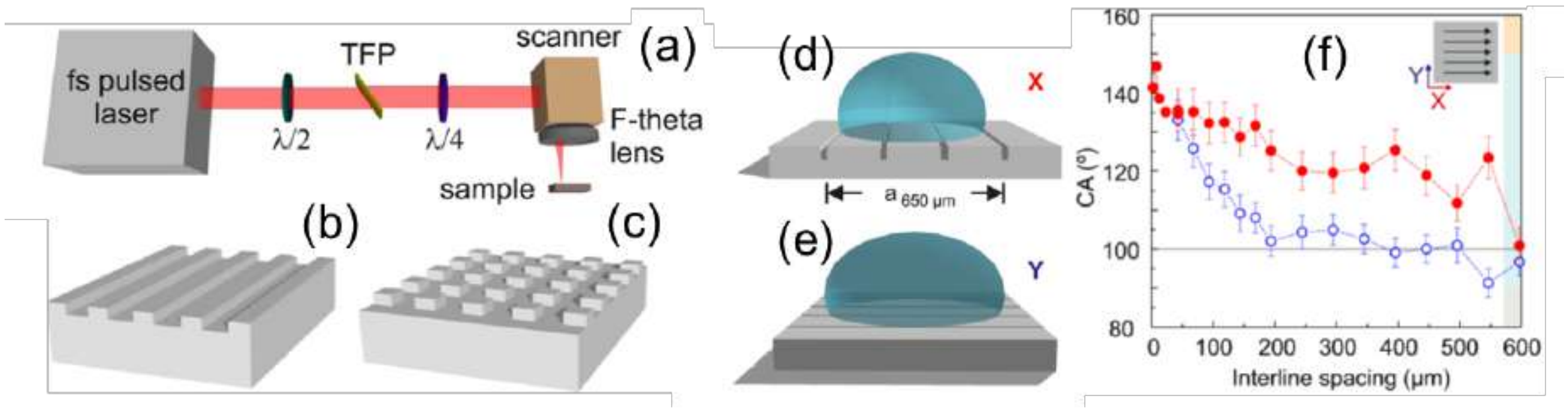

Fig.28: (a) Scheme of the experimental setup for fs-laser surface structuring. (b,c) Sketches of the simple structures produced, composed of parallel and crossed lines/trenches. (d,e) Sketches of a water droplet deposited over the structure fabricated with a certain interline spacing for parallel lines from two different viewing directions. (f) Contact angle (CA) of the water drop for parallel (b) line structures. Solid red circles (●) and empty blue circles (○) correspond to CAs measured along the X

and Y viewing positions, respectively. Adapted with permission from Ref. [272], Copyright (2018) American Chemical Society.

### 3.2.2 Complex patterns

The complexity of directly-written structures is only limited by the spatial resolution achievable. The most common implementation is the use of a laser scanning head, composed of two galvanometric mirrors combined with a telecentric lens to focus the beam. The desired structure is designed with computer software and the trajectories are sent to the laser scanner, which scans the beam over the static sample. Linear processing speeds in the m/s range are easily accessible in the case of planar samples [212]. For non-planar samples, an additional module is added that allows changing the focal position along the beam propagation direction. A well-known example for a highly complex 3D object fabricated that way is a metal stent inserted in a patient's blood vessel to prevent closure [273]. A recent example of a biomimetic structure fabricated by direct writing is a fluidic diode for passive unidirectional liquid transport, inspired by the topography of the spermathecal duct of fleas [59].

### 3.2.3 Processing in vacuum/gas atmosphere

An important factor that strongly influences both the morphology and the chemical composition of the laser-processed surface is the presence or absence of air or other types of gas during processing [242]. A gas environment of a given pressure enables chemical reactions (for instance oxidation in Ge [274]) at the surface of the laser-heated material, whereas working in vacuum increases the expansion velocity of the ablating material [275], strongly reducing debris formation around the laser processed region. Using non-oxygen based reactive gases allows the formation of chemically different surfaces with radically different properties [242,276]. It has to be said, though, that the complexity and thus cost of laser processing strongly increases when not working in air.

### 3.2.4 Interference patterning

As mentioned in the beginning of this section, the drawback of direct writing is the processing speed due to the need of writing each structure sequentially. Laser interference lithography (LIL) allows overcoming this limitation and writing a whole array of structures in parallel. In this sense, LIL is similar to self-organization based approaches, but does not rely on a material response to generate an interfering surface wave. Instead, the surface is irradiated by two or more coherent laser pulses [277] that interfere, generating a periodic intensity pattern that can be imprinted upon single pulse exposure into the material to form periodic structures over centimeter areas. For the case of two beams, the period $\Lambda$ is a direct function of the laser wavelength $\lambda$ and the half-angle $\theta$ formed between the beams, according to $\Lambda = \lambda / (2 \cdot \sin(\theta))$. If three or more beams interfere, complex intensity patterns can be produced, whose shapes are also a function of the relative intensities of the individual beams and the projected directions of incidence, defined by the angles $\alpha, \beta, \gamma$ (see Fig. 29(a)).

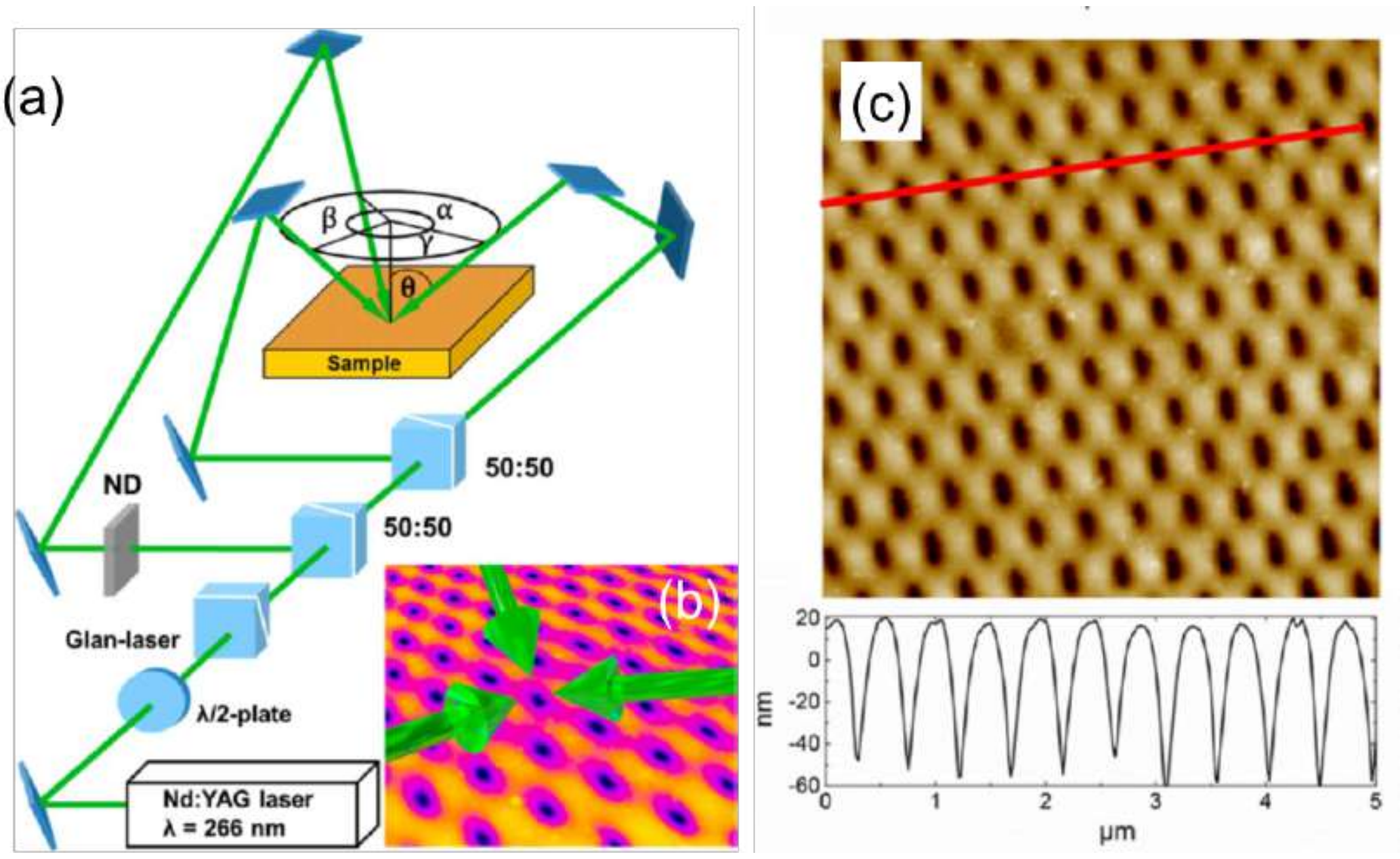

Fig. 29: (a) Scheme of the experimental setup for three-beam interference using a 266 nm laser. (b) Bird's-eye view of the sample surface, featuring the three beam directions and the imprinted pattern. (c) Topography image (5×5 $\mu m^2$) of the imprinted sub-micrometric cavities generated in a polymer film, with the depth profile along the red line depicted at the bottom. Adapted with permission from Ref. [278], Copyright (2014) American Chemical Society.

It should be mentioned, though, that the experimental implementation is rather complex compared to direct writing or self-interference based writing. Implementation is particularly a challenge when using ultrashort laser pulses [279], since it requires a perfect matching of the different beam path lengths up to a precision that is given by the product of the pulse duration and the speed of light, which amounts to approximately 30 μm for a 100 fs pulse. Moreover, the contrast of the interference patterns diminishes when femtosecond pulses with an inherently broad spectrum are involved. This is most likely the reason why this technique is mostly used with ns- or ps-laser pulses with a longer coherence length and suitably narrow spectrum, thus limiting the range of materials that can be processed over a large area. Impressive results have been obtained on thin polymer and metal films [280,281], but also on polymers [278,282], steel and Ti-based alloys [283].

## 3.3 Applications

The different biomimetic and self-organized structures discussed in the previous sections cover a broad range of surface morphologies and can be fabricated in most material classes, including metals, semiconductors, and polymers. It is this universality of both, the process and the structures that makes them ideal candidates for a vast range of technological applications that have already been proposed and are constantly being updated. In this section, we attempt to provide an overview over the most relevant applications to date. In many cases a certain structure fabricated in a given material has a dominant application in a given field, which changes upon fabricating the same structure in a different material.

Fig. 30 shows an overview of already-identified applications of the different structures, both self-organized and directly written, arranged into four major groups: Photonics, Biology/Medicine, Wetting/Microfluidics, as well as Other Technological Applications. The specific applications in each group will be reviewed briefly in the next sections. Due to the vast number of publications for each application, and in order to provide an overview of the state of art, we have preferred to cite recent works and reviews, rather than those that first reported the proof of concept, which are quoted in the recent references given. The reader can also refer to recent reviews focusing on specific types of applications [194,198,284–287].

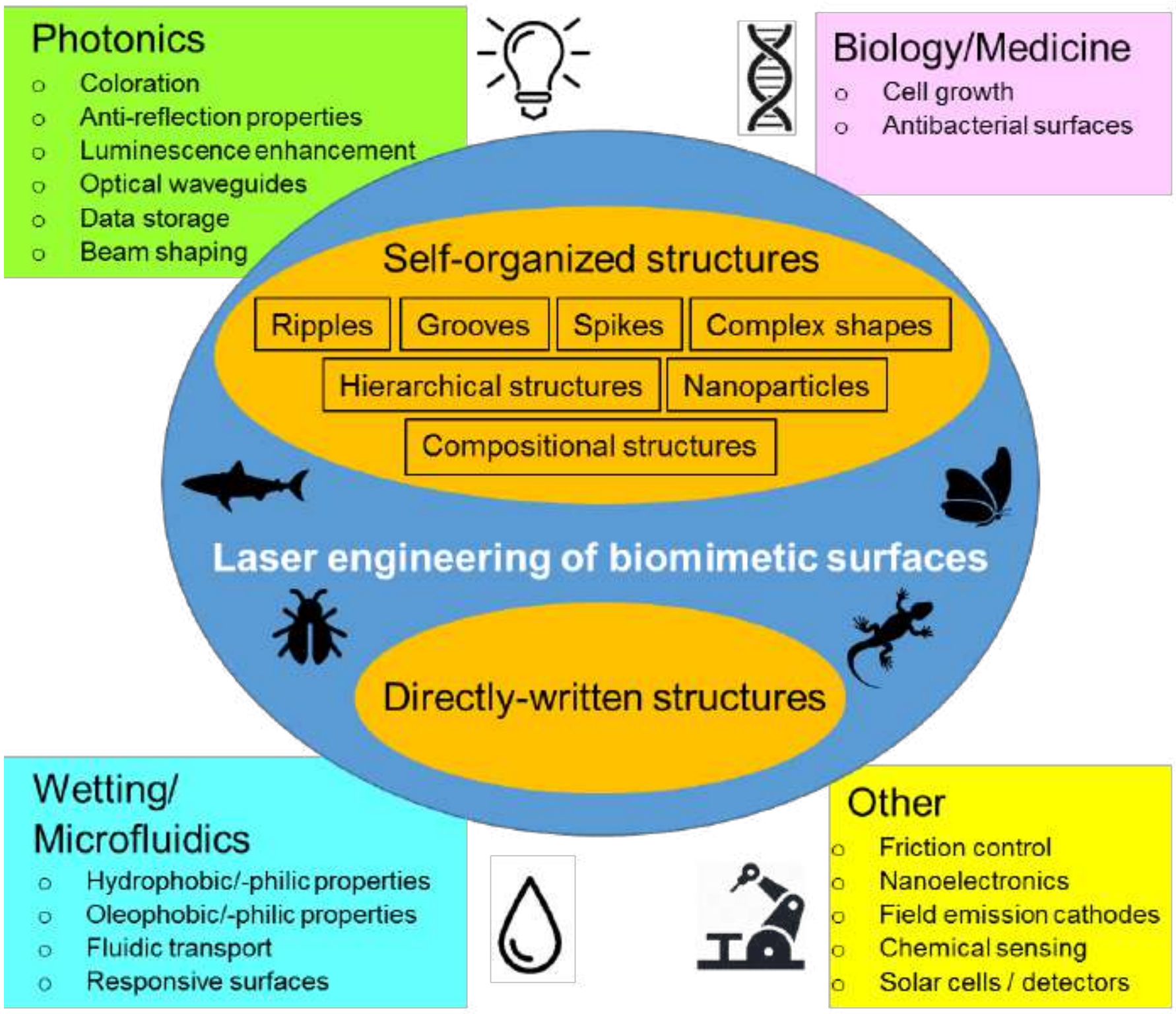

Fig. 30: Overview of different applications of laser-fabricated biomimetic surfaces.

### *3.3.1 Photonics*

***a) Coloration:*** The use of laser-induced self-organized biomimetic structures for coloring materials is arguably the most well known biomimetic application. Since the typical period of the structures matches the wavelength range of visible/near-infrared light, pronounced changes appear in the reflection or transmission spectra of a material, and in some cases diffraction of white light occurs into its different spectral components. It should be stressed that the coloration effect is caused by the surface structure/morphology rather than by wavelength-selective absorption of pigments. These so-called structural colors, as reported at section 2.1.1 has triggered a huge research activity aimed at transferring the impressive optical effects observed in many animals to technologically relevant materials. In this context, the work of Vorobyev and Guo [286] should be mentioned, who demonstrated for several metals a considerable control over the color appearance through generation of self-organized surface structures with femtosecond laser pulses. The authors reported not only angle–dependent colors caused by diffraction of periodic surface structures, but also angle-insensitive colors generated by irregular nanostructures with tuned morphological and statistical properties. The latter coloring strategy was further developed by Guay et al. [288], who were able to fabricate an extremely broad Hue color range of angle-insensitive colors on silver, gold, copper and aluminum surfaces. The authors were able to relate the spectral change to the size and distribution of the nanoparticles decorating the surface and causing plasmonic effects, as shown in Fig. 31.

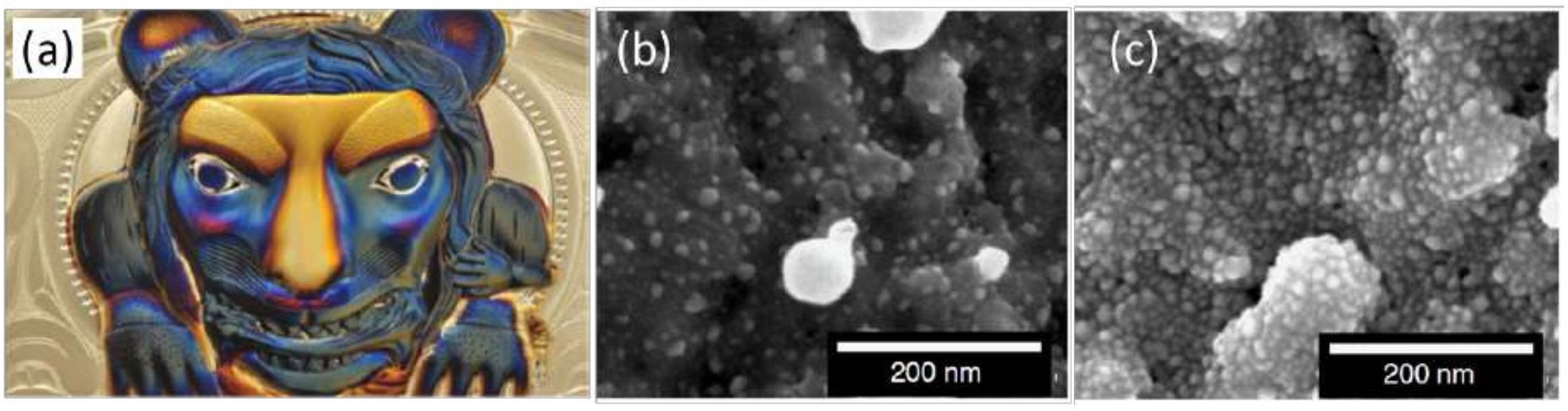

Fig. 31: (a) Photograph of a 14 cm x 9.2 cm region of a laser-colored 5 kg silver coin with significant topographic variations. SEM images of surface regions corresponding to laser-generated (b) blue and (c) red colors showing the different nanoparticle (bright spots) distributions at the surface. Adapted from Ref. [288], licensed under Creative Commons BY 4.0.

Plasmonic control through self-organization of nanoparticles for coloration applications is not limited to bulk metal systems but has been achieved in semitransparent nanocomposites. Remarkably, the processes can be triggered even with continuous wavelaser irradiation [289]. Recently, Liu et al. [290] demonstrated fs laser-induced growth and 3D of silver nanoparticles embedded in meso-porous $TiO_2$ thin films, caused by simultaneous excitation of independent orthogonal optical modes at different depths in the film. Such structures feature spectral dichroism and can be used in reflection and transmission, which enables multiplexed optical image encoding and offer great promise for applications in solar energy harvesting, photocatalysis, or photochromic devices.

***b) Anti-reflective properties:*** Inspired by the glass-wing butterfly *Greta oto* and various Cicada species, laser structuring has been employed to generate structural anti-reflective properties in glasses (Fig. 32). The randomly distributed nanopillar-like morphology of the butterfly wings could be reproduced by using circularly polarized fs laser pulses to fabricate omnidirectional transparent antireflective glass surfaces in the visible and infrared spectral ranges [7]. Moreover, it has also been demonstrated that fs laser nanostructuring of a SiC surface leads to an increase in visible light transmission by a factor of more than 60 [291].

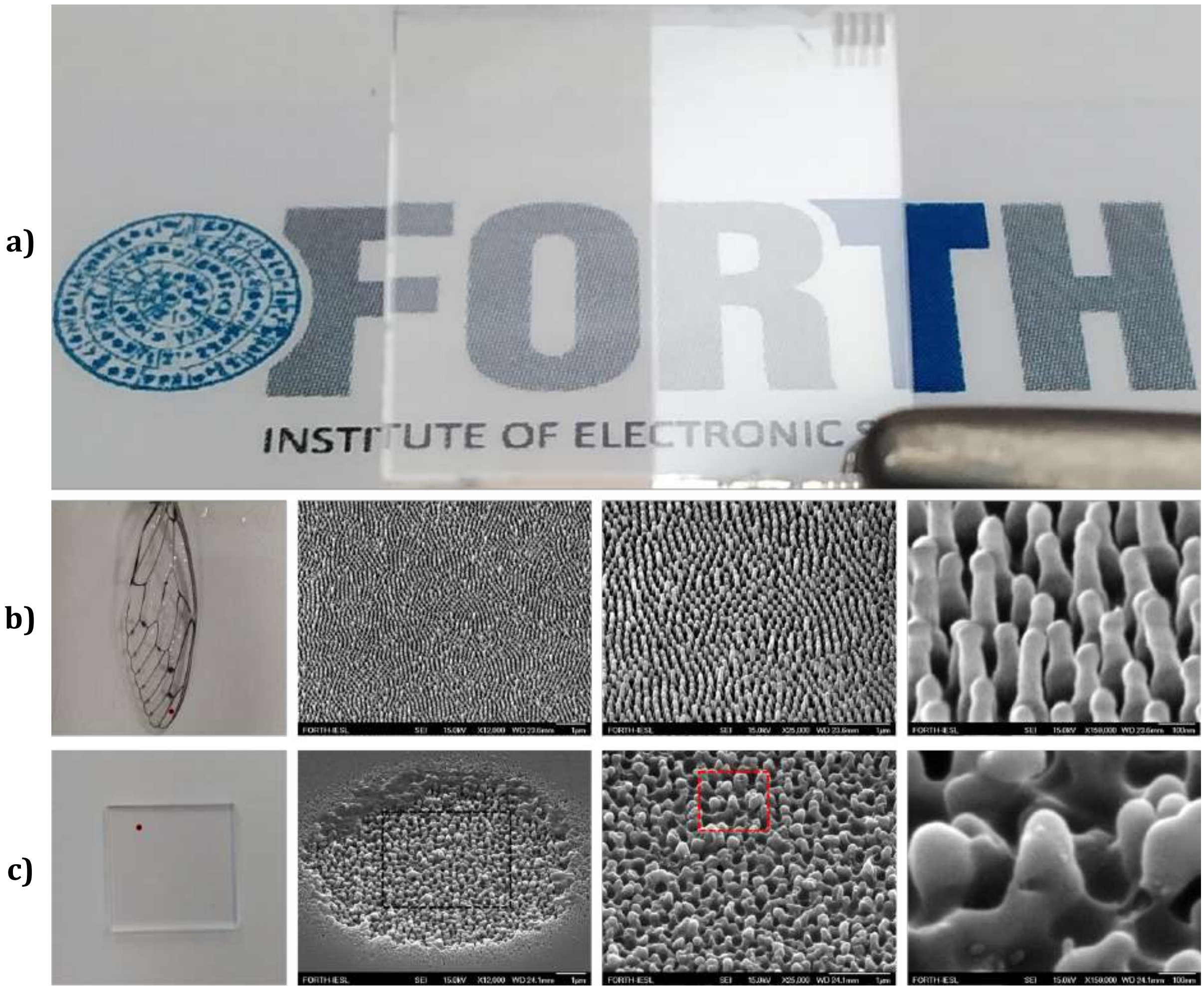

Fig. 32: (a) Photograph with the left-half laser treated biomimetic glass under ambient luminance. (b) Photograph of a *Cicada cretensis* wing, with respective SEM images (45° tilted) of the transparent antireflective area of the wing shown in (c) Photograph of a biomimetically laser-processed fused silica sample plate held before a printed paper. The central part of the plate was laser-treated to fabricate nanospikes resembling those of the wing; the black dashed rectangle indicates the processed area. Respective SEM images (45° tilted) of the laser-processed fused silica sample at different magnifications. (Reproduced with permission from Papadopoulos *et al* [7] and with permission from Wiley).

***c) Luminescence enhancement:*** In light-emitting diode (LED) applications based on GaN, Chen et al. showed that laser-induced periodic surface structures in the p-GaN layer caused a 30% increase in the output power, due to an increased surface area [292] achieved. Strong enhancement of UV luminescence in ZnO was also reported after nanostructuring the surface with interfering fs laser beams [293]. The authors attributed the effect to an increase in optical absorption accompanied by the formation of surface defect states. Also in ZnO, laser-fabricated, self-organized LIPSS have been used as a template for vapor solid growth of ZnO nanowires, with high aspect ratios and preserved luminescence properties [294].

***d) Data storage/beam shaping:*** It should be stressed that laser-induced self-organization processes are not limited to the surface of a material but can also be formed underneath its surface. One example are sub-wavelength periodic gratings inside transparent materials that share their electromagnetic formation mechanism with high spatial frequency LIPSS (HSFL) found at surfaces [216]. Such sub-surface self-organized gratings in form of voxels can be produced by focusing fs laser pulses inside fused silica glass [295]. These voxels can be erased and rewritten, which makes this strategy very interesting for high density 3D data storage (see Fig. 33). The information encoded in an individual voxel is not limited to binary information but can be multiplexed via the rotation of the nanograting, giving rise to optical anisotropy (form birefringence) and effectively adding a fourth dimension [296]. A fifth dimension can be added in form of the strength of the retardance, further increasing storage density [296]. The strong current interest in this type of application is motivated by the physical longevity of these structures, theoretically stable for up to 14 billion years, reason for which this concept has been chosen to encode information to be sent into space via SpaceX's rocket launch in 2018.

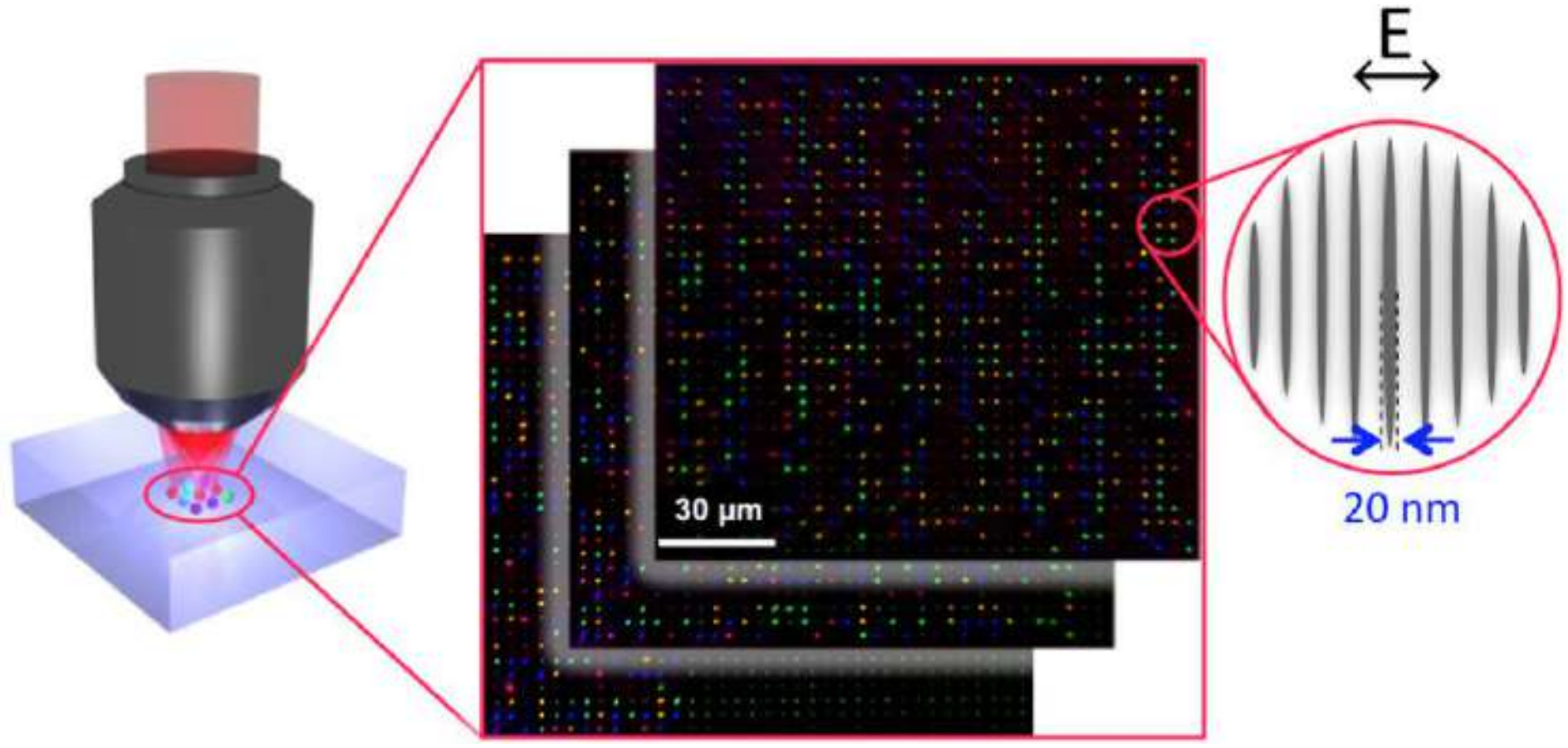

Fig. 33: 5D optical data storage. Voxels written inside a fused silica sample, using a focused femtosecond laser beam. Three spatial dimensions and two optical ones (the slow-axis orientation and the retardance) are exploited. Each voxel contains a self-assembled nanograting that is oriented in a direction perpendicular to the light polarization. The distance between two adjacent spots is 3.7 μm and the distance between each layer is 20 μm. E: Electric field of light wave. Arrow: Polarization direction. Reprinted with permission from [296].

The same self-organized grating structure in fused silica has also found applications in generating laser beams with exotic properties by means of inscribing complex 2D distributions of such gratings inside the material [297]. That way, continuous phase profiles of nearly any optical component can be fabricated in silica, which allows unparalleled control over the local phase of a laser beam that is transmitted by the element. This enables fabrication of advanced optical components (known as Q-plates) that are capable of generating beams with radial and azimuthal polarization, Airy beams and optical vortices, to name a few. The availability of such beam shaping elements for laser processing has already led to the fabrication of highly complex self-organized structures [167].

***e) Optical waveguides:*** Another self-organization process that can be triggered inside glasses is the self-organized re-distribution of different elements of the glass composition [287,298]. This ion migration process can be achieved on a micrometer scale, within the focal volume of a fs laser beam.

The driving force of this Soret-like effect has been related to the equilibration of the chemical potential inside the laser excited region [299]. This laser processing strategy has been further developed in order to fabricate optical waveguides by writing tracks inside a glass, whose self-organized cross sections feature a region enriched by heavy elements, locally increasing the refractive index and thus strongly confining light [287,300]. Using this technique, efficient optical waveguides, as well as optical amplifiers and integrated lasers with high net gain have been fabricated [299,301].

### *3.3.2 Wetting and fluid transport*

An excellent review of laser-induced wetting control of materials is found in Ref. [302], focusing on superhydrophobicity, underwater superoleophobicity, anisotropic wettability, and smart wettability. The following sections give a brief overview on the field and expand on microfluidics.

***a) Hydrophobic/hydrophilic properties:*** Another well-known field of applications for biomimetic surface structures is based on their interaction with water. Inspired by the extraordinary water-repellent properties of the lotus leaf (see Fig. 34), equivalent properties have been achieved in metals [303], [286], [250], [272], semiconductors [164], and dielectrics [304] employing fs laser irradiation to generate self-organized structures. For the specific case of dielectrics, wet-oxidation and coating post-processing adds another property to the surface, the strong adhesion of the water drop to the surface, despite being water repellant, a behavior which is equivalent to the rose petal effect found in nature [305]. Laser-induced hydrophobicity can also be achieved in natural stone, such as marble, as recently reported in [306] and currently done within the European Project BioProMarL [https://www.biopromarl.com/].

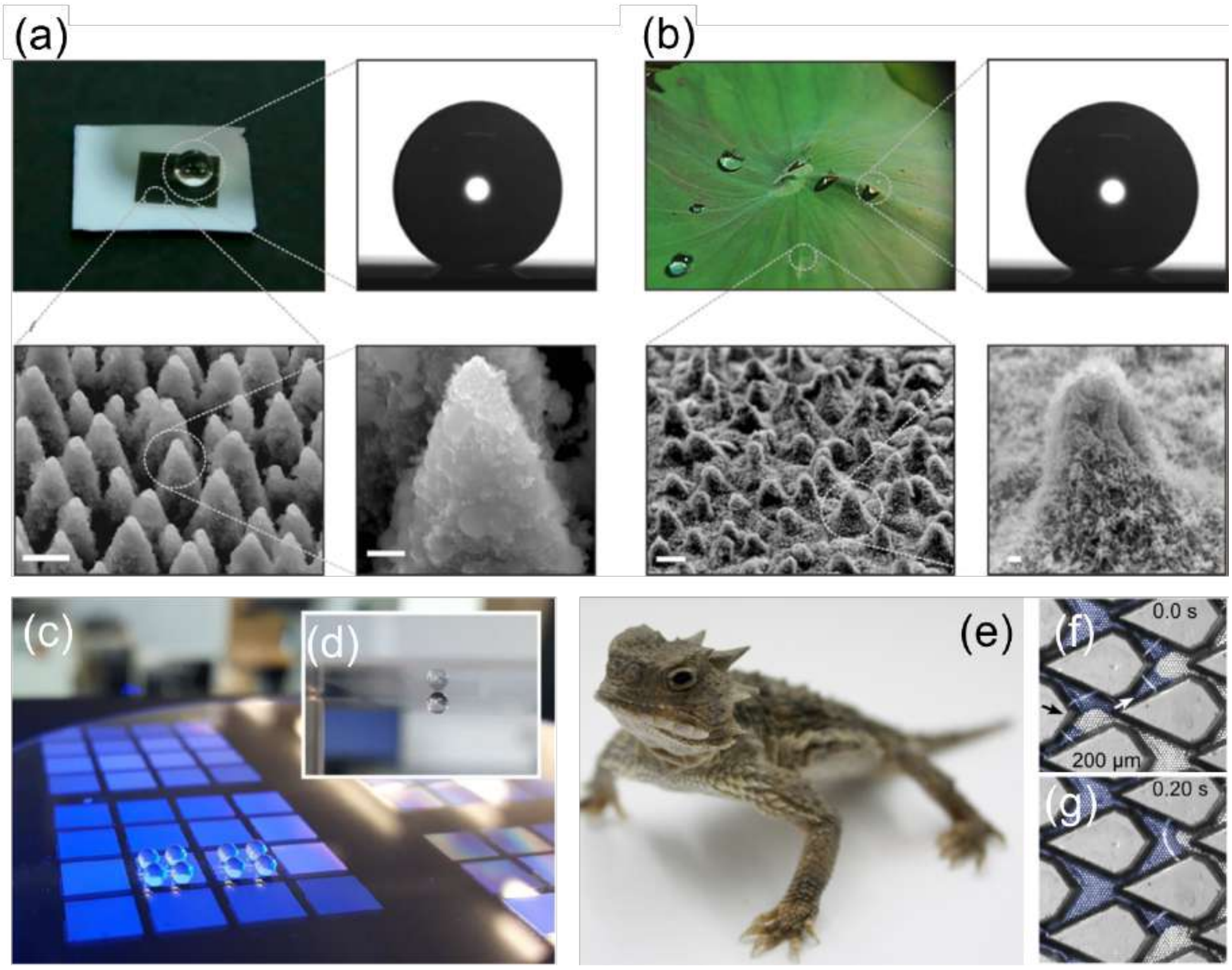

Fig. 34: (a) Top row: pictures of a water droplet on an artificial structured silicon surface (dark area). Bottom row: SEM images of the artificial surface (scale bars, left: 5 μm, right: 1 μm). (b) Top: pictures of water droplets on a *Nelumbo nucifera* Lotus leaf. Bottom: SEM images of the leaf surface (scale bars, left: 5 μm, right: 1 μm). (c) Water droplets placed on laser-processed, wet-oxidized and coated structures on fused silica. The high adhesion is demonstrated by the hanging drop in the embedded image (d). (e) Photograph of the lizard *Phrynosoma platyrhinos*, which has a hydrophilic skin that enables fast and directional water transport. (f-g) Sequence of images recorded with a time lap of 0.2 seconds, featuring directional capillary transport of a liquid in surface channels on steel, fabricated by laser and resembling the lizard skin. (a,b) Reproduced with permission from [164] Copyright 2008. Wiley-VCH Verlag. GmbH & Co. KGaA. (c,d) Reprinted with permission from Ref. [305], (e-g) Reprinted with permission from Ref. [64].

Other plants and animals feature specific hydrophilic or hydrophobic properties, including the bark bug, Texas horned lizard as well as the Namib Desert beetle. These properties have also been successfully conferred to metals [64,166,171,307], glass [305] and silicon [302,308] using laser processing.

***b) Oleophobic/oleophilic properties:*** In metals, the very same self-organized structures that feature hydrophobic behavior present oleophilic properties [171,309]. Such properties can be efficiently exploited in technical applications using oil-based lubricants, since they contribute to the fast and homogenous distribution of the lubricant, especially under conditions of starved lubrication [310]. The processing of "morphological gradients" featuring a continuous transition between LIPSS and hierarchical spikes in a processed surface region may even result in an mainly unidirectional oil transport from the rippled to spikes regions [171]. Underwater superoleophobicity, inspired by fish-scales that protect the fish from contamination in oil-polluted water, has been achieved in glass and silicon via fs laser processing [302].

***c) Fluidic transport:*** The Texas horned lizard is a role model for unidirectional water transport, since the channel network on its skin causes water to flow preferentially towards the snout, even against gravity. Transferring this bionic concept via direct laser writing to polymers and metals was recently demonstrated [311], [64] (see Fig. 34). Furthermore, a remarkable water transport velocity has been demonstrated on fs laser-fabricated silicon structures exhibiting a surface tension gradient [312]. Another straightforward method to fabricate complex 3D microfluidic channels in a transparent material is by focusing an ultrashort laser beam inside fused silica, writing the desired trajectories. As explained in the previous section on data storage, self-organized planar nanogratings can be formed in the focal volume. The structural change accompanied by the grating formation leads to a strong preferential chemical etching in certain acids of these structures compared to the unexposed material. The use of this technique has enabled laser-fabrication of optofluidic lab-on-chips, combining microfluidic with photonic devices [313].

***d) Control of condensation rates and impact on anti-icing properties:*** The fabrication of surfaces with superhydrophobic and ice-repellent properties constitutes nowadays a challenging task. These properties have an increased significance due to the abundance of potential applications where anti-icing properties, especially for aeronautic applications[314], given that the formation of supercooled water droplets can significantly reduce the aerodynamic performance and the operational capability. Most of the reported studies state that there is a clear relation between the spatial period and the surface microstructure depth, in the ice adhesion strength under low temperature conditions. Anti-icing surface properties are directly correlated with the wetting properties, i.e superhydrophobicity. So far there are prominent examples of surfaces that does not favor the ice accretion or extend the water freezing time for metal surfaces directly processed with laser pulses[179,315,316], as well as with combined laser texturing and chemical surface tailoring[317]. Furthermore, there are also studies of the ice-phobicity effect induced in Polytetrafluoroethylene (PTFE) with rapid and cost efficient with CW laser [318] and silicone rubber[319] achieved with picosecond laser sources. Were there was a correlation on the mechanical stability of the laser textured rubber with the forming ice freezing kinetics. It has been shown that droplet condensation occurs in the early stage of frosting before ice crystals appear; Experimental reports indicated that the presence of hierarchical microstructures on the patterned surface increased the available nucleation sites, resulting in a larger number of condensate site formation and leading to a superhydrophobic state [320,321]. The hydrophobic nature of the microstructured surface influence also the droplet geometrical shape, reduce droplet growth and delay droplet solidification which are an additional reason behind the ice-prevention [322].

***e) Cavitation dynamics and control of phase change-related properties:*** Controlling cavitation (i.e. transition from a liquid to a vapour phase) in fluid flows is of paramount importance to avoid unwanted effects of hydrodynamic cavitation such as erosion, noise and vibrations [323]. Hydrodynamic cavitation dynamics of droplets on hot surfaces is strongly correlated with various factors including the surface roughness of the heated surface that are kept in contact with droplets [324]. In recent reports, it was shown that micro-patterned surfaces are capable to significantly reduce the so-called Leidenfrost point (LFP), a critical temperature above which the vapor film between the droplet and hot surface is able to allow levitation of droplets [325,326]. Furthermore, results indicate that the patterned surface with hierarchical structures allows the development of hydrophobic states and results in longer droplet lifetimes [325,326]. The micro-textured surfaces could be therefore better suited for energy saving applications that take advantage of the Leidenfrost effect for droplet control or drag reduction.

***f) Responsive surfaces:*** The understanding and fabrication of surfaces with wetting properties that can be controllably altered on demand is important for a variety of potential applications, including micro/nanofluidics, lab-on-chip devices and sensor development. Responsive surfaces are able to reversibly switch their hydrophobicity/hydrophilicity in response to external stimuli. Functionalization of an artificially roughened surface with a responsive coating is expected to enhance the switching effect and result in a surface that can potentially switch between hydrophobicity/super-hydrophobicity and hydrophilicity/super-hydrophilicity in response to appropriate external stimuli. A unique methodology for creating responsive surfaces is based on adequate functionalization of fs laser micro/nano patterned Si substrates [191], see Fig. 35. Depending on the functional coating deposited onto such surfaces, different functionalities are attained including photo- [327], electro- [328] and pH- [329] responsiveness. In all cases, the principal effect of hierarchical roughness is to cause an amplification of the response to the external stimulus.

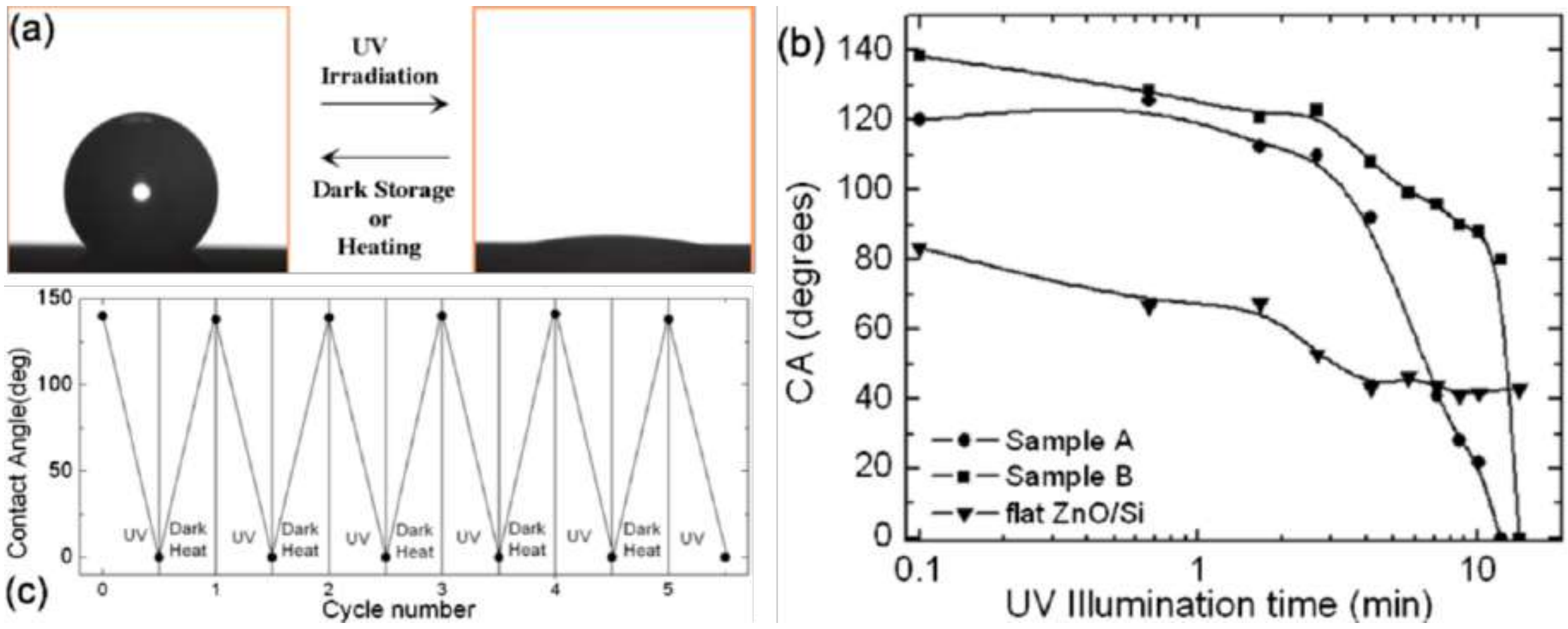

Fig. 35: (a) Photographs of the shape of a water droplet on laser-structured Si, coated with a nanostructured ZnO film, before and after UV illumination. The transition from hydrophobicity to superhydrophilicity is reversible upon dark storage or thermal heating. (b) Dependence of the water contact angle on the UV illumination for samples with different processing conditions. (c) Reversible

switch from hydrophobicity to superhydrophilicity for the sample shown in (a) under the alternation of UV irradiation and thermal heating. Reprinted with permission from Ref. [191].

### *3.3.3 Biology/Medicine*

***a) Cell growth, migration, patterning:*** The control over the surface morphology that is possible with self-organized laser structuring offers great potential for enhancing and directing cell growth on suitably structured substrates. An excellent overview on this topic is given in Ref. [194]. While much work has been performed on Ti-based materials [330], motivated by the excellent integrability of Ti prosthesis into the human body, polymers [331], Si [187,332] and steel [333] structured by fs laser pulses have also been used as templates for cell growth. For all materials, a considerable control of cell adhesion, migration and alignment to the fabricated structures was demonstrated at laboratory level. Besides this, Jeon et al. demonstrated control over cell migration and organization via laser-fabricated nanocrater-patterned cell-repellent interfaces [334]. Subsequently, Yiannakou et al. demonstrated the concept of cell patterning via direct laser structuring of cell-repellent and cell adhesive areas on Silicon, respectively [189].

A medical application, suggested by Heitz et al. [335], demonstrated in-vitro a fibroblast cell-repellant surface functionalization of titanium surfaces through the processing of hierarchical micro-spikes covered by nanoscale LIPSS (Fig. 36). Along with a subsequent electrochemical anodization, such surfaces may be used in miniaturized leadless pacemakers, which can be implanted directly into the heart. Furthermore, laser structured silicon surfaces have been successfully used for the development of implantable vaccines [336].

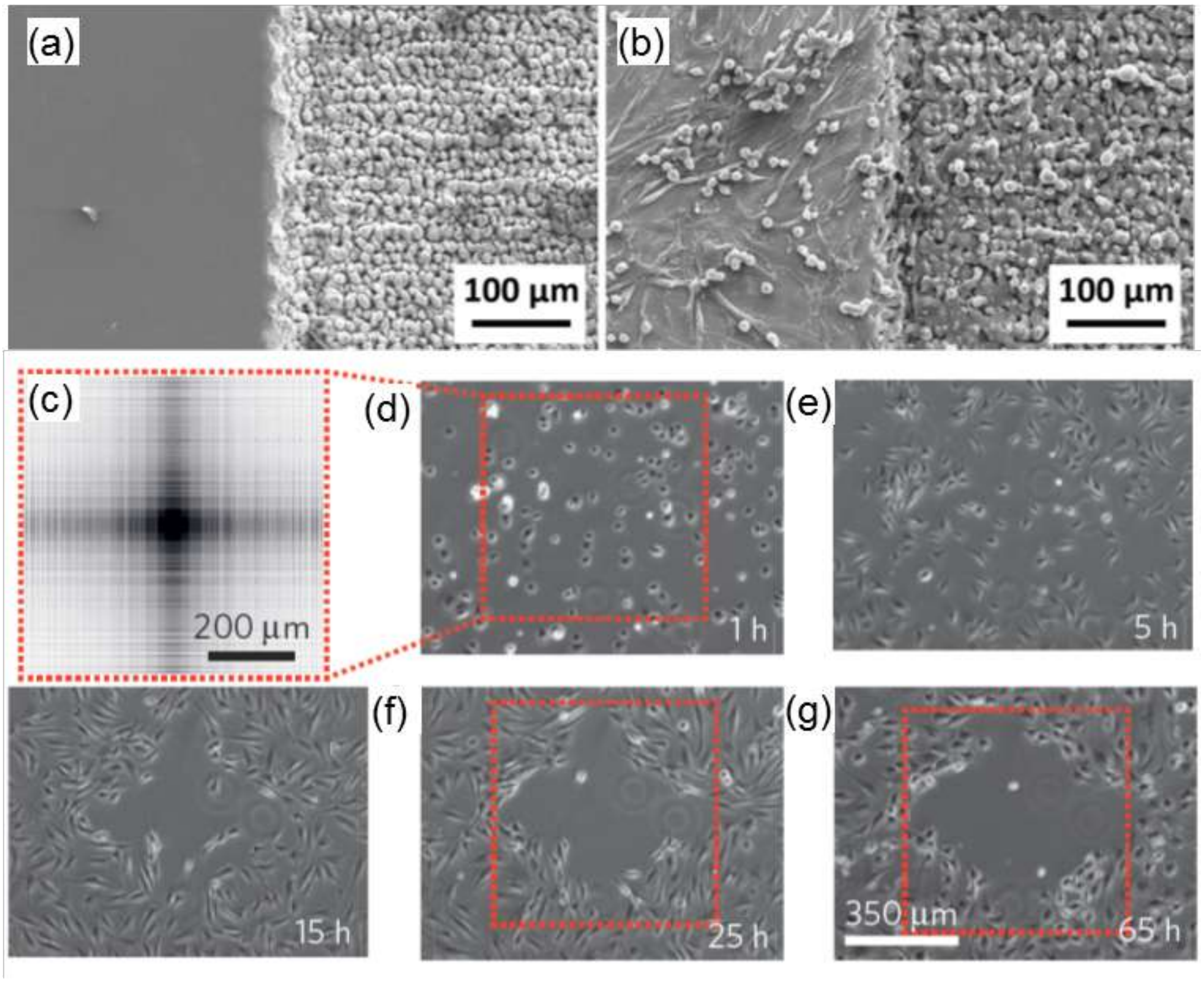

Fig. 36: (a,b) SEM images of the a sample region showing the flat untreated Ti-alloy surface (left) and the Ti-alloy surface with Ti:sapphire laser-induced spikes combined with electrochemical oxidation (right): (a) without cells and (b) with fibroblasts 1 week after seeding: Reproduced from Ref. [335], licensed under Creative Commons BY 4.0. (c-g) Time-lapse phase-contrast images of NIH3T3 cells cultured on a fused silica sample containing a spacing-gradient pattern. (c) Schematic diagram showing the gradient pattern design, which was fabricated on the red marked area on the sample surface. (d) After 1 h, most cells are able to attach homogeneously to the surface. (e–g) After several hours, the cells tend to migrate towards regions with outside the pattern. One day after cell seeding, clear boundary lines define a cell-repellent region. Reprinted with permission from [334]

***b) Antibacterial surfaces:*** The undesired formation of biofilms generates high risks in many industrial and medical applications. Such biofilms are assemblages of microbial cells that irreversibly adhere to a surface and that are enclosed in a stabilizing matrix of extracellular polymeric substances. Laser micro- and nanostructuring provides a promising method to reduce the growth of biofilms by altering topographical and chemical surface properties. A recent overview on this field is given in [194] and recent results for the specific cases of titanium and various steels and for different types of bacteria (*E. Coli, P. Aeruginosa, S. Aureus*) and their biofilm formation can be found in [337,338] and [181]. Cunha et al. [330] reported that the laser treatment of grade 2 titanium alloy resulting in LSFL reduces significantly the adhesion of *S. Aureus* and its biofilm formation as compared to polished surfaces. Similarly, LSFL with spatial periods of about 700 nm on corrosion resistant V4A steel clearly showed an anti-bacterial effect for *E. Coli* as test strain [330,331]. A similar finding was obtained for fs-laser structured polymeric (polyethylene) surfaces [339]. These results [338,339] indicate a dependency between the bacterial cell geometry and anti-adhesion effect, as the colonization characteristics of *S. Aureus* (spherical) and *E. Coli* (rod-shaped) differ significantly. Moreover, the choice of the biofilm cultivation method (static or flow conditions) and the cultivation time appear to be crucial [331].

### *3.3.4 Other*

***a) Friction control:*** Friction is a force that acts between two surfaces that are in relative motion. Many factors influence this force, including the texture of the surface. In nature, several species have developed a skin to reduce friction when moving in their environment. The probably best-known examples are the sandfish (a lizard) moving in the sand of deserts and the shark, whose skin exhibits riblet structures aligned in the direction of water flow, which are known to reduce skin friction drag in the turbulent-flow regime [117] and whose design has been adapted for several applications, including swim suits used in competitions. Another technological application developed recently within the European project LiNaBioFluid [http://www.laserbiofluid.eu/] aims at the reduction of friction and wear between the two metal surfaces of a bearing, exploiting biomimetic laser structuring. In this case, a lubricant is used, which adds, as an additional factor, the above-discussed oleophilicity of the treated surfaces. By means of a careful control of both factors, morphology and lubricant transport, friction has been reduced by 50% in a demonstrator bearing [http://www.laserbiofluid.eu/].

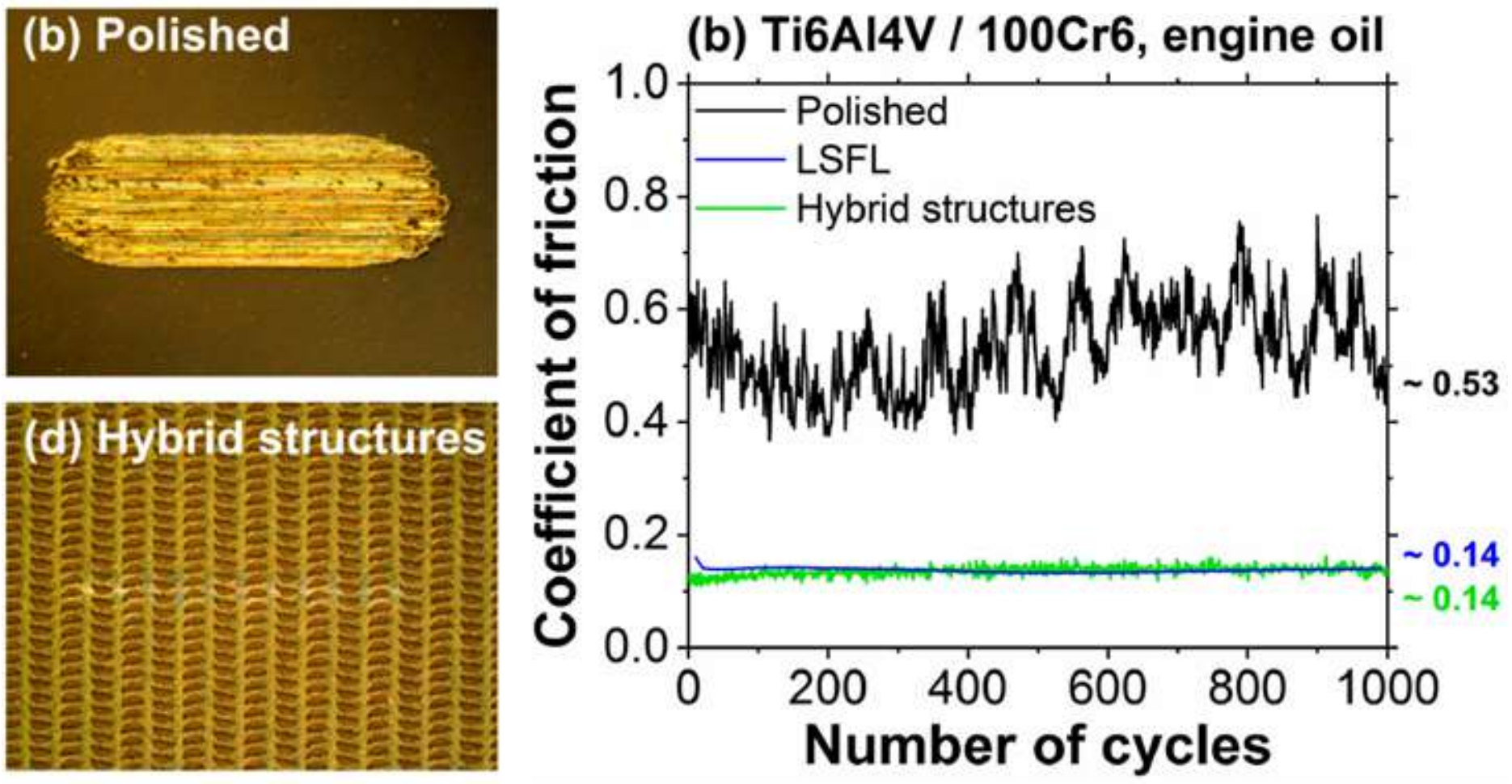

Fig. 37: (a,b) Optical micrographs (differential interference contrast) of the wear tracks on polished and laser-structured Ti6Al4V surfaces, associated with friction measurements using engine oil as lubricant. (c) Coefficient of friction as a function of the number of movement cycles as obtained during reciprocating sliding tests of the polished (black) and of the fs-laser structured (hybrid structures covered: green, LSFL-covered: blue) Ti6Al4V surface against a steel ball in engine oil. Adapted from Ref. [340], licensed under Creative Commons BY 4.0.

A recent review of the tribological performance of technical surfaces textured by various fs-laser-induced self-organized biomimetic structures (LIPSS, Grooves, and Spikes) can be found in [340] (see Fig. 37).

***b) Nanoelectronics:*** Fabricating self-organized structures with feature sizes less than 1 micrometer is also interesting for the field of nanoelectronics. Among the materials of interest for this application is, for instance silicon, which is the very pillar the industry is built on. The use of laser structuring for writing circuits offers not only a more flexible and mask-less method compared to lithographic approaches, it also allows switching the phase from crystalline to amorphous and back by using laser pulses of different durations. Recently, high-quality, self-organized amorphous fringe structures have been written in crystalline silicon, whose period can be tuned and which can be erased via laser irradiation [213]. Since the electrical properties of both phases are very different, the structures can be used to introduce anisotropy in the electrical conductivity. Finally, the fabrication of high regular, erasable and rewritable periodic surface patterns on silver metaphosphate glass by means of ultrashort pulsed laser processing, has been reported [341] which could potentially lead to ultrafast laser data storage on soft glass surfaces.

***c) Field emission cathodes:*** Field-electron emission refers to the quantum mechanical tunneling of electrons from a solid cathode (generally in the shape of a sharp tip) into vacuum under the influence of a strong electric field. Laser structured dense arrays of high-aspect ratio micro/nano spikes with controllability of density and roughness ratio has been demonstrated to serve as superior electron emitters for cold [342,343] as well as thermionic field emission applications [344].

***d) Sensing:*** The pronounced hierarchy of many types of self-organized structures in metals leads to strong electromagnetic field enhancement upon illumination by light. This effect is responsible for *surface enhanced Raman scattering* (SERS), which can be exploited for increasing the sensitivity in detection of certain chemicals. Microfluidic SERS chips have already been fabricated by all-femtosecond-laser processing [345].

***e) Solar cells / detectors:*** Some of the early applications of laser-induced self-organized structures are based on the efficient absorption light throughout the visible and near infrared range by spike structures in silicon processed in the presence of a gas containing sulfur hexafluoride and other dopants. Termed "black silicon", the group of E. Mazur has pioneered applications of these structures as sensitive layers in photodiodes, photodetectors, solar cells, field emission and other photoelectric devices [276].

# 4 Laser fabrication of biomimetic surfaces on soft materials and related applications

Table 3*: Biomimetic examples showing the capabilities for laser manufacturing of soft material surfaces.*

| Material | Natural archetype and functionality (-ies) | Laser parameters (Wavelength - Pulse duration, Repetition rate) | Fabrication Parameters (Polarization – Effective number of pulses – Fluence) | Structural features (type, periodicity, density, geometric characteristics) | Functionality (-ies) of processed material | Application/ Demonstrat ion | Ref |
|---|---|---|---|---|---|---|---|
| Polystyrene (PS) | Extra-cellular matrix (ECM) | 248 nm, 20 ns, 10 Hz | Linearly polarized, 6000 pulses, fluence 8 $mJ/cm^2$ | Ripples, periodicity 200 to 500 nm, dense, height 40 to 100 nm | Cell alignment | Tissue engineering | [331] |
| Polyethylene terephthalate (PET) | Extra-cellular matrix (ECM) | 1026 nm, 170 fs. 1 kHz | Linearly polarized, 11.9 $J/cm^2$, scan velocity of 7 mm/s | Grooves, width: ~29μm, depth: ~9μm | Cell alignment | Tissue engineering | [346] |
| Poly(methyl methacrylate) (PMMA) | Spilopsyllus cuniculi and Xenopsylla cheopis | 10.6 μm, CW | 95 W | Capillary channels, width: 0.3mm, Depth: 1mm – 0.3mm | Unidirectional Liquid Transport | Fluidic Diode | [59,62] |
| Poly(dimethyls iloxane, polyimide | Palomena prasina, Rhaphigaster nebulosi, Tritomegas bicolor | 248 nm, 20 ns, 1 Hz | Unknown polarization, 100–2200pulses, 255 mJ | Capillary channels 5x1$cm^2$, depth: 100mm | Fluid transport | Directional fluid spreading | [58] |
| Acrylic-based monomer liquid photoresist | *Maratus robinsoni* and *m. Chrysomelas* | 780 nm, 100 fs, 80 MHz | Unknown polarization, unknown number of pulses 0.3nJ | 2d nanogratings, period: 670 nm, thickness: 170 nm, spacing: 500 nm, depth 300 nm | Light diffraction | Light-dispersive component | [5] |
| Femtobond 4B | Morpho butterfly | 780 nm, 90 fs, 82 MHz | Unknown polarization, Unknown number of pulses, ~0.15 nJ | Grid structures, variable dimensions | Structural blue coloration | Sensing, anti-counterfeit protection | [347] |
| Polyethylene terephthalate (PET) | Extra-cellular matrix (ECM) | 248 / 193 nm, 20 ns, 10 Hz | Linearly polarized / unpolarized, 6000 / 30 pulses, fluence 9 / 30 $mJ/cm^2$ | Ripples / walls, periodicity 200 nm / 1.5 μm, dense, height 40 / 700 nm | Activation of β-catenin | Promotion of cell proliferation | [348,349] |
| Polyethylene terephthalate (PET) | Tropical flat bugs *Dysodius lunatus* and *Dysodius magnus* | 248 nm, 20 ns, 10 Hz | Unpolarized, 20 pulses, fluence 150 $mJ/cm^2$ | Naps, periodicity 3 μm, dense, Wenzel roughness r of 1.2 | Wetting | Camouflage | [55] |
| Polyimide (PI) | European true bug species (*Pentatomidae* and *Cydnidae*) | 248 nm, 20 ns, 1 Hz | Unpolarized, 2000 pulses, pulse energy 255 mJ, irradiation under 45° | Tilted cones, periodicity bout 10 μm, dense | Directional fluid transport | Fluid transport against gravity | [350] |
| Acryl-based photoresist | Tropical flat bug *Dysodius lunatus* | 800 nm, 150 fs, 80 MHz | Linearly polarized, power before NA 0.6 objective lens, 14 mW, spot size below 1 μm | Arrays of drop-shaped microstructures, 10 μm long, height about 5 μm, distance 15 / 35 μm | Directional fluid transport | Fluid transport of oily liquids in closed channels | [65] |
| Ormocer® photoresist | European true bug species (*Pentatomidae* and *Cydnidae*) | 800 nm, 150 fs, 80 MHz | Linearly polarized, power before NA 0.6 objective lens, 14 mW, spot size below 1 μm | 200 μm heigh microneedles ornamented with drop-shaped microstructures | Directional fluid transport | Loading of needles with pharmaceuti cal agents | [311] |
| Acryl/ methacryl-based photoresist | Spongeous bone | 800 nm, 150 fs, 80 mHz | Linearly polarized, power before NA 0.6 objective lens, 14 mw, spot size below 1 μm | Three- dimensional scaffold with 30 μm pore size | Cell-scaffold | Differentiati on of cells into osteogenic linage | [351] |
| Copolymer Filofocon A | *Planococcus halocryophilus* | 193 nm, 20 ns, 10 Hz | Unpolarized / linearly polarized, > 10 pulses, > | Ripples / entangled random network of | Under investigation | Under investigation | [352] |

| | | | | | | |
|---|---|---|---|---|---|---|
| (hydro-2) | bacteria | | 400 mJ/cm$^2$ | nanochains, periodicity 800 to 1000 nm, dense, height > 400 nm | | |
| Polytetrafluoro-ethylene (PTFE), polyimide (PI), polyethylene terephthalate (PET), EGF and collagen in starch | Extra-cellular matrix (ECM) | 193 nm, 20 ns, 248 nm, 30 ns | 9 mJ/cm$^2$, 35 mJ/cm$^2$, 2.2 J/cm$^2$ | Localized chemical surface modification combined with topographic changes | Immobilization of living cells along A required pattern | Production of biosensors, tissue engineering | [353] |
| Polystyrene film (PS) | Extra-cellular matrix (ECM) | 266 nm | P-polarized, fluence 2.98 mJ/cm$^2$ | Ripples, periodicity 250 nm, dense, height 60 nm | Cell alignment, directional migration | Cell migration and division studies | [354,355] |
| Polystyrene (PS) | Extra-cellular matrix (ECM) | 266 nm, 5 ns, 10 Hz | Linearly polarized, inclined irradiation, 30–50 mJ/cm$^2$ | Ripples, periodicity 250 nm, dense, height 50-60 nm | Cell alignment | Model substrate that mimics ECM for studies of cell biology | [356] |
| Polyethylene terephthalate (PET) | Extra-cellular matrix (ECM) | 193 nm, 10 ns, 10 Hz, 266 nm, 5 ns, 10 Hz | Linearly/elliptically/circularly polarized, inclined irradiation, 600/1200 pulses, fluence 3/7 mJ/cm$^2$ | Ripples/dots, periodicity 250-500 nm, height 50-120 nm | Activation of actin and β-catenin | Promotion of mesenchymal cell proliferation | [357] |
| Polystyrene (PS) | Extra-cellular matrix (ECM) | 266 nm, 5 ns, 10 Hz | Linearly polarized, inclined irradiation, 30–50 mJ/cm$^2$ | Ripples, periodicity 210 nm, depth 30–40 nm | Cell alignment | Platform for cell biology or biosensor research | [358] |
| Polystyrene (PS) | Extra-cellular matrix (ECM) | 266 nm, 5 ns, 10 Hz | Linearly polarized, inclined irradiation, 3 mJ/cm$^2$ | Ripples, periodicity 300–400 nm | Promotion of cell adhesion, growth, and Gene expression | Understanding of cell behavior at molecular level | [359] |
| Polyethylene (PE) | Extra-cellular matrix (ECM) | 790 nm, 30 fs, 1 kHz | Peak fluence 1.2 J/cm$^2$, linearly polarized | Laser-induced structures featuring valleys with 0.6 – 2 µm spacing | Reduced *E. Coli* biofilm growth | Anti-bacterial surfaces | [339] |
| Ormocer® photoresist | Three-dimensional (3D) tissues | 523 nm, <500 fs, 1 MHz | Power before NA 0.14 objective lens 3 mW | 330 µm long fibers with a Diameter of 6–9 µm. | Biocompatibility, cell morphologies | Fundamental studies, tissue engineering | [360] |
| Acryl-based photoresist | Extra-cellular matrix (ECM) | 800 nm, 140 fs | Power before NA 0.65 objective lens 10 mW | Nanopillars of different geometry, width 0.5-1.5 µm, length 2.5-6 µm | Controlled cell adhesion | Improved cell guiding | [361] |
| Modified fluorinated ethylene propylene (FEP) | Extra-cellular matrix (ECM) | 193 nm, 10 ns | Irradiation with contact mask | Lines, rhombs, squares with lateral dimension of 10 to 20 µm | Localized cell adhesion | Isolation of single cells | [362] |
| Acryl-based protein repellent photoresist, Orcomp® photoresist | Extra-cellular matrix (ECM) | Nanoscribe® system | Power before NA 1.4 objective lens 10-20 mW | 3D cell scaffold with consisting of two photoresists | Localized 3D ligands for cell adhesion | Systematical studies on cell behavior in 3D environments | [363] |
| Modified poly(lactic | Spongeous bone | 800 nm, <20 fs, 75 MHz | Power before NA 0.85 objective lens 15 mW | 3D woodpile structures | Various porosities | Bone tissue regeneration | [364] |
| acid) (PLA) | | | | | | | |
| Poly(methyl methacrylate) (PMMA) | Spermathecae of fleas | 10.64 µm, 42 kHz | Pulse energy 2.26 mJ, average laser power 95 W | Up-scaled biomimetic capillary channels (unit cell 2.6 mm) | Unidirectional fluid transport | Applications, e.g., microanalysis, lubrication | [365] |
| Teflon-like polymer layer on laser-structured glass | Fog-collecting Namib Desert beetles | 520 nm, 380 fs, 200 kHz | Linearly polarized, two-step ablation with 3.3 $J/cm^2$ and 2.1 $J/cm^2$ | Double-hierarchical surface structures of bumps and LIPSS | Water collection and transport | Biomedical And microfluidic devices | [182] |
| Positive photoresist | Lotus leaf | 366 nm, 10 ns | Two beam interference lithography, different exposure angles, 200-300 mw | Hierarchical periodic micro/nanostructures | Super-hydrophobic surfaces | Potential platform for bio-functionalization | [184] |
| Natural proteins | Soft tissue | 780 nm, 100 fs, 82 MHz | Power before NA 1.45 objective lens 100 mW | Several 10 µm wide grids | 3D structures, optionally including cells | Printing of tissues or organs | [366] |
| Polytetrafluoro-ethylene (PTFE) | Salvinia leaf And shark skin | 800 nm, 100 fs – 10 ps, 1 kHz | 40 mW to 100 mw, Spot overlaps between 70% and 80% | Hierarchical micro/nano structures combined with riblets | Super hydrophobicity | Drag reduction | [367] |

## 4.1 Self-organized surface structures

### *4.1.1 Ripples*

For many applications in medicine and biotechnology mammalian or human cells are adherent to an artificial surface, as in in-vitro to cell culture substrates or in-vivo to the surface of medical implants or prostheses. The adhesion of the cells to the substrate is a key factor for, e.g., cell multiplication (proliferation) and cell differentiation into various cell phenotypes. Cells adhere to a surface mainly by focal adhesion areas, which connect the cell skeleton with the surface which replaces the natural environment of the adherent cells consisting typically of extra-cellular matrix proteins, mineral constituents, and other cells. Theses junctions are influenced by various factors like the surface chemistry (including the presence of biological ligands), the wettability and elasticity module of the surface, as well as electrostatic and electrodynamic charges. Another very important aspect for the interaction between cells and artificial surfaces are specific nanotopographies [368–371]. Of special interest are periodic (or quasi-periodic) topographic fingerprint-like nanostructures [372], i.e., laser-induced periodic surface structures (LIPSS). Such structures on polymer surfaces can be employed for alignment of mammalian cells cultivated thereon, at least if the periodicity $\Lambda$ was larger than 200 to 300 nm [331,354–356,358,359]. This shown exemplarily in Fig. 38 for the alignment of Chinese hamster ovary (CHO) cells on LIPSS at a polystyrene (PS) substrate. The arrows indicate the directions of the ripples (solid line) and the cells (dashed line) derived by Fourier-transform image analysis.

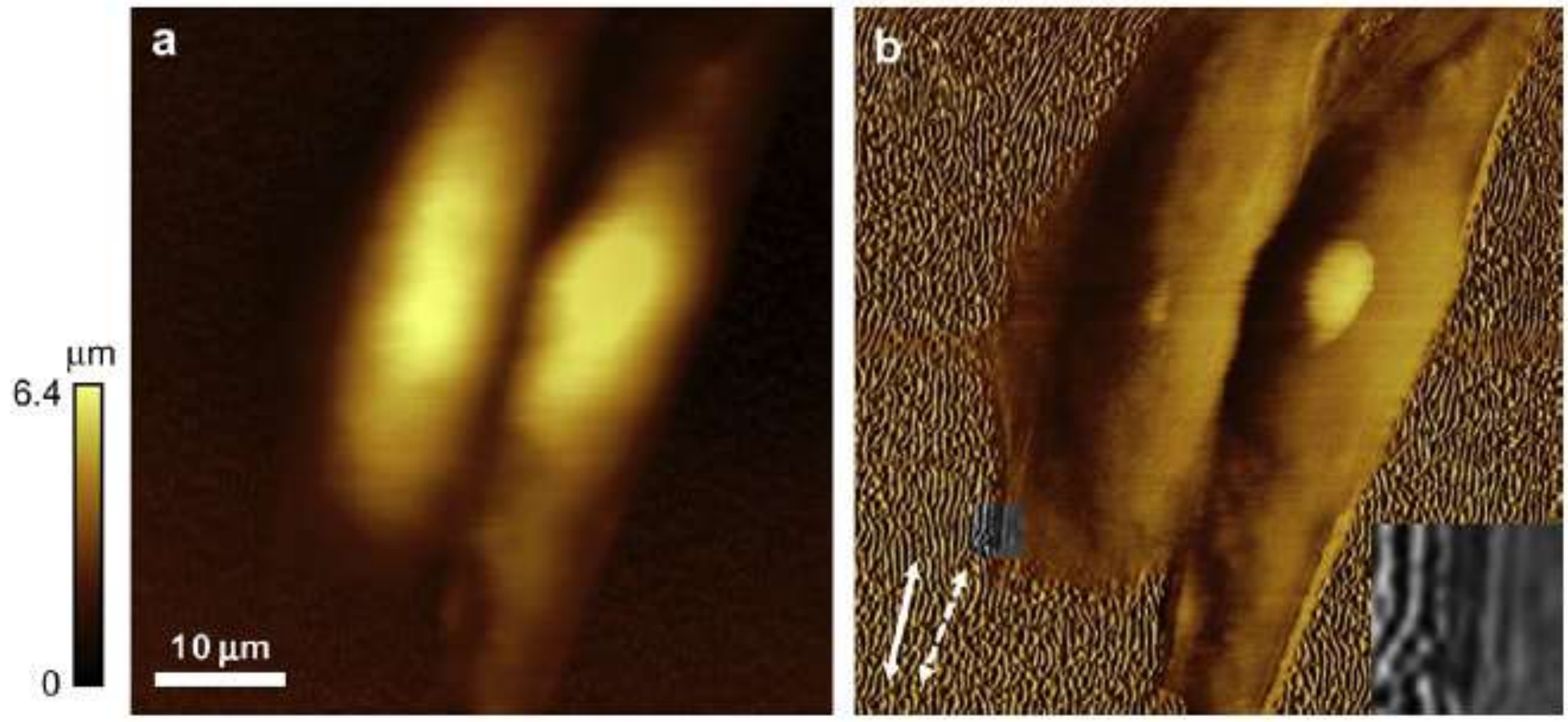

Fig. 38: LIPSS on polystyrene (PS) induce cell alignment: AFM images (height (a), deflection mode (b)) of aligned Chinese hamster ovary (CHO) cells. Reprinted with permission from [331]

Proliferative nuclear signaling in biological cells is profoundly activated by the nanotopography of the cell culture substrate, even if the LIPSS structures are too small to induce alignment [348,357,373,374]. If the periodic structures become even smaller, the focal adhesion points cannot attach anymore at the ridge of the nanostructures and the surface becomes cell-repellent [335]. On a microscopic scale, the contact between the nanostructures at the surface is realized by nanofibrous cell protrusions. As shown in Fig. 39, the interaction is restricted to the ridge of the nanostructures at the surface [349,374,375], resulting in reduced adhesion (Fig. 39A) or alignment of the protrusions (and the cells) (Fig. 39B). If the structure becomes too fine, no (cell) adhesion is possible anymore. Similar concepts should also apply for the adhesion of other microorganisms like bacteria, which exhibit nanofibrous pili or fimbriae for adhesion.

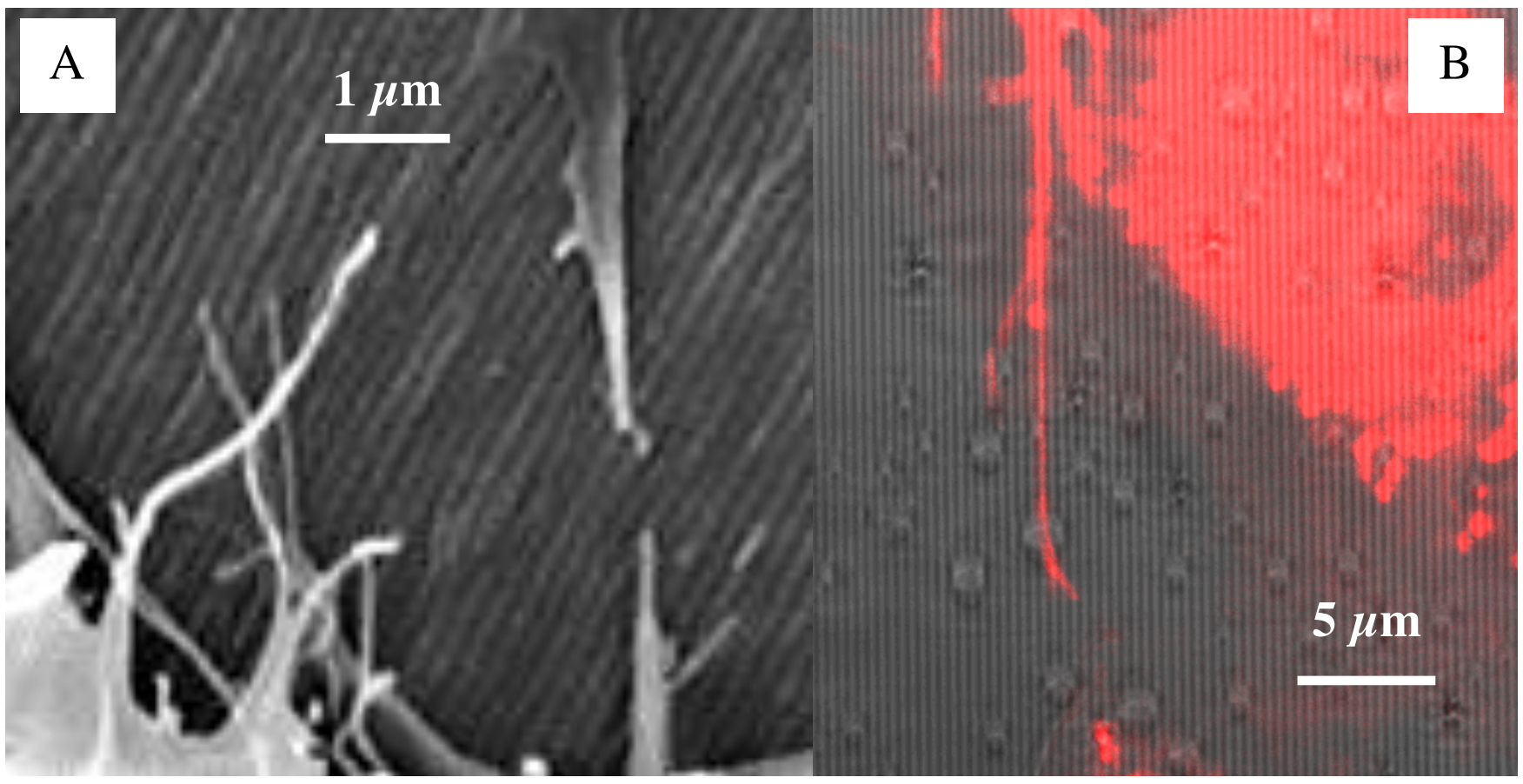

Fig. 39: Interaction with cell protrusions: (A, B) nanofibrous cell protrusions interacting with periodic surface structures on polymers.

Like for the case of hard materials, the interference of an incident laser beam with a surface wave can induce the formation of ripple structures [242], also denoted as laser-induced periodic surface structures (LIPSS). Ripple formation was first observed by Birnbaum [376] after ruby-laser irradiation of various semiconductor surfaces. Further investigations have shown that ripple formation is a general phenomenon, observed practically always on solid or liquid surfaces after laser irradiation with polarized light within certain ranges of laser parameters. Laser pulse lengths in the order of some ns, fluences well below the ablation threshold, and a large number of laser pulses have to be applied to induce LIPSS formation on polyethylene terephthalate (PET) or polystyrene (PS) surfaces. The formed nanostructures depend on the laser wavelength, λ, as well as the angle of incidence, θ, of the laser beam [377–379]. For s-polarization, the lateral periodicity Λ is given by

$$\Lambda = \lambda / (neff - sin\theta) \qquad \text{Eq. 9}$$

where $n_{\mathrm{eff}}$ is the effective refractive index, which lies between the index of air (≈1) and the index of the polymer. The direction of the ripples in case of polymers is parallel to the polarization direction. The structure periodicity Λ can be varied by means of different irradiation parameters, but not independently from the structure height, *h*. The aspect ratio of the ripples (i.e., *h* divided by half of Λ) is typically around 0.5.

Deviations from the scaling law of the ripple period shown in eq. 15 are not frequent but do occur. For instance, self-assembly of well-defined but complex surface structures has been observed in a commercial copolymer (Filofocon A) upon irradiation with unpolarized ultraviolet nanosecond laser pulses [352]. Instead of parallel ripples, the structures are composed of nanochains forming an entangled random network (Fig. 40) that resembles the surface morphology of *Planococcus halocryophilus* bacteria.

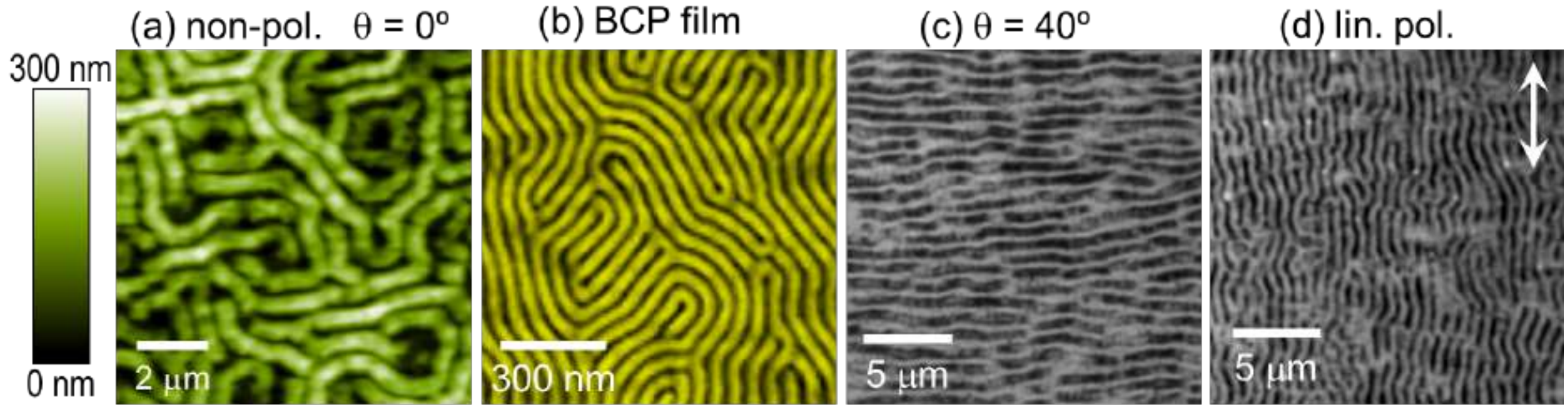

Fig. 40: Anomalous LIPSS in Filofocon A, self-organized upon excimer laser irradiation: (a) Topography map of an entangled random network structure obtained for non-polarized laser light; (b) SEM image of an annealed block copolymer film (BCP; PS-b-PMMA). (c-d) Optical micrographs of aligned lamellae induced by using (c) an angle of (non-polarized) laser incidence of 40º; (d) linear polarization (direction indicated by arrow). Source: Adapted from [352] with the permission of AIP Publishing.

These nanochains have periods that are a factor 4 longer than the laser wavelength at normal incidence, in evident discrepancy with equation 15, and are the result of a different formation mechanism. Analogies of these nanochains to lamellar structures fabricated on a smaller scale in block copolymers (BCP) are at hand, as shown in Fig. 40(b). This similarity suggests that the self-assembly process relies on a combination of chemical (BCP), optical (surface scattering) and thermal (melting, coarsening and ablation) effects. Both, the particular random network structure and the long periods accessible open new opportunities of fabricating periodic surface structures in polymers. As shown in Fig. 40(c-d), alignment of the nanochains to form a parallel distribution of lamellae can be achieved either by irradiation at an angle or by using linearly polarized light.

Also other types of micro- and nano-pattern surfaces produced by direct or indirect laser-based methods are suitable to control the adherence and growth of cells on a patterned surface. These effects are either due to the changed topography alone [380] or by a combination of topographical and chemical stimuli [353,361].

### *4.1.2 Microgrooves*

The formation of supra-wavelength groove-like structures, with a well-defined periodicity has also been reported in a few reports. Bartnik et al. observed the formation of supra-wavelength parallel features reaching a periodicity of about 5 μm after irradiating PET foils with EUV radiation. These surface patterns were found to enable Chinese hamster ovary cells to adhere to micropatterned PET samples and align along the grooves of the parallel fringes [381,382]. In this case, nature has again provided many examples as inspiration for the fabrication of functional materials. The neotropical flatbug species *Dysodius lunatus* and *Dysodius magnus* show a fascinating camouflage principle. Its appearance renders the animal hardly visible on the bark of trees. However, when getting wet due to rain, bark changes its color and gets darker. In order to keep the camouflage effect, the animal is covered with pillar-like microstructures which in combination with a surprising chemical hydrophilicity of the cuticle waxes, render the bug almost superhydrophilic: Water spreads immediately across the surface and the surface of the bug is getting darker in a similar fashion as the bark [58,362].

In order to see if the wetting (water spreading) behavior which can be seen on native *Dysodius lunatus* can be explained by the pillar-like microstructures, an attempt was made to mimic the effect. For this purpose, naps on polyethylene terephthalate (PET) foils were produced by irradiation with nanosecond UV laser pulses. The spontaneous formation of the quasi-periodic naps occurs self-organized due to the release of the biaxial stress-fields and crazing in the irradiated but not ablated surface. While the untreated PET is virtually flat, areas which were irradiated with the laser are covered by naps. These naps have a diameter of about 2 µm. The PET-naps are slightly denser (average distance 3.5 µm) than the naps observed on *Dysodius* but on the other hand they are not that high (about 1µm) resulting in similar surface enlargement when compared to the natural archetype [383]. In Fig. 41A, a comparison of the wetting by dyed water of the flat PET foil (upper drop, sessile) and of the laser-treated area at the PET foil (lower drop, spreading) is shown. Both areas were additionally coated with a thin gold layer before wetting to obtain the same surface chemistry and to avoid electrostatic charging effects. Even though gold does not occur at natural surfaces, the obtained contact angles are similar as for the natural role model. Fig. 41B shows the surface morphology of the laser-treated area at the PET-foil as seen in the SEM.

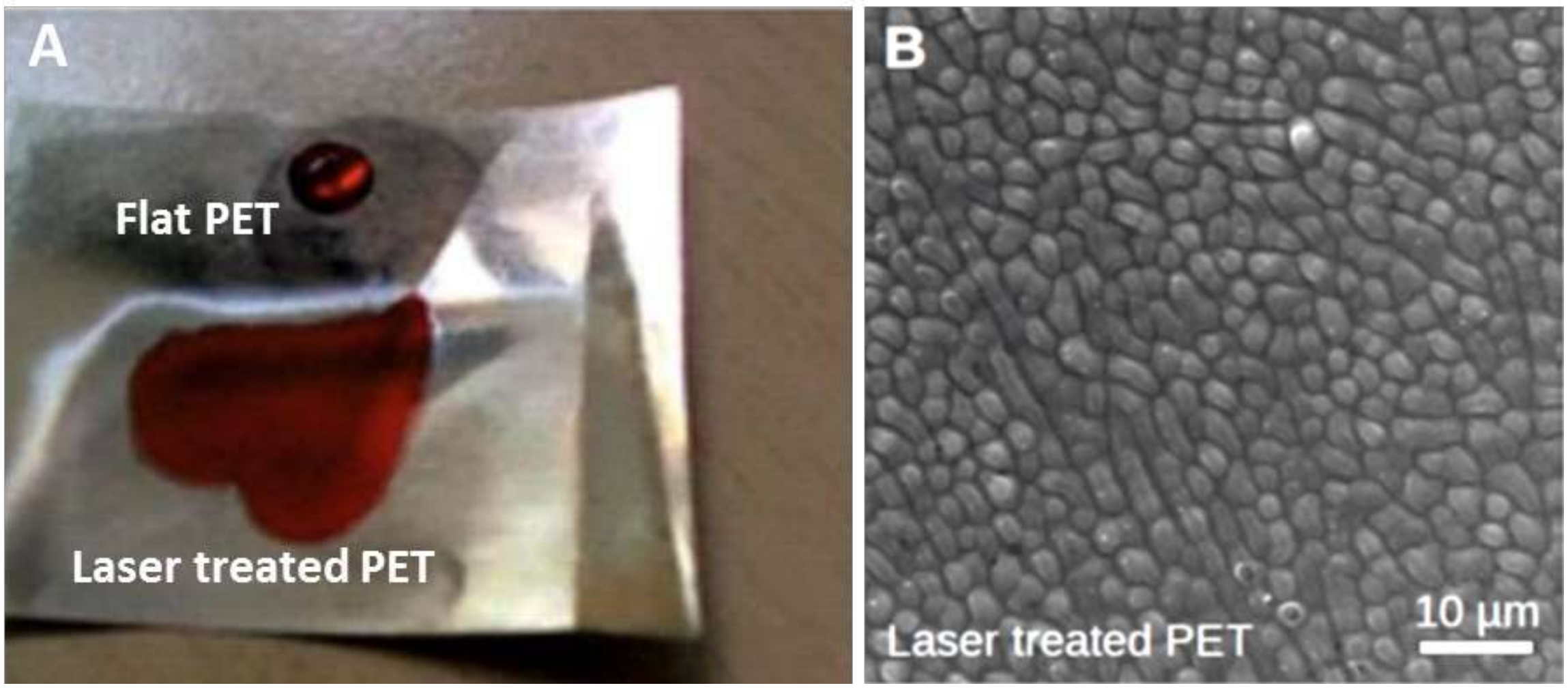

Fig. 41: Wetting by laser-induced microstructures: (A) comparison of the wetting by dyed water of the flat PET foil (upper drop, sessile) and of the laser-treated area at the PET foil (lower drop, spreading); (B) SEM image of laser-induced microstructures on PET. Adapted from Ref. [55], licensed under Creative Commons 4.0 license.

Not only superhydrophilic surfaces due to micro- and nanostructures are used by inspects and arachnids (like at the legs of wharf roach *Ligia exotica* [384]), but also superhydrophobic surfaces are used, as well as combinations of alternating areas with both properties. An example are fog-collecting Namib Desert beetles. Here high-contrast wetting, which were reproduced by femtosecond laser-processing, are reported to be responsible for the fog collection and transportation properties [182]. Hydrophobic surfaces with micro- and nanostructures can show the Lotus effect [60], which is can be increased by hierarchical micro- nanostructures. This has been shown for structures in positive photoresists produced by two beam laser-interference patterning with two different exposure angles [184].

### *4.1.3 Microspikes and Hierarchical structures*

The appearance of micro-spikes has been reported on a plethora of polymers ranging from thermoplastics such as polysulfone (PES) and polycarbonate (PC) to thermosetting polymers like Polyimide (PI) and aromatic polymers such as Polyether-ether-ketone (PEEK) and Polyethylenimine (PEI). The irradiation conditions, which are required for formation of micro-cones, have been extensively investigated and range both in laser wavelength and pulse duration. Micro cones have been reported to appear after irradiation with pulses in the UV as well as IR spectrum [385–387]. The pulse duration effect has also been studied with reported structures appearing with pulses in the ns, ps and fs regime [385,388,389]. Furthermore, it has been reported that their shape also varies depending on the wavelength of the incident pulses as well as the incidence angle [387,390]. The formation of such structures has been exploited to tailor a variety of material properties derived from surface topography. For example, Pazokian et al. fabricated functional surfaces on PES surfaces decorated with micro-cones which were able to tune the wettability response of the material, from hydrophilic to hydrophobic [385].

Structures that would facilitate the directional transport of the liquids were inspired by the external scent efferent system of European true bugs [58], which make use of small oriented microstructures for the unidirectional transport of the defensive liquids on their body surfaces (from the places where the liquids are secreted to the places where they are evaporated). Similar bug-inspired structures can be created on polyimide foils by irradiating with a KrF excimer laser, with 248 nm wavelength. The obtained microspike-like structures (tilted 45° cones) had dimensions comparable to the bug structures. When wetting tests were performed with a soap-water solution, with a suitable contact angle, to assess the liquid's behavior on the structured area – the liquid moved in a unidirectional manner. It can be seen in Fig. 42, that the soapy liquid is moving upwards, against gravity, reaching the upper rim of the structured area, while the liquid front was halted in the downwards direction [350]. These results show that it is possible to obtain an upwards directional liquid transport on a tilted surface, against gravity. Another type of "fluidic diodes" for passive unidirectional liquid transport are structures bioinspired by the spermathecae of fleas [59]. Unidirectional liquid spreading is also reported by grooves and cavities with arc-shaped edges and gradient wedge corners mimicking the surface of the peristome of the pitcher plant *Nepenthes alata*. Here the structures were produced in a photoresist by two-step inclined UV exposure photolithography [365].

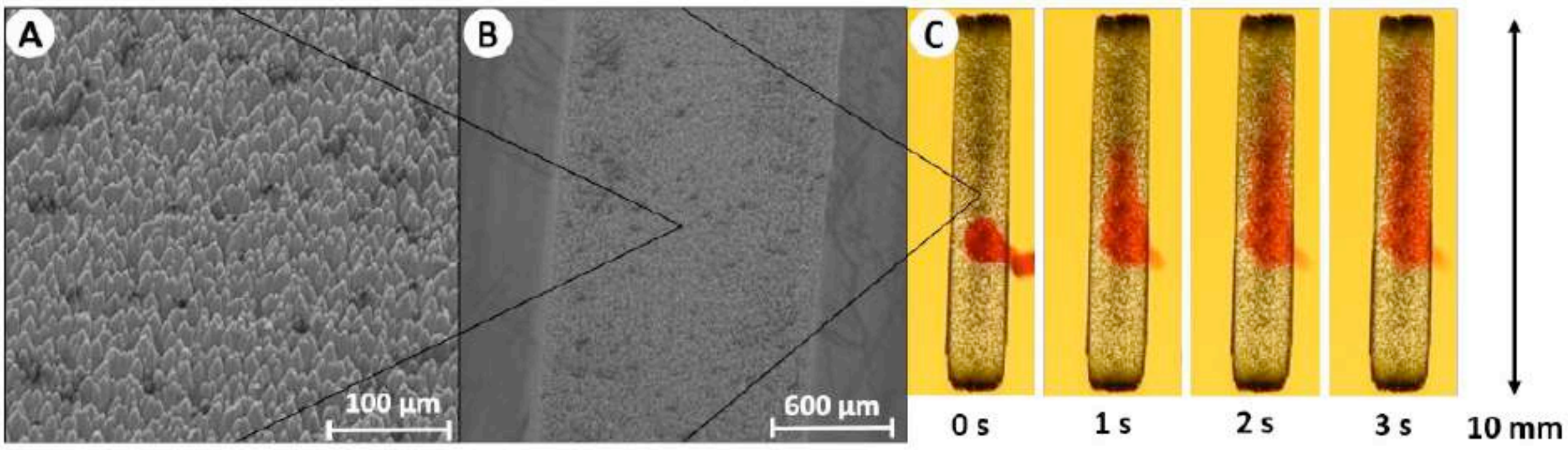

Fig. 42: Unidirectional liquid movement on laser-induced microstructures: (A, B) Scanning electron microscope images of the structures within the irradiated surface of the polyimide foil; (C) Directional liquid movement of the soapy solution starting from liquid deposition (0 s) and ending when the liquid has reached the upper rim of the structured area (3 s). Reprinted with permission from [350].

### *4.1.4 Replication from laser structured hard molds*

Laser micro- and nano-structuring of soft materials is a challenging task During irradiation, soft materials undergo a fast degradation, chemical modification and surface contamination [391,392]. An alternative technique of the direct laser processing is the replication procedure. The main concept of replication is to produce a mold material with the desired surface texture. Then the surface texture of the mold material will be transferred and replicated to the surface of a soft material. Polymers are mainly used as the replica material due to their ability to deform easily and adapt to different kinds of surface textures. The main advantage of the replication process is the fast production of large quantities, since mold substrates can be used multiple times compared to laser processing of individual samples. For mass production, the replication process is advised by the majority of the scientific community [393–398].

First experiments were performed using excimer lasers to process the mold substrates to create various shapes [396,399–402]. With the development of ultrafast laser sources, both directly written structures as well as LIPSS were also used as templates in the replication process. Structures such as grids [272], ripples [403–405], grooves[406], spikes [393,397,405,407–409] and a combination of the above mentioned [410] were replicated in polymer for a variety of applications as shown in Fig. 43. The fact that LIPSS structures vary depending on the material, gives the flexibility to choose different types and complex surfaces structures

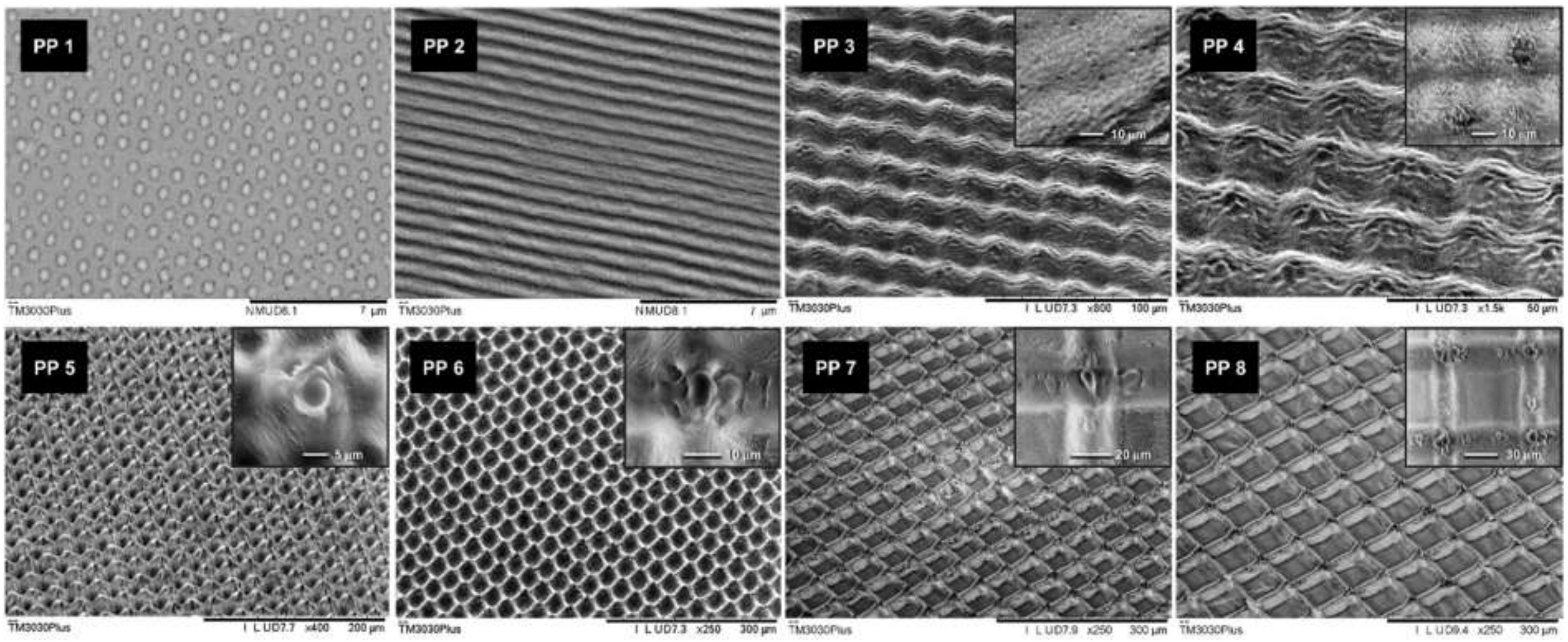

Fig. 43: SEM images polypropylene replicas from various mold textures. Reprinted from [406], Copyright (2019), with permission from Elsevier.

Numerous replication techniques were developed to date in order to optimize the process depending on the replica polymer type. Heating during the replication, as shown in Fig. 44B, is usually required for thermoplastic polymers such as Poly-methyl methacrylate (PMMA), while in the other hand room temperature replication is also possible for organosilicon polymers such as Polydimethylsiloxane (PDMS), see Fig. 44A. Table 4 summaries the replication techniques where the most effective and widely used method is the hot embossing replication which offers a fast, simple, flexible and high quality replication process [411].

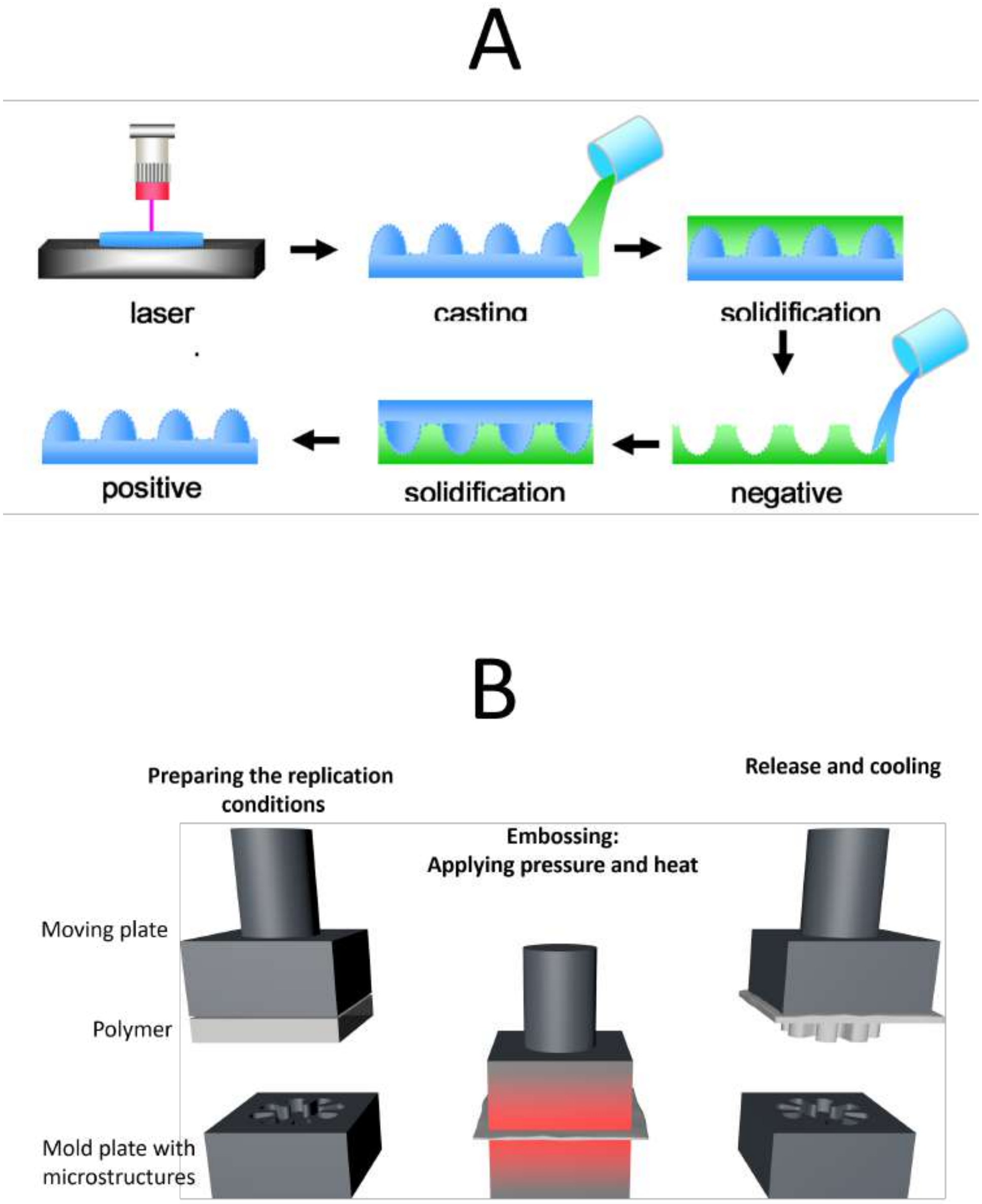

Fig. 44: Replication Procedure. A) cold casting replication. Reprinted with permission from [412] © The Optical Society. B) Hot embossing replication

Table 4: Replication methods

| Replication method | Reference |
|---|---|
| Hot embossing | [393,397,401,403,407,408,413–419] |
| Cold casting | [395,410,420] |
| Injection molding | [405,406,421] |
| Hot casting | [399,412] |
| Compression molding and UV curing | [422,423] |
| LIGA process | [400] |
| Imprinting process | [404] |
| Photomolding | [402] |
| Phase separation micro-molding | [424] |
| Micro-replication | [409] |
| | |

The majority of the scientific community focuses on the production of hydrophobic and superhydrophobic replicas with micro surface texture [395,397,398,405,407,408,410,412–415,417,421–423]. Nevertheless other applications such as tissue engineering [416], anti-reflective surfaces [393,403,408], micro-fluid channels [418], anti-icing [419], biomimetic superoleophobic [407], capillary-electrophoresis chips [12], diffraction gratings [404] and fluorescent patterns [399] were also exploited. An interesting biomimetic example is the replication of filefish *N. septentrionalis* as presented in Fig. 45, where its surface is decorated with microstructures enabling a superoleophobic feature.

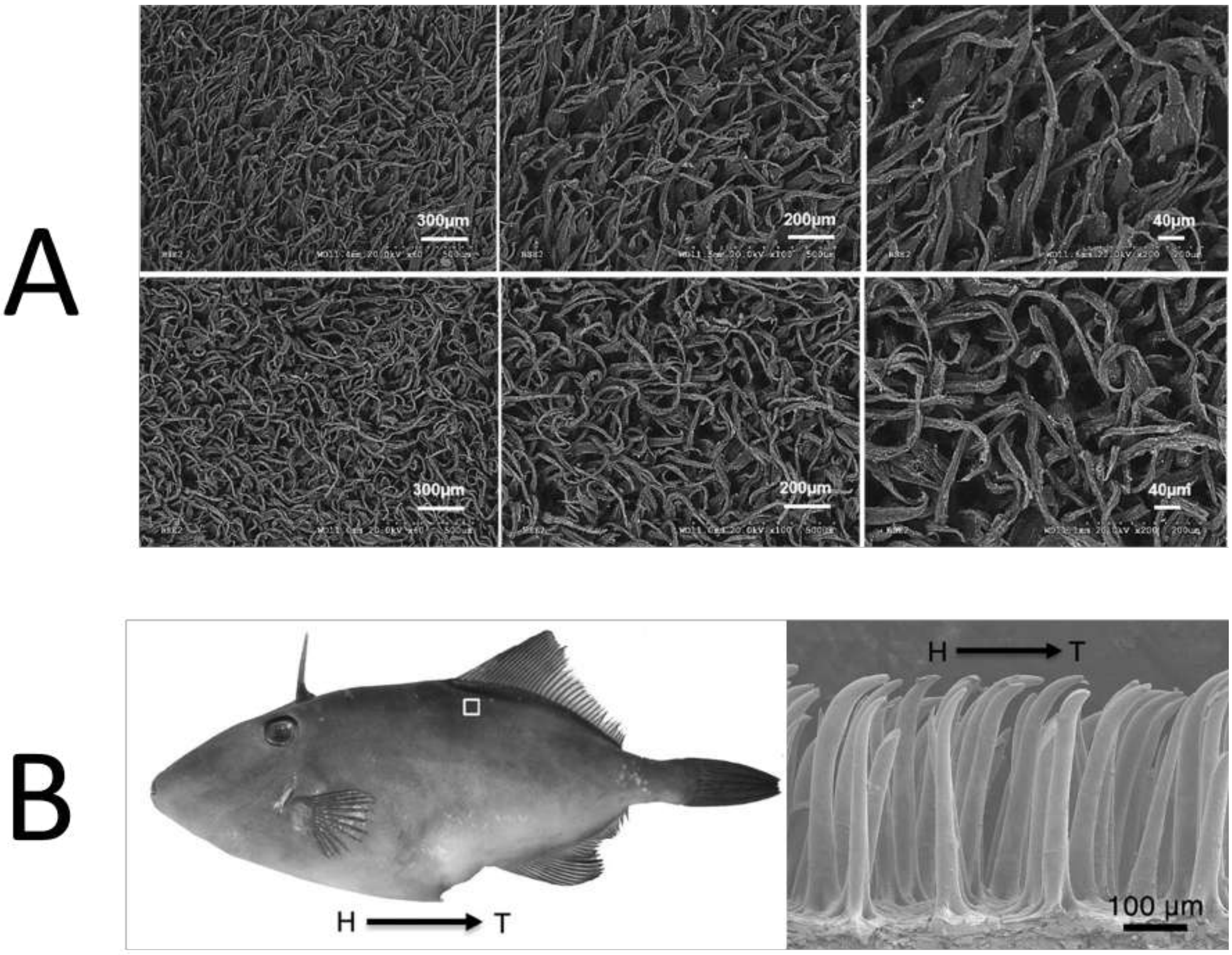

Fig. 45: A) SEM images of superoleophobic surface of high density polyethylene replicas. Reprinted from [407], Copyright (2016), with permission from Elsevier. B) Picture of filefish *N. septentrionalis* (left) and the SEM picture of the superoleophobic skin of the fish (right) where arrows directing from head (H) to tail (T) indicates the oriented direction of hook-like spines. Reprinted with permission from [425].

The durability and the thermal stability of the replicas is an important factor for a robust material to be used in the desired application. The effect of mechanical abrasion that simulates the real-life cleaning procedure was tested on superhydrophobic polypropylene replicas. The submicron single-design textured replicas were found to be vulnerable to cleaning cycles as the wetting properties of the surface after the mechanical abrasion were considerably reduced. On the other hand, replicas with hierarchical surface structures proved to be durable to cleaning cycles. [406] Also the thermal stability was tested on polytetrafluoroethylene replicas. The results reveal that the micro–nano texture polytetrafluoroethylene replicas remained stable to heat exposure up to 250 °C for 2 hours, with minimum structural changes up to 340 °C for 2 hours. The untreated polytetrafluoroethylene withstands heat exposure up to 430 °C for 2 hours. [415]

## 4.2 Direct writing via two-photon polymerization

Bio-inspired microstructures are also produced by using the technique of two-photon lithography [65] an intrinsic three-dimensional (3D) microfabrication method, used to produce microscale devices with resolutions below the resolution limit [426]. To write features by two-photon polymerization (2PP), a laser beam is focused by the objective lens into a liquid photoresist containing photo-initiators [427]. In the focus, the photo-initiators are activated by two-photon absorption starting a chain reaction to crosslink the photoresist. 3D features are obtained by sample or beam scanning. The not polymerized material can be later washed off by a solvent. State-of-the-art high-resolution 3D printers with lateral resolution below 100 nm are based on 2PP using ultra-short pulsed femtosecond laser beams. Current systems suffer from the fact that the laser focus has an ellipsoidal shape with a considerable larger focus length than focus diameter. In a proof of principle, it has been shown that tailored laser beams can overcome this obstacle by employing stimulated-emission depletion (STED) lithography, which allows to write freestanding 3D features in a photoresist with a size of about 50 nm [428].

The bark bug *Dysodius lunatus* also possess caudally oriented micro-ornamentation underneath the wings, around the glands that secrete an oily defensive liquid, as an anti-predator adaptation. These microstructures contribute to the transport of the oily substance from the gland system to the area where they are evaporated. The micro-ornamentation consists of a periodical array of droplet-like structures of around 10 μm in length. Arrays of similar polymer microstructures were produced by the 2PP technique that mimic the micro-ornamentation from the bug's cuticle. A good directionality of oil transport was achieved, directly controlled by the direction of the pointed microstructures at the surface [65]. If the tips of the drop-like microstructures are pointing towards the left side, the liquid front moves to the right (as is shown in Fig. 46) and vice versa. Similar effects could be expected for the transport of oily lubricants.

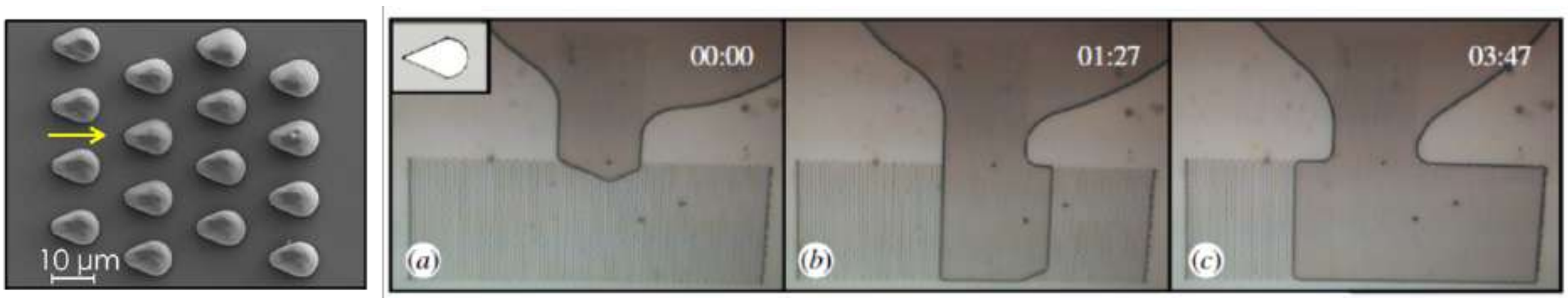

Fig. 46: SEM images of bioinspired drop-shaped microstructures produced by 2PP (left); time series of directional oil transport at these microstructure (right) in the direction indicated by the yellow arrow in the image at the right. Adapted from Ref. [65], licensed under Creative Commons 4.0 license.

Similar structures inspired by the external scent efferent systems of some European true bugs as shown Fig. 41, were suggested in a new design for microneedles and microneedle arrays, intended for rapidly coating the MNs with a drug or vaccine [350]. The biomimetic approach consists in ornamenting the lateral sides of pyramidal MNs with structures inspired by the external scent efferent systems of some European true bugs, which facilitate a directional liquid transport. To realize these MNs, the 2PP technique was used as is shown exemplarily in Fig. 47. Liquid coating capabilities of structured and non-structured microneedles were compared.

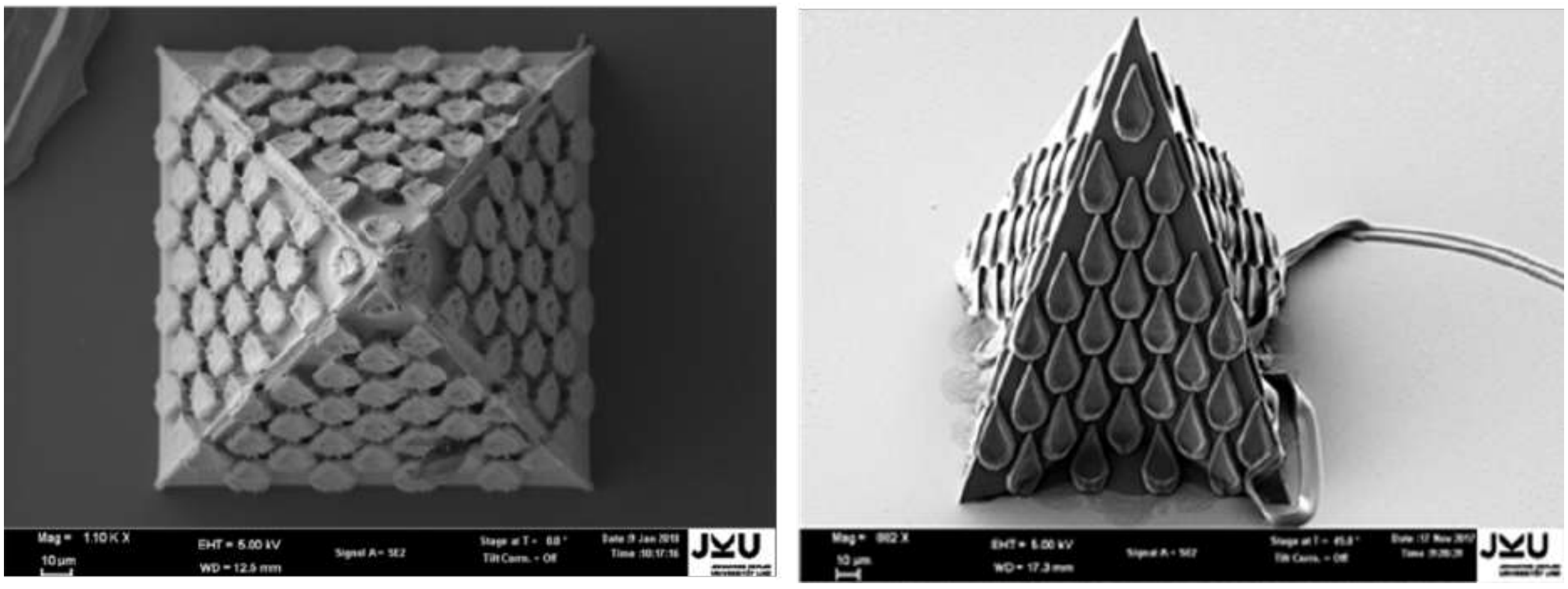

Fig. 47: SEM images of microneedles produced by 2PP with bioinspired microstructures for directional fluid transport.

Another application of the 2PP technique is the fabrication of 3D scaffolds for the cultivation of cells for tissue engineering applications [360]. The laser written structures replace here the 3D surrounding of the cells in the tissue for instance, the trabecular network of spongeous bones [335]. As is shown exemplarily in Fig. 48, progenitor cells, differentiated into an osteogenic lineage by the use of medium supplemented with biochemical stimuli, can be seeded on to the hydrophilic three-dimensional scaffolds. Due to confinement to the microstructures and/or mechanical interaction with the scaffold, the cells are stimulated to produce high amounts of calcium-binding proteins, such as collagen type I, and show an increased activation of the actin cytoskeleton. The best results were obtained for quadratic pore sizes of 35 μm.

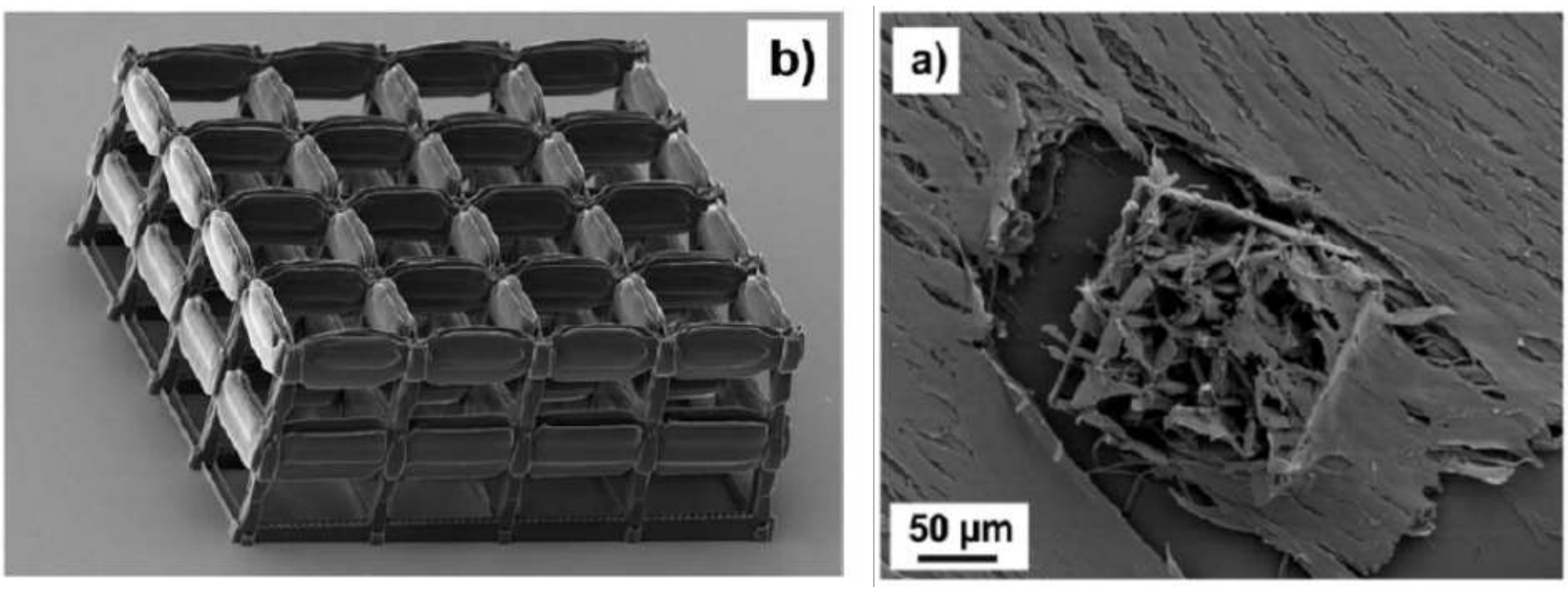

Fig. 48: SEM images of cell-scaffold with 30 μm pore size produced by 2PP (left) and of preosteogenic cells growing 3-dimensionally in those structures (right). Adapted from Ref. [351], licensed under Creative Commons Attribution-NonCommercial-NoDerivs License.

In recent years, numerous printing and writing techniques have evolved producing 3D microstructures that mimic the extra-cellular matrix [429,430]. For tissue engineering, many techniques are employed in which a scaffold — usually consisting of a polymer or polymer composite material — is produced, on to which cells are seeded and adhere, and where they then proliferate. These scaffolds can be produced by a variety of techniques including direct laser-writing methods, such as stereo lithography and the 2PP technique. By choice of appropriate materials, additional coating, or surface treatment, the formed structures can be made bio- and cytocompatible. For example, microstructures can be produced from the commercial photomaterial ormocer® (an organic–inorganic hybrid polymer), which enables good cell adherence and proliferation [431]. Further, cell-repellent structures produced by two-photon polymerization and combinations of cell-adherent and cell-repellent areas on the same microstructures have been described in the literature [363]. It has been recently demonstrated that microstructures with quadratic pores perform well as cell scaffolds for mural preosteogenic cells [364]: these structures consist of biodegradable poly(lactic acid) and are produced by the 2PP technique. Numerous other publications describe the use of two-photon polymerization for the production of cell-scaffolds, for instance [432–439]. Interestingly, natural proteins like collagen can also be used directly to form by fs laser-writing 3D structures which may contain even living cells similar as in soft tissues [366,440].

# 5 Conclusions and perspectives

The application of laser processing methodologies for the fabrication of bioinspired micro- and nano-patterned surfaces with extraordinary optical, mechanical, wetting, tribological, biological adhesion and antibacterial properties has been reviewed. At the same time, the physical and biological principles behind the exceptional functionalities exhibited by model natural surfaces, used for inspiration, were analyzed and discussed. It was highlighted that; biological surfaces present a virtually endless potential of technological ideas for the development of new materials and systems. In this context, the innovative aspect of laser functionalized biomimetic surfaces for numerous current and potential applications was particularly demonstrated.

Although current laser fabrication techniques are valuable within a limited range of applications, the present effort is still far from the ultimate goal of realizing large-scale artificial structures that fully mimic the architecture complexity and functionality of natural surfaces, in a convenient, rapid and cost-effective manner. Accordingly, there are still many challenges to overcome. Indeed, to mimic the complex hierarchical structure of natural surfaces, future fabrication methods must incorporate precise control over different size scales in a single step, i.e. tailored fabrication of nano-features imprinted on the surface of micro-features. Another challenge is the precision fabrication of highly complex textures and the subsequent creation of meta-surfaces incorporating micro-/nano- features of multiple spatial frequencies and orientations. Finally, a big challenge for laser technology is to combine both artificially generated surface topologies with *ad-hoc* designed surface chemistry, as it is commonly found in nature, in a single processing step.

To date, laser-based surface processing methods can manufacture structures at scales down to 100 nm, so there is a demand for controllable fabrication at the nanometer scale. This is one of the biggest challenges towards engineering truly biomimetic hierarchically structured artificial surfaces and to take full advantage of the potential of nanofeature incorporation. A key issue towards improving laser structuring resolution is the minimization of thermal damage effects including melting, burr formation and cracking that limit the resolution of the fabricated structures. Femtosecond (fs) lasers present unique capabilities in this respect and it is encouraging that recent significant improvements in the fs sources available have accelerated the acceptance of such lasers as a strong option for large-scale and high-throughput fabrication. The lasers available today offer vastly improved peak powers and reliability, offering the possibility to fs processing technology into a range of surface engineering applications. Furthermore, high repetition rate fs laser systems are extensively developed and attract much attention as new light sources due to the potential increase of the fabrication rate and thus production throughput.  Accordingly, fs laser texturing presents unique capabilities for large-scale nanostructures production, opening new opportunities for innovation in surface functionalization and a new paradigm in surface coatings industry.

There are also emerging aspects of laser based fabrication techniques, which may be exploited for expanding the complexity and novelty of biomimetic textures. For example, the use of temporally shaped ultrafast pulses may provide an additional route for controlling and optimizing the outcome of processing. Likewise, the exploitation of filamentation effects, produced by ultrafast lasers. No doubt, all these techniques require further development before they can become competitive. However, the wealth of arising possibilities in laser based micro and nanofabrication and the number of new approaches to bioinspiration prescribe a future where biomimetic control of biomimetic artificial surfaces and subsequent functionality can be accomplished with a level of sophistication that we cannot presently envisage.

With regard to future prospects, the coming years may primarily belong to the laser fabrication of bioinspired multifunctional surfaces, i.e, surfaces that combine multiple functionalities. Multifunctional surfaces are important for a broad range of applications in engineering sciences, including adhesion, friction, wear, lubrication, filtering, sensorics, wetting phenomena, self-cleaning, antifouling, thermoregulation, optics, to name a few. Research on responsive and intelligent materials, combined with the use of the principles of evolution for optimization, is expected to give rise to multifunctional biomimetic surfaces that currently remain beyond our grasp. No doubt, research in this field involves synergies from biology, physics, chemistry, materials science and engineering and therefore represents an excellent example of modern interdisciplinary science.

# 6 Acknowledgements

This work was supported by the European Union's Horizon 2020 research and innovation program through the project "BioCombs4Nanofibers" [grant agreement No. 862016] and "BioProMarL" [grant agreement No. 852048], the Excellence Initiative of the German federal and state governments and the Spanish Research Agency (AEI, Ministry of Research and Innovation) jointly with the European Regional Development Fund (ERDF) through project "UDiSON [grant No. TEC2017-82464-R].